\documentclass[pre,11pt, amsmath, amssymb,floatfix]{revtex4}

\usepackage{graphicx}
\usepackage{dcolumn} %
\usepackage{bm}      %

\usepackage{setspace}
\usepackage{enumitem}

\usepackage{subfigure}
\usepackage{amsfonts}
\usepackage[usenames]{color}
\usepackage{rotating}
\usepackage{fontenc}
\usepackage{cancel}
\usepackage{amsthm}
\usepackage{hyperref}
\usepackage[normalem]{ulem}

\hypersetup{CJKbookmarks,%
       bookmarksnumbered,%
              colorlinks,%
               linkcolor=black,%
               citecolor=black,%
              plainpages=false,%
            pdfstartview=FitH}

\newcommand{\be}{\begin{equation}}
\newcommand{\ee}{\end{equation}}
\newcommand{\bea}{\begin{eqnarray}}
\newcommand{\eea}{\end{eqnarray}}

\definecolor{darkred}{RGB}{139,0,0}
\definecolor{chartreuse}{RGB}{127,255,0}
\definecolor{goldenrod}{RGB}{218,165,32}
\definecolor{gray}{RGB}{127,127,127}

\definecolor{Magenta}{RGB}{255, 0,255}
\definecolor{Orange}{RGB}{255,165, 0}
\definecolor{Gray}{RGB}{127,127,127}

\mathchardef\mhyphen="2D
\newcommand{\mm}{\mathrm}

\begin{document}

\title{
Inducing Effect on the Percolation Transition in Complex Networks
}

\author{Jin-Hua Zhao$^1$,
        Hai-Jun Zhou$^1$\footnote{Corresponding author. Email: {\tt zhouhj@itp.ac.cn}},
 and Yang-Yu Liu$^{2,3}$}

\affiliation{$^1$State Key Laboratory of Theoretical Physics,
Institute of Theoretical Physics,
Chinese Academy of Sciences,
Zhong-Guan-Cun East Road 55, Beijing 100190, China
}
\affiliation{$^2$Center for Complex Network Research and Department of
Physics, Northeastern University, Boston, Massachusetts 02115, USA}

\affiliation{$^3$Center for Cancer Systems Biology, Dana-Farber Cancer Institute,
Boston, Massachusetts 02115, USA}

\date{\today}

\begin{abstract}
  Percolation theory concerns the emergence of connected clusters that
  percolate through a networked system.
  Previous studies ignored the
  effect that a node outside the percolating cluster may actively induce its
  inside neighbours to exit the percolating cluster.
  Here we study this inducing effect on the classical site percolation and
  $K$-core percolation, showing that the
    inducing effect always causes a discontinuous
    percolation transition.
  We precisely predict the percolation threshold and core size
  for uncorrelated random networks with arbitrary degree distributions.
  For low-dimensional lattices the percolation threshold fluctuates
  considerably over realizations, yet we can still predict the
  core size once the percolation occurs. The core sizes of real-world
  networks can also be well predicted using degree distribution as the only input.
  Our work therefore provides a theoretical framework for quantitatively
  understanding discontinuous breakdown phenomena in various complex
  systems.
\end{abstract}

\maketitle

Percolation transition on complex networks occurs in a wide range of natural,
technological and socioeconomic systems
\cite{Stauffer-Aharony-1994,Dorogovtsev-Goltsev-Mendes-2008,Buldyrev-etal-2010}.
The emergence of macroscopic network connectedness, due to either gradual
addition or recursive removal of nodes or links,
can be related to many fundamental network properties, e.g., robustness and
resilience
\cite{Albert-Jeong-Barabasi-2000,Cohen-etal-2000},
cascading failure
\cite{Watts-2002,Buldyrev-etal-2010,Li-etal-2012},
epidemic or information spreading
\cite{PastorSatorras-Vespignani-2001,Kitsak-etal-2010,Gleeson-2011},
and structural controllability
\cite{Liu-Slotine-Barabasi-2011,Liu-etal-2012}.
Particularly interesting are the emergence of a giant connected component
\cite{Erdos-Renyi-1960,Callaway-etal-2000,Bollobas-2001,Achlioptas-DSouza-Spencer-2009,Riordan-Warnke-2011,Nagler-Levina-Timme-2011,Nagler-Tiessen-Gutch-2012,Boettcher-Singh-Ziff-2012,Cho-etal-2013},
the $K$-core (obtained by recursively removing nodes with degree less than $K$)
\cite{Chalupa-Leath-Reich-1979,Pittel-etal-1996,Dorogovtsev-etal-2006,Baxter-etal-2010},
and the core (obtained by recursively removing nodes of degree one and
their neighbours) \cite{Karp-Sipser-1981,Bauer-Golinelli-2001,Liu-etal-2012}.

These classical percolation processes are \emph{passive} in the sense
that whether or not a node belongs to the percolating cluster
depends only on its number of links to the percolating
cluster.
However, in many physical or information systems,  each
  node has an intrinsic state and
after a node updates its state, it can  \emph{actively induce}
its neighbours to update their states too.
One example is the frozen-core
formation in Boolean satisfiability problems \cite{Mezard-Zecchina-2002},
where non-frozen nodes can induce its frozen neighbours into
the non-frozen state (the so-called whitening process
\cite{Parisi-2002,Seitz-Alava-Orponen-2005,Li-Ma-Zhou-2009}).
In the glassy dynamics of
kinetically constrained models,
a spin in a certain state facilitates the flipping
of its neighbouring spins \cite{Ritort-Sollich-2003}.
In inter-dependent networks, a collapsed node of one network causes
the failure of the connected dependent node in the other network
\cite{Buldyrev-etal-2010,Brummitt-DSouza-Leicht-2012},
resulting in a damage cascading process.
The inducing effect can also be related to information or
opinion spreading, e.g., an  early adopter of a new product or
  innovation might  persuade his or her friends to adopt
it either.

Despite its implications on a wide range of
important problems, the inducing effect on percolation transitions has
not been fully understood. In this work, we study the inducing
effect on the classical site percolation and
$K$-core percolation in complex networks.
We analytically show that the inducing effect always causes a
discontinuous percolation transition, therefore
providing a
new perspective on abrupt breakdown phenomena in
complex networked systems.
Our analytical calculations are confirmed by extensive numerical
simulations.

\section{Results}

\subsection{Description of the model.}

We assume each node of the network
has a binary internal state: protected or unprotected.
We allow an initial $p$ fraction of nodes
randomly chosen from the network to be protected. If $p=1$,
all the nodes are initially protected.
As time evolves,
a protected node \emph{spontaneously} becomes unprotected
if it has less than $K$ protected neighbours.
(In case of $K=0$, a protected node will never
spontaneously become unprotected.)
A protected node with $K$ or more protected  neighbours will be \emph{induced} to
the unprotected state if at least one of its unprotected
neighbours has less than $K^\prime$ protected neighbours.
(In case of $K^\prime=0$ or $1$, the inducing effect is absent and
our model reduces to the classical site percolation or $K$-core percolation.)
Note that once a node becomes unprotected it  will remain unprotected.
%

% figure 1
\begin{figure}
\begin{center}
\includegraphics[width=55mm]{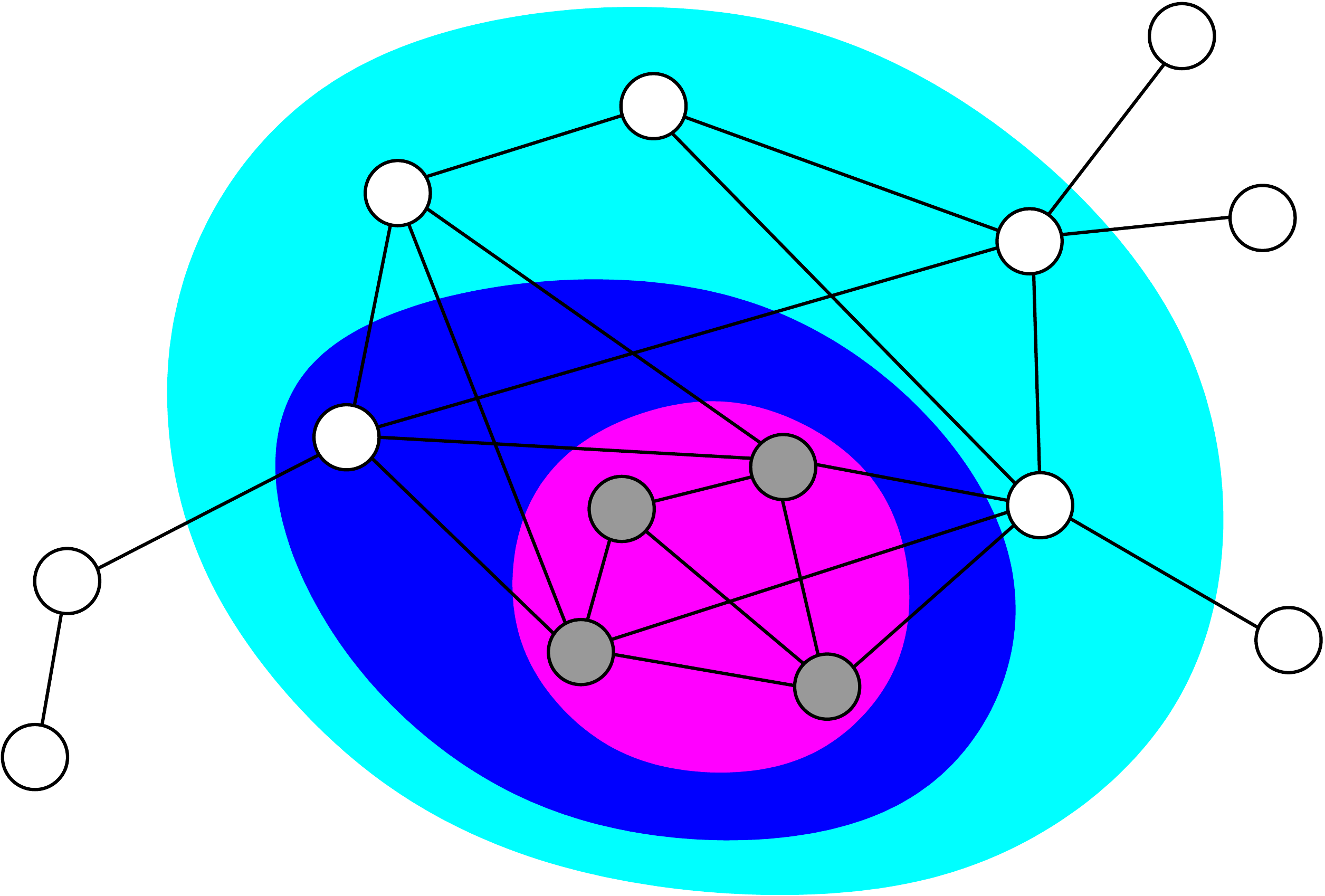}
\end{center}
\caption{\label{fig:example}
{\bf The $(2,2)$-protected core of a small network.}
The $(2,2)$-protected core (magenta region) is contained in  the
core (blue), which is then contained in the $2$-core (cyan).
Protected and unprotected nodes are colored in gray and white, respectively.
}
\end{figure}

We refer to the above-mentioned evolution process as the
$(K, K^\prime)$-protected core percolation.
The $(K, K^\prime)$-protected core, or simply, the
   protected core is the subnetwork formed by all the
surviving protected nodes and
the links among them  (see Fig.~\ref{fig:example} for an
example).
We denote the total number of nodes in
the protected core as
$N_\mm{p\mhyphen core}$.
We can prove that the protected core is independent of
the particular state evolution trajectory of the
nodes and hence is well defined (see Supplementary note 1).

In the context of opinion spreading or viral marketing, the $(K,
K^\prime)$-protected core percolation can be described as follows:
Consider a population of users to adopt a new product (or idea,
opinion, innovation, etc.). Initially there is a $p$ fraction of users in
the ``protected'' (or conservative) state and refuse to adopt the new
product. The other $(1-p)$ fraction of users are in the
``unprotected'' state, i.e., they are early adopters.
A conservative user will automatically adopt the new product if he/she
has less than $K$ conservative friends.
An adopted user with less than $K^\prime$ conservative friends will persuade
all his or her conservative friends to adopt the new product.
Then the protected core, if exists, can be viewed as the subnetwork of the
most conservative individuals, who will never adopt the new
product.

\subsection{Analytical approach.}

Consider a large uncorrelated random network
containing $N$ nodes, with arbitrary degree distribution $P(k)$ and
mean degree $c=\sum_{k \geq 0} k P(k)$
  \cite{Newman-Strogatz-Watts-2001,Molloy-Reed-1995}.
We assume that if any node $i$ is still
in the protected state, its neighbours do not mutually
influence each other and therefore their states are independently
distributed. This is a slight extension of the Bethe-Peierls approximation
widely used in spin-glass theory and statistical inference
\cite{Mezard-Montanari-2009}.
Note that a closely related approximation in network science is the tree
approximation~\cite{Callaway-etal-2000,Cohen-etal-2000,Newman-Strogatz-Watts-2001},
which assumes the neighbours of node $i$ become disconnected if
$i$ is removed from the network.
Under our assumption of state independence we can calculate the normalized size
$n_\mm{p\mhyphen core}$ ($\equiv N_\mm{p\mhyphen core}/N$) of the
protected core as
\begin{equation}
\label{eq:ncore}
n_\mm{p\mhyphen core}
= p  \sum\limits_{s\geq K} \sum\limits_{k\geq s} P(k)
C_{k}^{s}        (1-\alpha -\beta )^{s} \beta^{k-s}\; ,
\end{equation}
with $C_{k}^{s} \equiv k! /[s! (k-s)!]$ being the binomial coefficient
(Supplementary note 2).
The parameter $\alpha$ denotes the probability that,
starting from a node $i$ that is still in the protected state, a node $j$
reached by following a randomly chosen link $(i, j)$
is in the unprotected
state and having at most $K^\prime-1$
protected neighbours (including $i$).
The parameter $\beta$ is the probability that such a node $j$ is in the unprotected
state but having at least $K^\prime$ protected neighbours.
We further define $\gamma$ as the probability that
such a node $j$ is in the protected state and having
exactly $K$ protected neighbours.
Note that if initially we randomly choose a finite $p$ fraction of nodes to
be protected, then $(1-p)$ fraction of the nodes will be and remain
unprotected.
Let us define $\eta$ as the probability that, starting
from such an initially unprotected node
$m$, a node $n$ reached by following a randomly chosen link $(m, n)$
will eventually be in the unprotected state
even if the inducing effect of node $m$ is not considered.

Because of the inducing effect, each node $j$ mediates strong
correlations among the states of its neighbouring nodes if it is in
the unprotected state.
After a careful analysis of all the possible microscopic inducing patterns
following the theoretical method of \cite{Zhou-2005a,Zhou-2012},
we obtain a set of self-consistent
equations for the  probabilities $\alpha$, $\beta$, $\gamma$ and $\eta$:
\begin{eqnarray}
\alpha
&=&
  (1-p)
  \sum\limits_{s=0}^{K^\prime -2}
      \sum\limits_{k\geq s+1} Q(k) C_{k-1}^{s}  (1-\eta)^s \eta^{k-1-s}
\nonumber \\
& & + p \ \biggl\{
         \sum\limits_{s=0}^{K-2} \sum\limits_{k\geq s+1}
         \sum\limits_{r=0}^{\min(s,K^\prime -2)}
         Q(k) C_{k-1}^{s} C_{s}^{r} (\alpha + \beta)^{k-1-s}
         (1-\alpha-\beta-\gamma)^{r} \gamma^{s-r} \nonumber \\
& &
        + \sum\limits_{s\geq K-1}\sum\limits_{k\geq s+2}
          \sum\limits_{r=0}^{\min(s,K^\prime -2)}
          Q(k) C_{k-1}^{s} C_{s}^{r} \bigl[
          (\alpha+\beta)^{k-1-s}-\beta^{k-1-s} \bigr]
          (1-\alpha-\beta-\gamma)^{r} \gamma^{s-r}
        \biggr\} \; , \nonumber \\
& &
 \label{eq:alphaprob1}
 \end{eqnarray}
 \begin{eqnarray}
\beta & = &
 (1-p)
  \sum\limits_{s \geq K^\prime -1}
  \sum\limits_{k\geq s+1} Q(k) C_{k-1}^{s}  (1-\eta)^s \eta^{k-1-s}
\nonumber \\
&  + & p\ \biggl\{
         \sum\limits_{s = K^\prime -1}^{K-2} \sum\limits_{k\geq s+1}
         \sum\limits_{r=K^\prime -1}^{s}
         Q(k) C_{k-1}^{s} C_{s}^{r} (\alpha + \beta)^{k-1-s}
         (1-\alpha-\beta-\gamma)^{r} \gamma^{s-r} \nonumber \\
&   + & \sum\limits_{s\geq \max(K,K^\prime)-1}\sum\limits_{k\geq s+2}
          \sum\limits_{r=K^\prime -1}^{s}
          Q(k) C_{k-1}^{s} C_{s}^{r} \bigl[
          (\alpha+\beta)^{k-1-s}-\beta^{k-1-s} \bigr]
          (1-\alpha-\beta-\gamma)^{r} \gamma^{s-r}
        \biggr\} \; , \nonumber \\
& &
\label{eq:betaprob1} \\
\gamma &= & p \sum\limits_{k \geq K} Q(k)
C_{k-1}^{K-1}         (1-\alpha -\beta )^{K-1} \beta^{k-K} \; ,
\label{eq:gammaK1} \\
\eta & = & 1- p +
  p \ \biggl\{ \sum\limits_{s=0}^{K-1} \sum\limits_{k\geq s+1} Q(k) C_{k-1}^{s}
  (1-\alpha-\beta)^s (\alpha+\beta)^{k-1-s} \nonumber \\
  & & \quad\quad \quad\quad \quad\quad
  + \sum\limits_{s\geq K} \sum\limits_{k\geq s+2} Q(k)
  C_{k-1}^{s} \bigl[(\alpha+\beta)^{k-1-s} - \beta^{k-1-s} \bigr]
  (1-\alpha-\beta)^{s}
  \biggr\} \; ,
  \label{eq:eta}
\end{eqnarray}
%\endwidetext
%
\noindent
where $Q(k) \equiv kP(k)/c$ is the degree distribution for
the node at an end of a randomly chosen link.
These equations can be understood as follows.
The first term on the r.h.s. of
Eq.~(\ref{eq:alphaprob1}) is the probability that a node $j$ reached by
following a link $(i,j)$ is initially unprotected and having at most $K^\prime-2$ protected neighbours (excluding
 node $i$)  without considering its inducing effect. The other two
  terms in the r.h.s. of Eq.~(\ref{eq:alphaprob1}) yield
the probability that an initially protected node $j$
at the end of a link $(i,j)$ will either spontaneously transit to or be induced to
the unprotected state and,  when it is still in the protected state,
at most $K^\prime-2$ of its protected neighbours
(excluding node $i$) have more than $K$ protected neighbours
themselves.
The terms in Eqs.~(\ref{eq:betaprob1})-(\ref{eq:eta}) can be understood similarly
(see Supplementary note 2 for more explanations).

The above self-consistent equations can be solved using a simple
iterative scheme (see Supplementary note 3). When $K, K^\prime \geq 2$,
these equations always have a trivial solution
$(\alpha,\beta,\gamma,\eta)=(1,0,0,1)$, yielding no protected core
($n_\mm{p\mhyphen core}=0$).
This solution is always
locally stable,
and it is the only solution if the mean degree $c$ of the network is small or the
initial fraction $p$ of protected nodes is small
(see Supplementary note 4).
As $c$ (or $p$) increases, another stable solution of Eqs.~(\ref{eq:alphaprob1})-(\ref{eq:eta})
appears at the critical mean degree $c= c^*$
(or the critical fraction $p=p^*$),
  corresponding to the percolation transition.
In the limiting cases of $K \in \{0, 1\}$, Equations~(\ref{eq:alphaprob1})-(\ref{eq:eta}) also change from having
only one stable solution to having
two distinctive stable solutions at certain critical value $c=c^*$ or $p=p^*$
(see Supplementary note 4).

\subsection{The minimal inducing effect.}

% figure 2
\begin{figure}[b]
\begin{center}
\includegraphics[width=85mm]{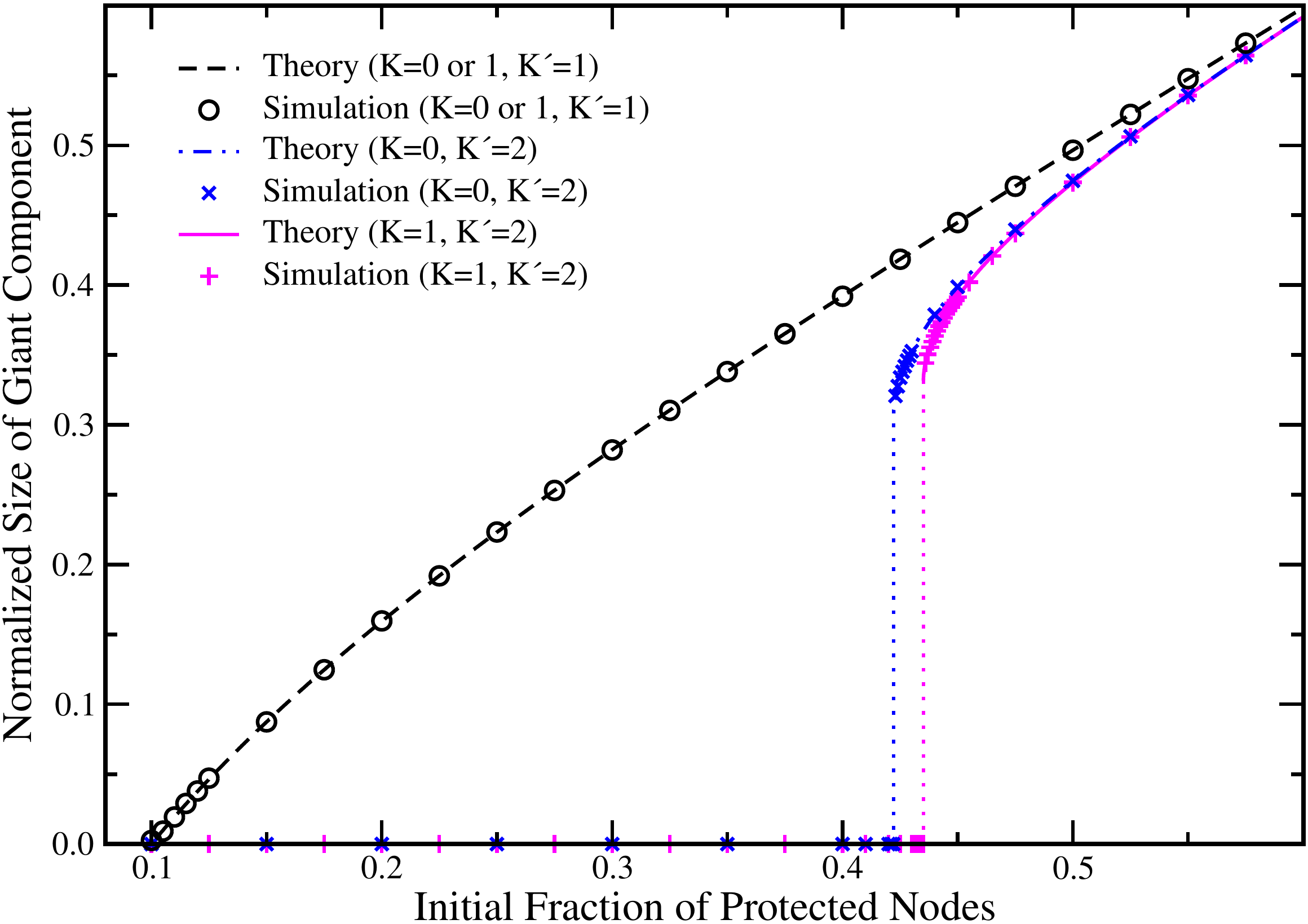}
\end{center}
\caption{ \label{fig:K1Kp2Giant}
  {\bf Normalized size of the giant connected component of protected nodes}.
  Symbols are simulation results on a single ER random network of $N=10^6$
  nodes and mean degree $c=10$, while the lines are theoretical predictions at $N=\infty$. The giant connected component of protected nodes continuously
  emerges in the $(0,1)$- and $(1, 1)$-protected core
  percolation problems (without inducing
  effect), but it emerges discontinuously in the $(0,2)$- and
  $(1,2)$-protected core percolation
  problems (with minimal inducing effect).
 }
\end{figure}

The minimal inducing effect on percolation transitions can be
demonstrated by comparing $(0,1)$- and $(1,1)$-protected core percolation
transitions with $(0,2)$- and $(1,2)$-protected core
percolation transitions as we tune the initial fraction of protected
node $p$.
Note that the $(K,1)$-protected core percolation with $K\in \{0,1\}$
is essentially the classical site
percolation~\cite{Stauffer-Aharony-1994,Cohen-etal-2000,Callaway-etal-2000},
because a protected node will
remain protected if it has at least one protected
neighbour and there is no inducing effect at all.
In this case, a giant connected component of protected nodes gradually
emerges in the network as $p$ exceeds $p^*=1/[\sum_{k\geq 1} (k-1) Q(k)]$ (see
Fig.~\ref{fig:K1Kp2Giant}).
The minimal inducing effect is naturally present
in the $(0,2)$- and $(1, 2)$-protected core percolation problems, namely if an
unprotected node has only one protected neighbour, this neighbour will
be induced to the unprotected state.
In this case our analytical calculation shows that both the normalized size
of the protected core and that of its giant connected component
 will jump from zero to a finite positive value at certain
critical value $p^*$
(see Supplementary notes 4 and 5). For Erd\"{o}s-R\'{e}nyi
(ER) random networks \cite{Albert-Barabasi-2002,He-Liu-Wang-2009}
with mean degree $c=10$, this
threshold fraction is $p^* \approx 0.44$ (for $K=1$) and
$p^* \approx 0.42$ (for $K=0$), which are much larger than the threshold value
$p^* = 0.1$ of the classical continuous site percolation
transition (see Fig.~\ref{fig:K1Kp2Giant}).
Note that in case $K=0$, a protected node will never
  spontaneously become unprotected, hence the discontinuous
  $(0,2)$-protected core percolation transition is solely due to the
  inducing effect.

\subsection{Inducing effect on $K$-core percolation.}

The inducing effect can also be demonstrated by comparing the $K$-core
percolation
and the $(K,K^\prime)$-protected core percolation as we tune the
mean degree $c$. In the following discussions we
set $p=1$ and focus on the representative case of $K^\prime = K$
(the results for $p<1$ and $2 \leq K^\prime \neq K$ are qualitatively the same).
And we refer
to $(K, K)$-protected core simply as $K$-protected core.

% figure 3
\begin{figure}[b]
\begin{center}
\includegraphics[width=85mm]{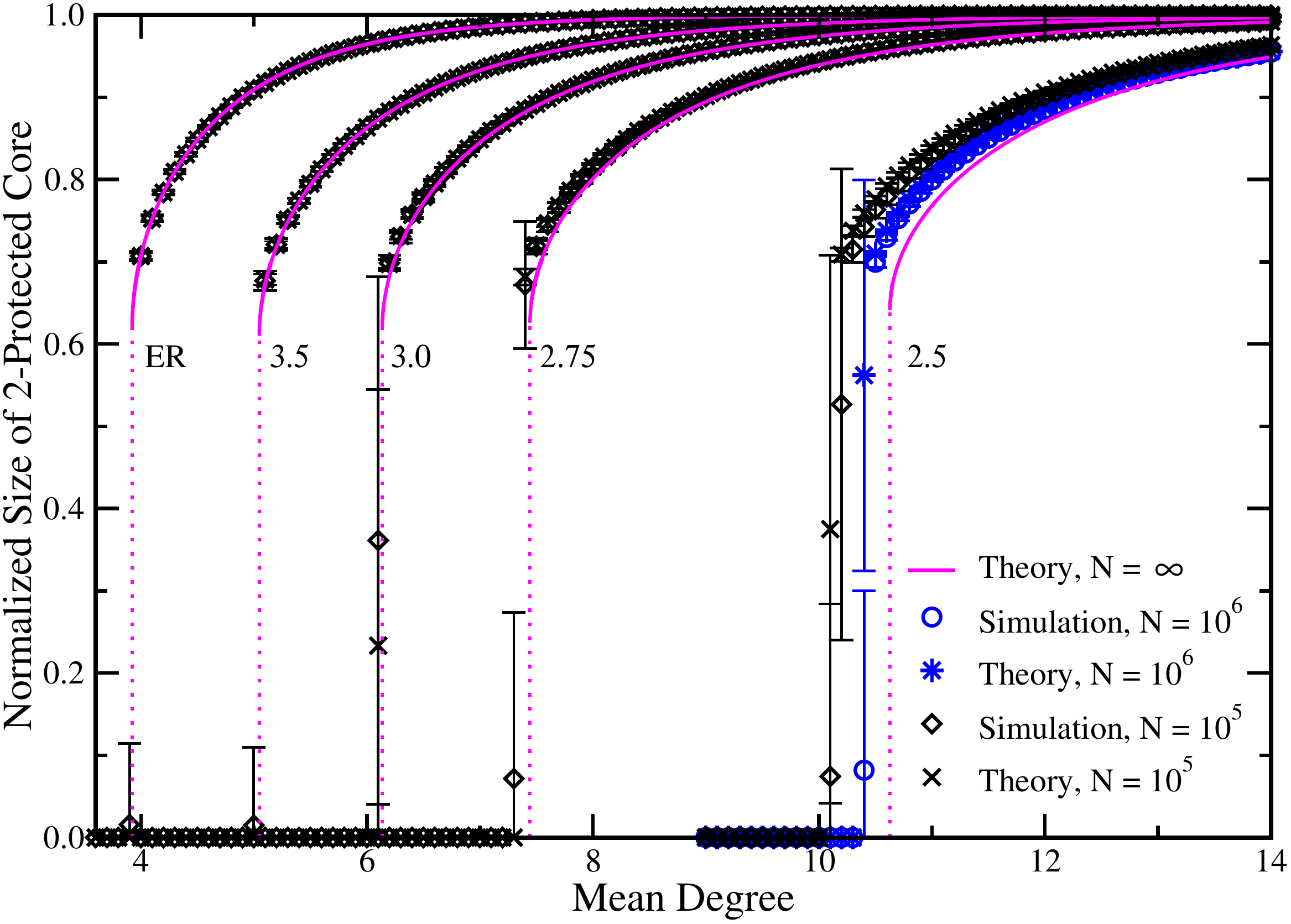}
\end{center}
\caption{\label{fig:KJammCore2}
  {\bf Normalized size of $2$-protected core for ER
  networks and SF networks.} The degree exponents of the SF networks
  are $\lambda = 3.5, 3.0, 2.75, 2.5$ (from left to right).
  Lines are analytic predictions for infinite system ($N=\infty$),
  circles and diamonds are exact results obtained through the state evolution process;
  star and cross symbols are the analytic results using the exact degree sequences of
  the constructed networks. Each simulation point is obtained by averaging over $80$
  independent network instances.
Note that for $\lambda < 3$, especially when $\lambda \to 2$,
significant finite-size effect is observed. This is rooted in the intrinsic degree correlations in the static model when $\lambda < 3$~\cite{Goh-Kahng-Kim-2001,Catanzaro-PastorSatorras-2005,Lee-etal-2006}.
}
\end{figure}
%

% figure 4
\begin{figure}
\begin{center}
\includegraphics[width=85mm]{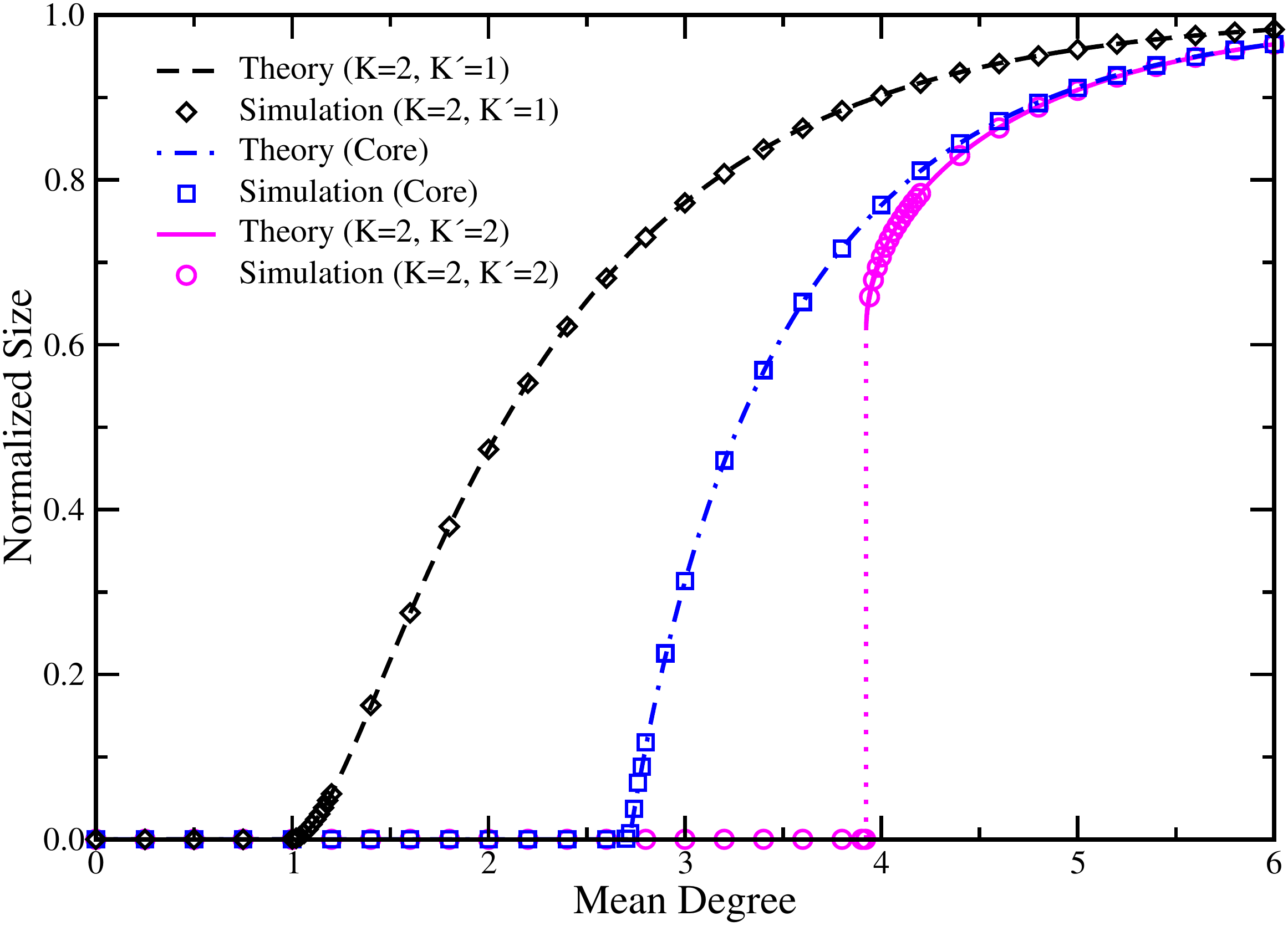}
\end{center}
\caption{\label{fig:Core2Core2pCoreER}
  {\bf Comparing the percolation transitions.}
  Symbols are simulation results on a single ER random network of
  $N=10^6$ nodes, while the lines are theoretical predictions at
  $N=\infty$. Both the $2$-core (equivalent to the $(2,1)$-protected core)
  and the core emerges continuously  but
  the $2$-protected core emerges discontinuously.
}
\end{figure}

We find that for any $K\ge 2$, as $c$ reaches the
critical value $c^*$, $n_\mm{p\mhyphen core}$ jumps from $0$ to a
finite value $n_\mm{p\mhyphen core}^*$ (see Supplementary note 4),
indicating a discontinuous percolation transition.
 We also find that for any $K\geq 2$ and independent of
network types,
$n_\mm{p\mhyphen core} - n_\mm{p\mhyphen core}^*  \propto (c-c^*)^{1/2}$
in the supercritical regime  where $c-c^* \to 0^+$ (see Supplementary note 6).
Such a hybrid phase transition and the associated critical exponent $1/2$
were also observed in  $K$-core percolation and core
percolation~\cite{Chalupa-Leath-Reich-1979,Pittel-etal-1996,Dorogovtsev-etal-2006,Liu-etal-2012}.

In the following, we study the discontinuous $2$-protected core percolation in
a series of random networks with specific degree distributions.
We first consider the ER random network with
Poisson degree distribution
$P(k) = e^{-c} c^k / k!$.
We find that the discontinuous $2$-protected core
percolation transition occurs at $c=c^* \approx 3.92$, with a jump of $n_\mm{p\mhyphen core}$ from
$0$ to $n_\mm{p\mhyphen core}^* \approx 0.62$ (see Fig.~\ref{fig:KJammCore2}).
Note that for ER random networks the classical $2$-core and core percolation transitions
occur at $c^*=1$  and $c^*=e \approx 2.72$,
respectively, and they are both continuous
 \cite{Liu-etal-2012,Dorogovtsev-etal-2006}.
Hence, allowing unprotected nodes to induce other nodes
not only delays the occurrence of the percolation
transition to
a larger value of $c$ but also makes it discontinuous
(see Fig.~\ref{fig:Core2Core2pCoreER}).

Scale-free (SF) networks characterized by a power-law degree
distribution $P(k) \sim k^{-\lambda}$
with degree exponent $\lambda$ are ubiquitous in
  real-world complex systems \cite{Albert-Barabasi-2002}.
Interestingly, we find that for purely scale-free networks with $P(k)
= k^{-\lambda}/\zeta(\lambda)$ and $\zeta(\lambda)$ the Riemann $\zeta$
function, the $K$-protected core does not exist for any $\lambda>2$
(see Supplementary note 7).
If the smallest degree $k_{\rm min} \geq K$ and a
fraction $\rho$ of the links are randomly
 removed from the purely SF network, then a discontinuous
 $K$-protected core percolation transition will occur (see Supplementary note 7).
 For asymptotically SF networks generated by the static model with
 $P(k) \sim k^{-\lambda}$ for large $k$ only \cite{Goh-Kahng-Kim-2001,Catanzaro-PastorSatorras-2005,Lee-etal-2006},
 the $K$-protected core develops when the mean degree $c$ exceeds a
 threshold value $c^*$.
For this type of random networks with different values of $c$ and
$\lambda$, we compare the theoretical and simulation results
 and find that they agree
well with each other (see Fig.~\ref{fig:KJammCore2}).

\begin{figure}
\begin{center}
\includegraphics[width=85mm]{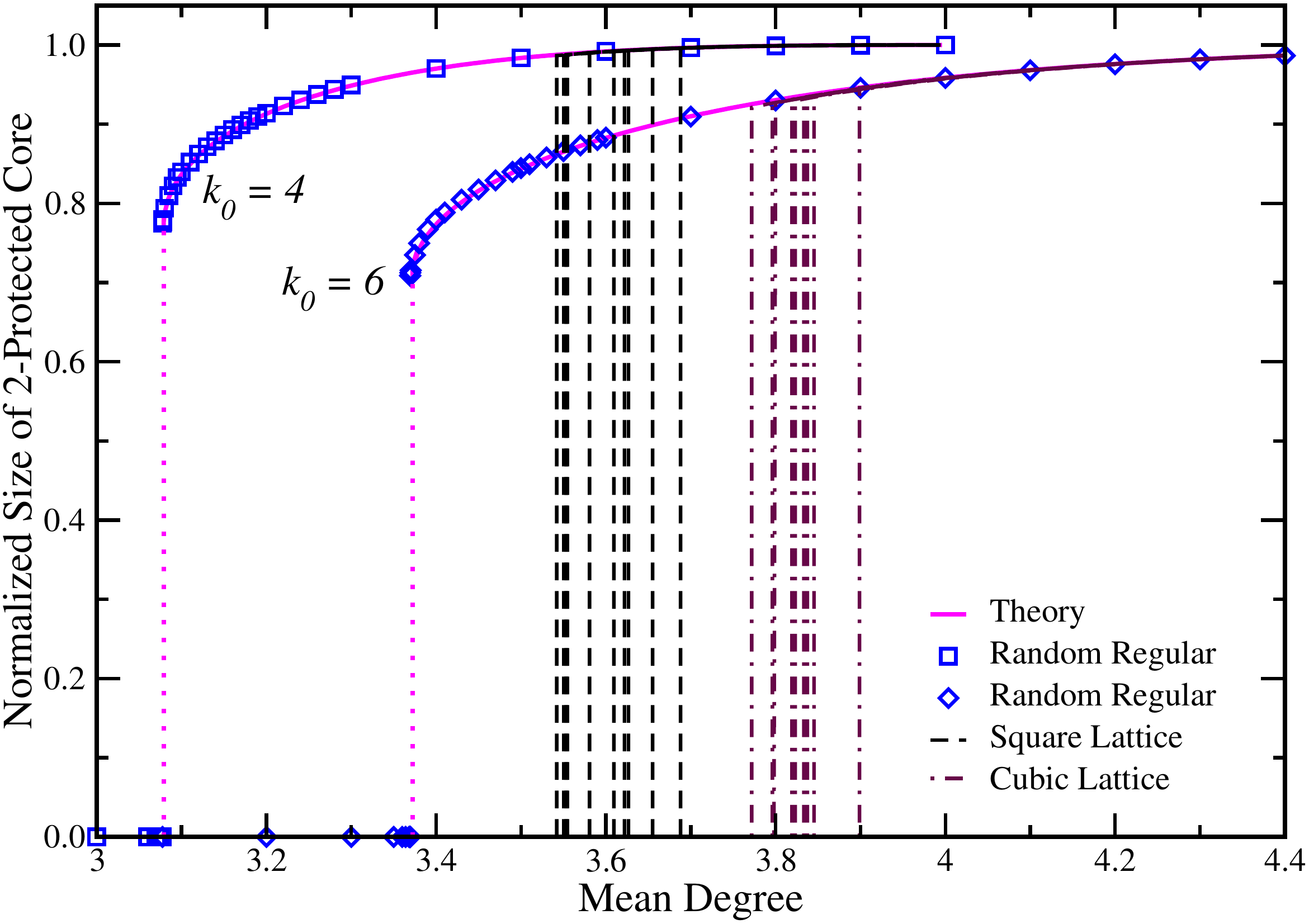}
\end{center}
\caption{\label{fig:KJammCoreRegularK2}
  {\bf Normalized size of $2$-protected core for RR networks and regular lattices.}
  Solid and dotted lines are
  analytic predictions for infinite system.
  Squares and diamonds are simulation results obtained on a
  diluted RR network instance with node degree $k_0=4$ and $k_0=6$,
  respectively, while dashed and long-dashed lines are the
  simulation results obtained on $20$ independent diluted network instances
  of the square and cubic lattice.
  Each simulated network has $N=10^6$ nodes.
}
\end{figure}

For random regular (RR) networks, all the nodes have the same degree
$k_0$, and  the $K$-protected core contains the whole
network when $k_0  \geq K$. If a randomly chosen fraction $\rho$
of the links are
removed, the degree distribution of the diluted network is given by
$P(k) = [k_0! / k! (k_0 - k)!] (1-\rho)^k \rho^{k_0 - k}$ with
mean  degree  $c= (1-\rho) k_0$.
We predict that $n_\mm{p\mhyphen core}^* \approx 0.77$ (for
$k_0=4$) and  $n_\mm{p\mhyphen core}^* \approx 0.71$ (for $k_0=6$)
at  the $2$-protected core
percolation transition, with
$c^* \approx 3.08$ and $c^* \approx 3.37$, respectively.
These predictions are in
full agreement with simulation results
(see Fig.~\ref{fig:KJammCoreRegularK2}).

% figure 5

\begin{figure}
\begin{center}
\includegraphics[width=85mm]{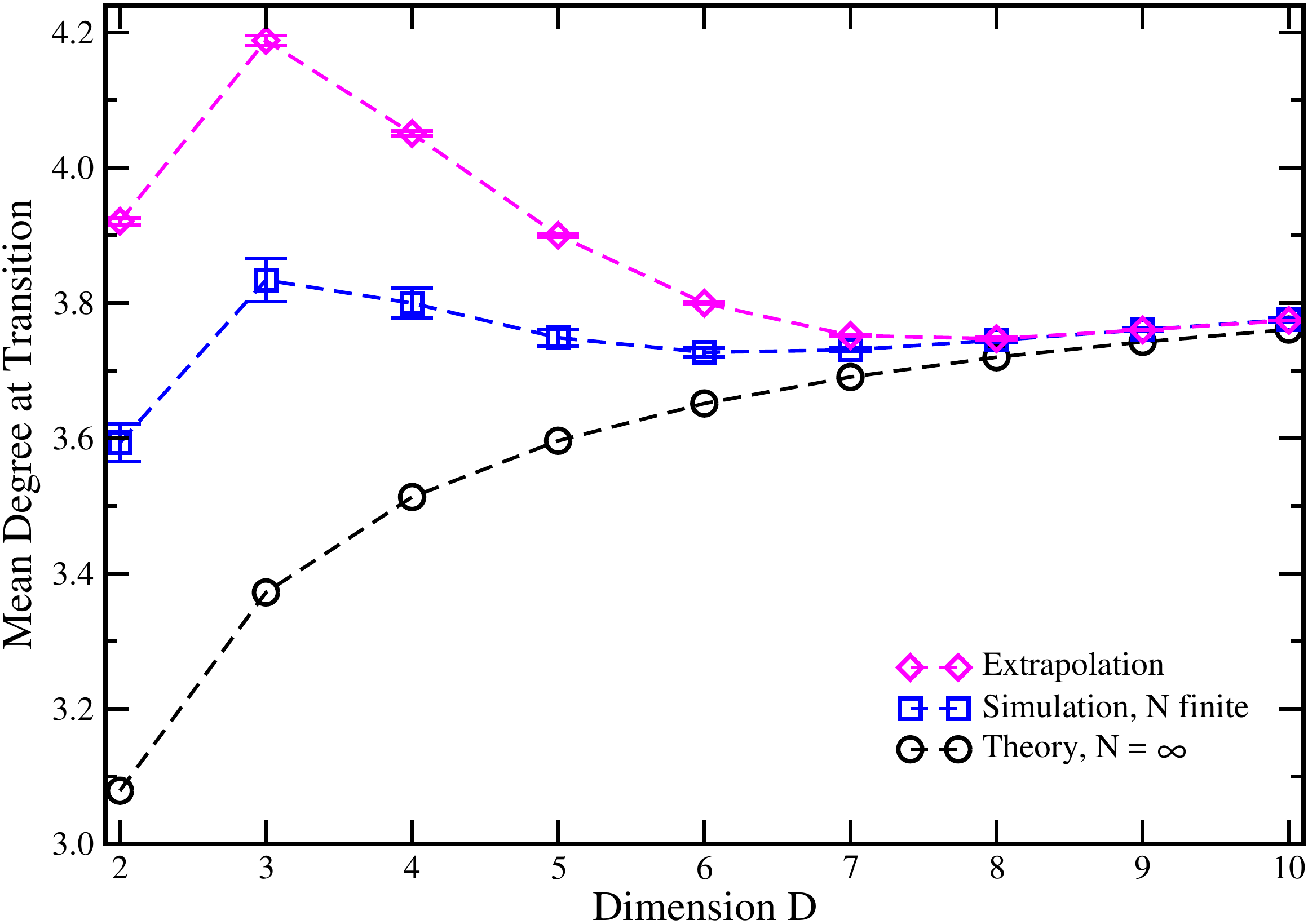}
\end{center}
\caption{ \label{fig:Dimension}
  {\bf The $2$-protected core percolation transition point $c^*$
  for the $D$-dimensional
  hypercubic lattice.} Each square is the value of $c^*$ obtained by averaging
  over $1600$ independent diluted network instances with $N\simeq 2\times 10^6$ nodes,
  diamonds are extrapolated simulation results to $N=\infty$, and
  circles are analytical predictions of $c^*$ for an infinite RR network with
  vertex degree $k_0 = 2 D$.
  The differences between the extrapolated simulation results and the
  theoretical predictions are due to the ignorance of lattice structures in
  the theory.
 }
\end{figure}

 We also study the $2$-protected core percolation in
  diluted $D$-dimensional hypercubic lattice and again find a
  discontinuous transition.
Interestingly, in low dimensions the numerically observed transition
point $c^*$ is remarkably larger than the theoretical prediction
(see Figs.~\ref{fig:KJammCoreRegularK2} and \ref{fig:Dimension}).
We find that
this difference is not a finite-size effect but intrinsic
(it remains in the $N\rightarrow
\infty$ limit),
 and the difference decreases quickly  as
$D$ increases.  The transition point $c^*$ fluctuates
considerably for low dimensions (especially for $D=2, 3$) and
 depends considerably on the system size $N$ (for $D\leq 7$, see
 Supplementary note 8).
 Moreover, there is no critical scaling behavior in
the supercritical regime (similar absence of critical scaling
was also observed in $4$-core percolation on $D=4$ lattices \cite{Parisi-Rizzo-2008}).
Surprisingly, the value of $n_\mm{p\mhyphen core}$ at
and after
the percolation transition agrees well with our
theoretical prediction
(see Fig.~\ref{fig:KJammCoreRegularK2}).

% figure 6

Finally, we apply our theory to a wide range of real-world networks of
different sizes and topologies, and find that for most of these
networks the normalized sizes of the
$2$-protected core can be precisely predicted using
the degree distribution as the only input
(see Supplementary Tables 1 and 2 and Supplementary note 9).

\section{Discussion}

Inducing effect plays an important role in many complex networked
systems. Yet, little was known about how it will affect classical percolation
transitions in complex networks. Here we develop analytical
tools to address this problem for arbitrary network topologies. Our
key finding, that the local inducing effect causes
discontinuous site percolation and
$K$-core percolation (for any $K\ge 1$),
suggests a simple local mechanism to better understand and
ultimately predict many abrupt breakdown phenomena observed in various
systems, e.g., the global failure of a national-wide power grid, the
sudden collapse of a governmental system or a network of financial
institutions.

The results presented here also raise a number of questions, answers
to which could further deepen our understanding of complex networked
systems.
First of all, we can improve the local inducing mechanism
  to be more realistic, e.g., by considering
that the parameters $K$ and $K^\prime$ might be different for
different nodes, an unprotected node may only be able to induce some
particular neighbours (e.g., in a directed network),
or an unprotected node may recover to the
protected state with certain rate, etc..
Secondly,
for low-dimensional lattice systems, the
lattice structures and the associated
short loops cause strong local and long-range correlations among the
states of the nodes, which should be properly
considered in a future refined theory, e.g., by
changing the form of $Q(k)$ to include local degree-degree
correlations and by exactly computing the effects of short loops up to
certain length.
Finally,
an interesting optimization problem consists of identifying a minimal
set of nodes such that  perturbing these nodes to the unprotected state will
cause the protected core of the whole
network to breakdown.
In the context of opinion dynamics or viral marketing,
  this amounts to identifying a minimal set of users for targeted
  advertisement so that we can dissolve the protected core and
  eventually all the users will adopt the new opinion or product.
We hope our work will stimulate further
  research efforts on these and other related
interesting and challenging questions.

\subsection*{Acknowledgement}
J.-H. Zhao and H.-J. Zhou thank Prof. Zhong-Can Ou-Yang for support and
Hong-Bo Jin for technical assistance on computer simulation.
J.-H. Zhao and H.-J. Zhou were supported by the National Basic Research Program of China (No. 2013CB932804), the Knowledge Innovation Program of Chinese
Academy of Sciences (No.~KJCX2-EW-J02), and the National Science Foundation of China
(grant Nos.~11121403, 11225526).
Y.-Y. Liu was supported by the Network Science Collaborative
Technology Alliance under Agreement Number
W911NF-09-2-0053,
the Defense Advanced Research Projects Agency under Agreement Number 11645021,
 the Defense Threat Reduction Agency-WMD award numbers HDTRA1-08-1-0027 and HDTRA1-10-1-0100,
 and the generous support of Lockheed Martin.

{\bf Author Contributions}
H.-J. Zhou conceived research;
H.-J. Zhou, J.-H. Zhao and Y.-Y. Liu performed research;
H.-J. Zhou and Y.-Y. Liu wrote the paper.

{\bf Competing Interests} The authors declare that they have no
competing financial interests.

{\bf Correspondence} Correspondence should be addressed to H.-J. Zhou~(email: zhouhj@itp.ac.cn).

\clearpage

\begin{center}
\begin{table*}[h!]
\caption{
{\bf List of $37$ real-world networks analyzed in this work.}
For each network, we show its type, name, reference and brief description.}
\begin{tabular}{lll}
\hline
\hline
& Name & Description\\
\hline%
%\color{red}
Regulatory
&  TRN-Yeast-1 [51]  & Transcriptional regulatory network of \emph{S. cerevisiae} \\
&  TRN-Yeast-2 [52]     & Same as above (compiled by different group).  \\
&  TRN-EC-1 [53]      & Transcriptional regulatory network of \emph{E. coli} \\
&  TRN-EC-2 [52]           & Same as above (compiled by different group).   \\
&  Ownership-USCorp [54]      & Ownership network of US corporations.\\
\hline%
%\color{green}
Trust
&  College student [55,56]  & Social networks of positive sentiment (college students).\\
&  Prison inmate [55,56]  & Same as above (prison inmates). \\
&  Slashdot [57]         & Social network (friend/foe) of Slashdot users. \\
&  WikiVote [57]     &  Who-vote-whom network of Wikipedia users.\\
&  Epinions [58]   & Who-trust-whom network of Epinions.com users. \\
\hline%
%\color{blue}
Food Web
&  Ythan [59]    &  Food Web in Ythan Estuary.\\
&  Little Rock [60]   & Food Web in Little Rock lake.\\
&  Grassland [59]      & Food Web in Grassland.\\
&  Seagrass [61]      &  Food Web in St. Marks Seagrass.\\
\hline%
%\color{Magenta}
Power Grid
&  TexasPowerGrid [62]     & Power grid in Texas. \\
\hline%
%\color{cyan}
Metabolic
&  \emph{E. coli} [63]              &  Metabolic network of \emph{E. coli}.  \\
&  \emph{S. cerevisiae} [63]         &  Metabolic network of \emph{S. cerevisiae}.  \\
&  \emph{C. elegans} [63]          &  Metabolic network of \emph{C. elegans}.  \\
\hline
%\color{goldenrod}
Electronic
&  s838 [52]          & Electronic sequential logic circuit. \\%(from ISCAS89 benchmark set)   \\
%\color{goldenrod}
Circuits
&  s420 [52]         &  Same as above.  \\
&  s208 [52]           &  Same as above. \\
\hline
%\color{black}
Neuronal
&  \emph{C. elegans} [64]         & Neural network of \emph{C. elegans}. \\
\hline
%\color{Orange}
Citation
&   ArXiv-HepTh [65]   &  Citation networks in HEP-TH category of Arxiv.   \\
&   ArXiv-HepPh [65]    &  Citation networks in HEP-PH category of Arxiv.  \\
\hline
%\color{Gray}
WWW
&  nd.edu [66]     & WWW from nd.edu domain.  \\
&  stanford.edu [57]  &WWW from stanford.edu domain.   \\
&  Political blogs [67]   &Hyperlinks between weblogs on US politics.    \\
\hline
%\color{red}
Internet
&  p2p-1 [68]    & Gnutella
peer-to-peer file sharing network.\\
&  p2p-2 [68]  &  Same as above (at
different time). \\
&  p2p-3 [68]    &  Same as above (at
different time). \\
\hline
%\color{green}
Social
&  UCIonline [69] & Online message network of students at UC, Irvine. \\
%\color{green}
Communication  &  Email-epoch [70]      &  Email network in a university.\\
&  Cellphone [71]      & Call network of cell phone users.\\
\hline
%\color{blue}
Intra-
&  Freemans-2 [72]              &  Social network of network researchers.\\
%\color{blue}
organizational&  Freemans-1 [72]     & Same as above (at different time). \\
&  Manufacturing [73]       & Social network from a manufacturing company.\\
&  Consulting [73]           & Social network from a consulting company. \\
\hline
\hline%
\end{tabular}
\label{tab:realnetwork}
\end{table*}
\end{center}

\newpage

\begin{center}
\begin{table*}[h!]
\caption{{\bf $2$-protected core size of $37$ real-world networks.}
For each network, we show its type, name; number of nodes
  ($N$) and links ($M$); number of protected nodes in the $2$-protected core ($N_\mm{p\mhyphen core}^{\rm real}$); normalized size of
  $2$-protected core ($n_\mm{p\mhyphen core}^{\rm real}$), and
  theoretical prediction of normalized size of $2$-protected core
  ($n_\mm{p\mhyphen core}^{\rm theory}$). In the last two columns, if
  empirical value and theoretical prediction are significantly different,
  these items are highlighted by bold type.
}
\begin{tabular}{llccccccc}
\hline
\hline
& Name & $N$ & $M$ & $N_\mm{p\mhyphen core}^{\rm real}$ &
$n_\mm{p\mhyphen core}^{{\rm real}}$ & $n_\mm{p\mhyphen core}^{{\rm theory}}$ \\
\hline%
%\color{red}
Regulatory
&  TRN-Yeast-1 &  4,441    &    12,864  &  0 & 0     & 0 \\
&  TRN-Yeast-2 &    688    &    1,078   &  0 & 0      & 0 \\
&  TRN-EC-1    &  1,550    &    3,234   & 24 & 0.0155 & 0 \\
&  TRN-EC-2    &    418    &    519     & 0  & 0      & 0 \\
&  Ownership-USCorp &7,253 &    6,711   & 8  & 0.0011 & 0 \\
\hline%
%\color{green}
Trust
&  College student &    32 &    80        &  32 & 1 & 1 \\
&  Prison inmate &    67      &    142     &  58 & 0.8657 & 0.7970 \\
&  Slashdot      &    82,168   &   504,230  & 229 & 0.0028 & 0 \\
&  WikiVote  &    7,115    &    100,762   &  0& 0 & 0 \\
&  Epinions &    75,888   &    405,740   & 261 & 0.0034 & 0 \\
\hline%
%\color{blue}
Food Web
&  Ythan &    135     &   596   &  0 & 0 & 0 \\
&  Little Rock  &    183     &    2,434   & 181 & 0.9891 & 0.9889 \\
&  Grassland &    88      &    137      & 0 & 0  & 0 \\
&  Seagrass &    49      &    223      &  49 & 1 & 1 \\
\hline%
%\color{Magenta}
Power Grid
&  TexasPowerGrid &    4,889    &    5,855     & 4 & 0.0008 & 0 \\
\hline%
%\color{cyan}
Metabolic
&  \emph{E. coli} &    2,275    &    5,627     & 148 & {{\bf 0.0651}} &
{{\bf 0}} \\
&   \emph{S. cerevisiae} &    1,511    &   3,807     &
 311 & {{\bf 0.2058}} & {{\bf 0}} \\
&  \emph{C. elegans} &    1,173    &    2,842     &  806 &
{{\bf 0.6871}} & {{\bf 0}} \\
\hline
%\color{goldenrod}
Electronic
&  s838 &    512     &    819      & 0 & 0 & 0 \\
%\color{goldenrod}
Circuits
&  s420 &    252     &    399      &  0 & 0 & 0 \\
&  s208 &    122     &    189      &  0 & 0 & 0 \\
\hline
%\color{black}
Neuronal
&  \emph{C. elegans} &    297     &    2,148     & 272 & 0.9158 & 0.8962 \\
\hline
%\color{Orange}
Citation
&   ArXiv-HepTh &    27,770   &    352,285   & 23379 & 0.8419 & 0.8477 \\
&   ArXiv-HepPh &    34,546   &    420,877   & 30704 & 0.8888 & 0.9075 \\
\hline
%\color{Gray}
WWW
&  nd.edu &    325,729  &    1,090,108  & 85,469 & {{\bf 0.2624}} &
{{\bf  0}} \\
&  stanford.edu &    281,903  &    1,992,636  &210,612 & {{\bf 0.7471}} &{{\bf 0}} \\
&  Political blogs  &    1,224    &    16,715   & 0 & {{\bf 0}} &
{{\bf 0.5763}} \\
\hline
%\color{red}
Internet
&  p2p-1 &    10,876    &    39,994    & 0 & 0 & 0 \\
&  p2p-2 &    8,846    &    31,839    & 0 & 0 & 0 \\
&  p2p-3 &    8,717   &    31,525    & 0 & 0 & 0 \\
\hline
%\color{green}
Social
&  UCIonline &    1,899    &    13,838 & 0 & 0 & 0\\
%\color{green}
Communication
&  Email-epoch &    3,188 &    31,857    & 0 & 0 & 0 \\
&  Cellphone &    36,595   &    56,853    & 804 & 0.0220 & 0\\
\hline
%\color{blue}
Intra-
&  Freemans-2  &    34      &    474      & 34 & 1 & 1 \\
%\color{blue}
organizational
&  Freemans-1 &    34      &    415 & 34 & 1 & 1\\
&  Manufacturing &    77      &    1,341     & 77 & 1 & 1 \\
&  Consulting &    46      &    550 & 46 & 1 & 1 \\
\hline
\hline%
\end{tabular}
\label{tab:realnetwork2}
\end{table*}
\end{center}

\clearpage

{\bf Supplementary Note 1}

The $(K,K^\prime)$-protected core of a network only depends on the initial
states of the nodes. Here we prove that
the same final
$(K, K^\prime)$-protected core will be reached independent of
the particular state evolution trajectory of the nodes.

{\it Proof}:
Let us suppose
the contrary is true, namely there exist two different patterns (say $P_1$ and $P_2$)
of final states for a given network. Denote by $S_{pp}$ the set of nodes that are in
the protected state in both patterns $P_1$ and $P_2$, by $S_{uu}$ the set of nodes that are in the unprotected state in both patterns, and by $S_{pu}$ the set of nodes that
are in the protected state in one pattern but are in the unprotected state in the other pattern. These three sets are mutually exclusive and their union contains all the nodes of the network.
Because $P_1$ and $P_2$ are two different state patterns,  the set $S_{pu}$ must be non-empty.

Consider a node $i\in S_{pu}$. By definition, node $i$ is protected in one
pattern (say $P_1$) and unprotected in the other pattern ($P_2$). Since $i$ has the
final protected state in pattern $P_1$, its final unprotected state in $P_2$ must not
be induced by any of the unprotected nodes of set $S_{uu}$. Therefore, the
protected-to-unprotected flipping of node $i$ in pattern $P_2$ must be preceded by at
least one protected-to-unprotected flipping in pattern $P_2$ of another node $j \in S_{pu}$.
But not all nodes of the set $S_{pu}$ can have this property. Therefore the set
$S_{pu}$ must be empty and the two state patterns $P_1$ and $P_2$ must be identical.
This proves the uniqueness of the $(K, K^\prime)$-protected core.

\clearpage

{\bf Supplementary Note 2}

We give more explanations on the mean field equations (1)-(5) of the main text.
First consider a randomly chosen node $i$ that is initially in the protected state. The
probability that this node has $k$ neighbours is just $P(k)$.
We assume that, given the central node $i$ is still in the protected state,
the states of its neighbouring nodes are completely independent of each other.
Under this assumption we then obtain that, given node $i$ being in the protected state, the probability that $s$ of its neighbours are also in the protected state and $r$
of its neighbours are in the unprotected state with less than $K^\prime$
protected neighbours while the remaining $k-s-r$ neighbours are in the unprotected state with at least $K^\prime$ protected neighbours is expressed as
$$
\frac{k!}{s!r!(k-s-r)!}(1-\alpha-\beta)^s \alpha^r \beta^{k-s-r} \; .
$$
For node $i$ to remain in the protected state, the number $r$ must be
zero and the number $s$ must be equal or greater than $K$. In other words,
the probability for an initially protected node $i$ to keep its initial state is
$$
\sum_{s=K}^k C_{k}^{s} (1-\alpha-\beta)^s \beta^{k-s} \; ,
$$
with $C_k^s \equiv k!/[s!(k-s)!]$ being the binomial coefficient.
After considering all the possible values of the node degree $k$, we then obtain the expression (1) concerning the normalized size $n_\mm{p\mhyphen core}$ of the protected
core, which is also the probability that a randomly chosen node $i$ is in the protected state.

We continue to discuss the expressions for $\alpha$, $\beta$ and $\gamma$ following
the theoretical approach of Refs.~[37, 38].
Consider a node $j$ that is neighbouring to a protected node $i$.
In an uncorrelated network, the probability of node $j$ to have $k$ neighbours is expressed as
\begin{equation}
\label{eq:Qk}
Q(k) = \frac{k P(k)}{c} \; ,
\end{equation}
with $c \equiv \sum_k k P(k)$ being the mean node degree of the network.
For node $j$ to be in the protected state with exactly $K$ protected neighbours, it
must have $K-1$ other protected neighbours besides node $i$, and it must not be
connected to any unprotected node with less than $K^\prime$ protected neighbours.
These conditions lead to the expression (4) for the probability $\gamma$.

If node $j$ is initially in the unprotected state but is not able to induce other
protected nodes, it must have $s \geq K^\prime-1$
other protected neighbours besides node $i$. This leads to the first term on the
r.h.s. of Eq.~(3). If node $j$ is initially in the protected state but having less than
$K$ protected neighbours, it will spontaneously become unprotected.
If node $j$ is initially in the protected state but having at least
$K$ protected neighbours,
it will be induced to the unprotected state if $j$ is connected to
at least one unprotected node (say node $l$)
which has less than $K^\prime$ protected neighbours when node $j$ is still in the protected state. Some of the protected neighbours (excluding node $i$) of node $j$ may have exactly $K$
protected neighbours when $j$ is still in the protected state. Such nodes are
 referred to as critical protected nodes.
All these critical protected nodes will spontaneously become
unprotected after node $j$ becomes unprotected.
If the remaining
number of protected neighbours of node $j$ is still
at least $K^\prime$ after all these critical protected nodes become
unprotected,  node $j$ will not be able to induce these remaining
protected nodes. The second term on the r.h.s. of Eq.~(3) is just the
total probability for this situation to occur.

If node $j$ is initially in the unprotected state and having less than $K^\prime$
protected neighbours, it will be able to induce node $i$ and all its other
protected neighbours to the unprotected state.
This situation corresponds to the first term on the r.h.s. of Eq.~(2).
If node $j$ is initially in the protected state but having less than $K$
protected neighbours,
it will spontaneously become unprotected. If node $j$ is initially
in the protected state and having $K$ or more protected neighbours,
it will be induced to the unprotected state if at least one of its unprotected neighbours has less than $K^\prime$
protected neighbours. Some of the protected neighbours (excluding node $i$)
of node $j$ may have exactly $K$ protected neighbours when $j$ is still in
the protected state. All these critical protected nodes
will spontaneously become
unprotected after node $j$ has changed to the unprotected state.
The number of protected neighbours of node $j$ may become less than $K^\prime$
after all these critical protected nodes have transited to the unprotected state.
If this is the case, node $j$
will then be able to induce its remaining protected neighbours
 to the unprotected state. The second term on the r.h.s. of Eq.~(2) is just the
total probability for this situation to occur.

To understand the expression (5) for the probability $\eta$, let us consider a neighbouring node $n$ of an initially unprotected node $m$.
If $n$ is initially unprotected (with probability $1-p$),
it will remain in the unprotected state.
If $n$ is initially protected but having less than $K$ protected neighbours,
it will spontaneously become
 unprotected.
If node $n$ is initially protected and having $K$ or more protected neighbours,
it will be induced to the unprotected state if at least
one of its unprotected neighbours (excluding node $m$)
has less than $K^\prime$ protected neighbours.
The two terms inside the curly brackets of
Eq.~(5) are the probabilities for the initial protected node $n$ to spontaneously
become unprotected and to be induced to the unprotected state, respectively.

\clearpage

{\bf Supplementary Note 3}

The mean field equations (1)-(5) work
both for finite networks and infinite networks
($N=\infty$).
The only input to these equations is the degree distribution $P(k)$.
The other degree distribution $Q(k)$ is determined from $P(k)$ through
Eq.~(\ref{eq:Qk}).
Here we introduce a simple numerical scheme for determining the values of
$\alpha$, $\beta$, $\gamma$ and $\eta$.

From Eqs.~(2) and (3) we obtain that
\begin{equation}
\beta =  g(\beta) \equiv
1- \alpha  - p
\sum\limits_{k \geq K} Q(k) \sum\limits_{s\geq K-1}^{k-1} C_{k-1}^{s}
\beta^{k-1-s} (1-\alpha - \beta)^s \; .
\label{eq:betaalpha}
\end{equation}
The value of $\beta$ can be obtained by solving $\beta = g(\beta)$
at each fixed value of $\alpha$. When $K\geq 2$, this equation always has a
solution $\beta=1-\alpha$.
After the value of $\beta$ is determined from $\alpha$, then the value
of $\gamma$ and the value of $\eta$ can be obtained through Eq.~(4)
 and Eq.~(5), respectively, using the values of
$\alpha$ and $\beta$ as inputs. Notice that for $K\geq 2$, if $\beta=1-\alpha$
then $\gamma=0$ and $\eta=1$.

After the values of $\beta$, $\gamma$ and $\eta$ are obtained at a given
value of $\alpha$, then we can obtain a new value of $\alpha$ through
Eq.~(2). In this way a mapping
$\alpha \leftarrow f(\alpha)$
from $\alpha$ to $\alpha$ is constructed for
$0 \leq \alpha \leq 1$.  The mapping function $f(\alpha)$ is just the
r.h.s. of Eq.~(2), namely
\begin{eqnarray}
f(\alpha)  & = &
  (1-p)
  \sum\limits_{s=0}^{K^\prime -2}
      \sum\limits_{k\geq s+1} Q(k) C_{k-1}^{s}  (1-\eta)^s \eta^{k-1-s}
\nonumber \\
& & + p \ \biggl\{
         \sum\limits_{s=0}^{K-2} \sum\limits_{k\geq s+1}
         \sum\limits_{r=0}^{\min(s,K^\prime -2)}
         Q(k) C_{k-1}^{s} C_{s}^{r} (\alpha + \beta)^{k-1-s}
         (1-\alpha-\beta-\gamma)^{r} \gamma^{s-r} \nonumber \\
& &         + \sum\limits_{s\geq K-1}\sum\limits_{k\geq s+2}
          \sum\limits_{r=0}^{\min(s,K^\prime -2)}
          Q(k) C_{k-1}^{s} C_{s}^{r} \bigl[
          (\alpha+\beta)^{k-1-s}-\beta^{k-1-s} \bigr]
          (1-\alpha-\beta-\gamma)^{r} \gamma^{s-r}
        \biggr\} \; , \nonumber \\
& &
\label{eq:falpha}
\end{eqnarray}
in which $\beta$, $\gamma$ and $\eta$ are all regarded as functions
of $\alpha$.
By solving the equation $\alpha = f(\alpha)$ we obtain the value of
$\alpha$, which then fixes the values of $\beta$, $\gamma$ and $\eta$.
For $K\geq 2$ and $K^\prime \geq 2$, it can be easily checked that
$\alpha =1$ is always a solution of the equation $\alpha = f(\alpha)$.

Denote a generic solution of the equation $\alpha = f(\alpha)$ as $\alpha_0$.
If starting from a value of $\alpha$ slightly different from $\alpha_0$, the
iteration $\alpha \leftarrow f(\alpha)$ can drive $\alpha$ back to $\alpha_0$, then
we regard $\alpha_0$ as a locally stable solution of $\alpha = f(\alpha)$. Otherwise
$\alpha_0$ is regarded as an unstable solution of this equation.

\clearpage

{\bf Supplementary Note 4}

We explain that the mean field equations (1)-(5) predict
the $(K, K^\prime)$-protected
core percolation transition to be discontinuous if $K^\prime \geq 2$, independent of
the value of $K$. We begin with the simpler cases of $K=2$  and $K\geq 3$, and
then discuss the more difficult cases of $K=1$ and $K=0$.

\subsection*{ (a) $K=2$}

% figure S1

\begin{figure}[b]
\begin{center}
\resizebox{8.5cm}{!}{\includegraphics{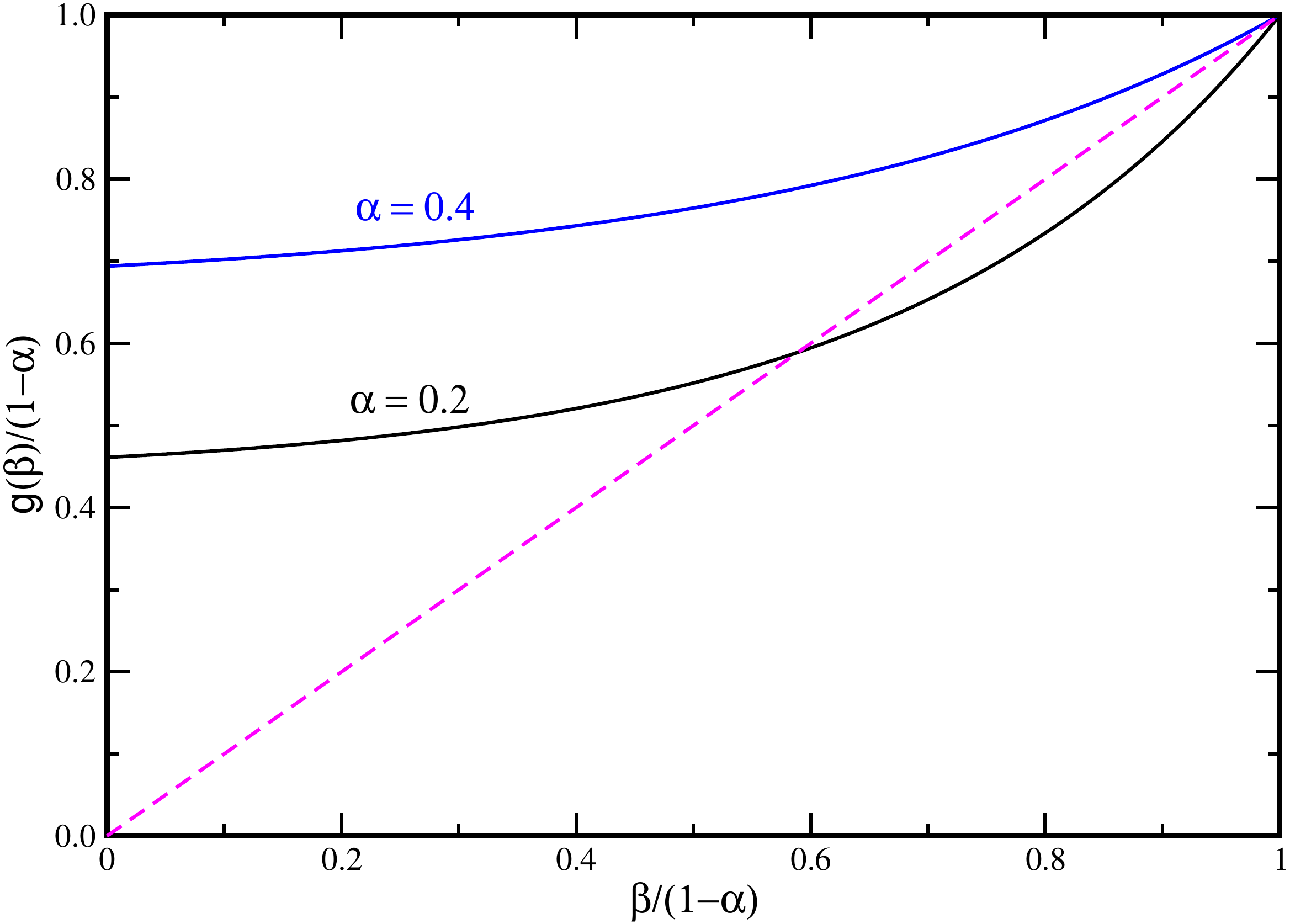}}
\caption{
\label{fig:gbetaK2}
{\bf Function $g(\beta)$ at $K= K^\prime=2$ for the infinite
ER random network with mean degree $c=4$.}
The initial fraction of protected nodes is $p=1$, and
the probability $\alpha$ is $\alpha=0.2$ or $\alpha=0.4$.
Both $g(\beta)$ and $\beta$ are
normalized by  $(1-\alpha)$ so that all the curves meet at the point $\beta/(1-\alpha)=1$.
The solutions of the equation $\beta = g(\beta)$ can be read off from the
crossing points of the $g(\beta)$ curve with the dashed line representing
$\beta=\beta$. At $\alpha = 0.4$ there is only one solution, while at $\alpha=0.2$ there are two solutions.
}
\end{center}
\end{figure}\noindent
%

%\clearpage

A protected node will spontaneously
become unprotected if it has less than two protected neighbours.
The function $g(\beta)$  is expressed as
\begin{equation}
g(\beta) = 1-\alpha - p \sum\limits_{k \geq 1} Q(k)
 \bigl[ (1-\alpha)^{k-1} - \beta^{k-1} \bigr] \; .
\end{equation}
The first derivative of $g(\beta)$ with respective to $\beta$ is
simply
\begin{equation}
\frac{\partial g(\beta)}{\partial \beta}
= p \sum\limits_{k \geq 2} (k-1) Q(k) \beta^{k-2} \; ,
\end{equation}
which increases with $\beta$. Therefore $g(\beta)$
as a function of $\beta$ is convex in the interval
$0 \leq \beta\leq 1-\alpha$, see Fig.~\ref{fig:gbetaK2} for some examples obtained
from the Erd\"{o}s-R\'{e}nyi (ER) random network of mean degree $c=4$.
If $p \sum_{k\geq 2} (k-1) Q(k) > 1$, then there exists a value of
$\alpha_t$ such that
$$
\left.\frac{\partial g(\beta)}{\partial \beta}\right|_{\beta = 1-\alpha}
\equiv  p \sum_{k\geq 2} (k-1) Q(k) (1-\alpha)^{k-2} < 1
$$
for
$\alpha >  \alpha_t$. Notice that $\alpha_t$ is strictly less than $1$.
Then for $\alpha \in [\alpha_t, 1]$, the equation
$\beta = g(\beta)$ has only a unique
solution $\beta = 1 -\alpha$.
If  $p \sum_{k\geq 2} (k-1) Q(k) \leq 1$, then
$\left.\frac{\partial g(\beta)}{\partial \beta}\right|_{\beta = 1-\alpha} \leq 1$
at any value of $\alpha \in [0,1]$.
This means the solution of
$\beta = g(\beta)$ is always $\beta = 1-\alpha$ for any $\alpha_t \leq \alpha \leq 1$
with $\alpha_t=0$.

% figure S2
\begin{figure}[t]
\begin{center}
\resizebox{8.5cm}{!}{\includegraphics{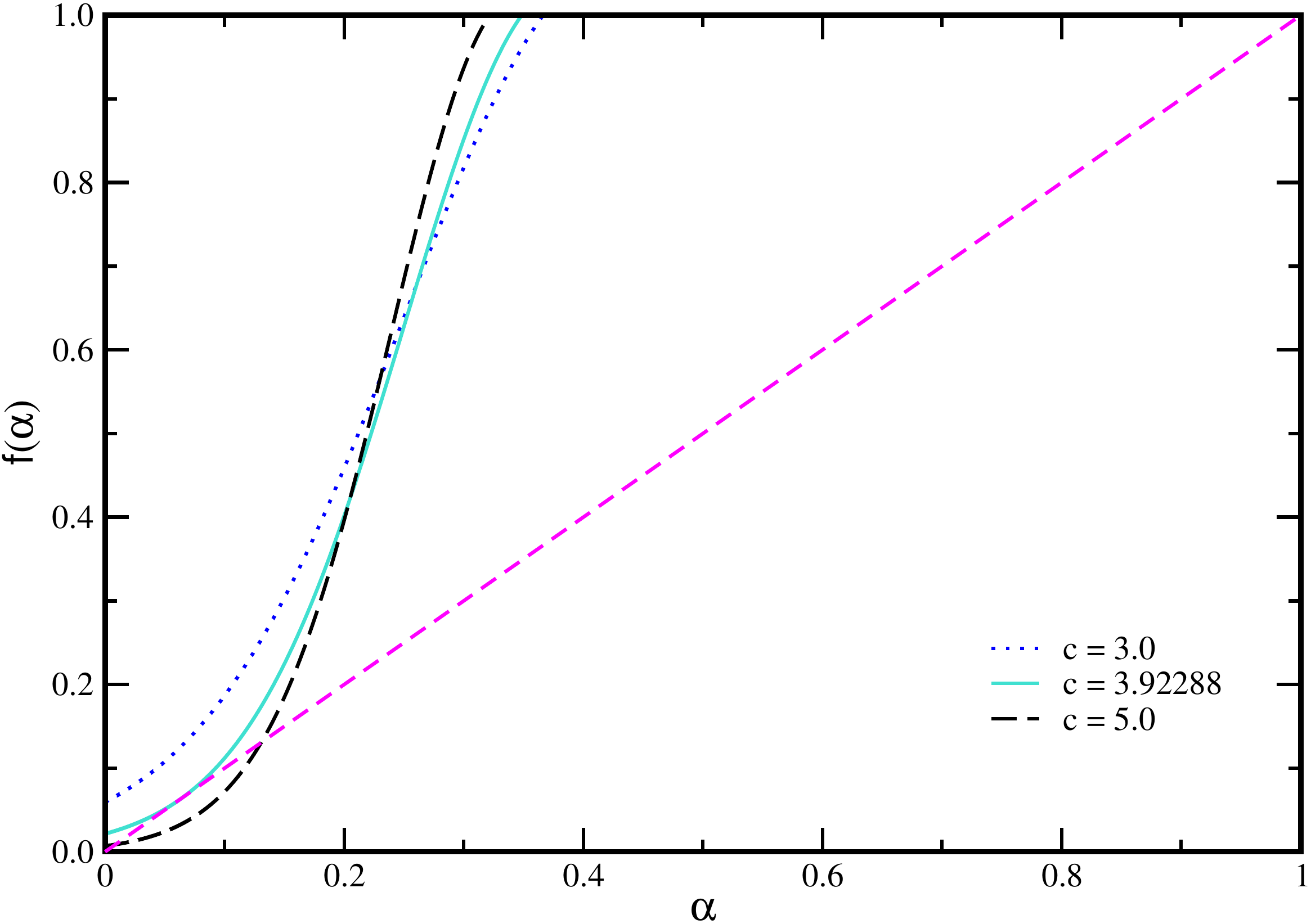}}
\caption{
\label{fig:falphaK2}
{\bf Function $f(\alpha)$ at $K= K^\prime = 2$
for the infinite ER random network.}
The initial fraction of protected nodes is set to $p=1$,
and the mean degree is $c=3.0$ (dotted line),
$c=3.92288$ (solid line), or $c=5.0$ (long-dashed line).
The function $f(\alpha)=1$ for $\alpha$ sufficiently close to $1$.
When $c<3.92288$ the equation $\alpha=f(\alpha)$ has only the solution $\alpha=1$.  When $c>3.92288$, there are three solutions of $\alpha=f(\alpha)$ as can be
read off from the crossing points of $f(\alpha)$ with the dashed line representing
$\alpha=\alpha$:
 the solutions $\alpha=1$ and
the one with $\alpha\approx 0$ are stable, while the middle one is unstable.
}
\end{center}
\end{figure}\noindent
%
%\clearpage

From these discussions, we know that there exists a value $\alpha_t$
strictly less than $1$ such that $f(\alpha)=1$ for $\alpha_t \leq \alpha \leq 1$.
Some example curves  of   $f(\alpha)$ are shown in Fig.~\ref{fig:falphaK2} for
infinite ER random networks.
If $\alpha_t=0$, then $\alpha=1$ is the only solution of $\alpha = f(\alpha)$.
If $0 < \alpha_t <1$, then
$f(\alpha_t) = 1 > \alpha_t$ (notice that
$\beta = 1-\alpha_t$ at $\alpha = \alpha_t$). Consequently if
$\alpha = f(\alpha)$ has another solution different from $\alpha = 1$, this  solution
$\alpha$ must be strictly less than $\alpha_t$, and  $\alpha + \beta$ at this solution must be strictly less than $1$.

According to Eq.~(1), the normalized size of the protected core
$n_\mm{p\mhyphen core} = 0$ if $\alpha + \beta=1$.
On the other hand,
if $\alpha + \beta < 1$ then $n_\mm{p\mhyphen core}$ will
be strictly positive. We therefore conclude that, if the $(2,K^\prime)$-protected
core percolation transition occurs in a network, the normalized size of the
$(2,K^\prime)$-protected core will have a finite jump at the transition point.

\subsection*{(b) $K\geq 3$}

% figure S3
\begin{figure}[b]
\begin{center}
\resizebox{8.5cm}{!}{\includegraphics{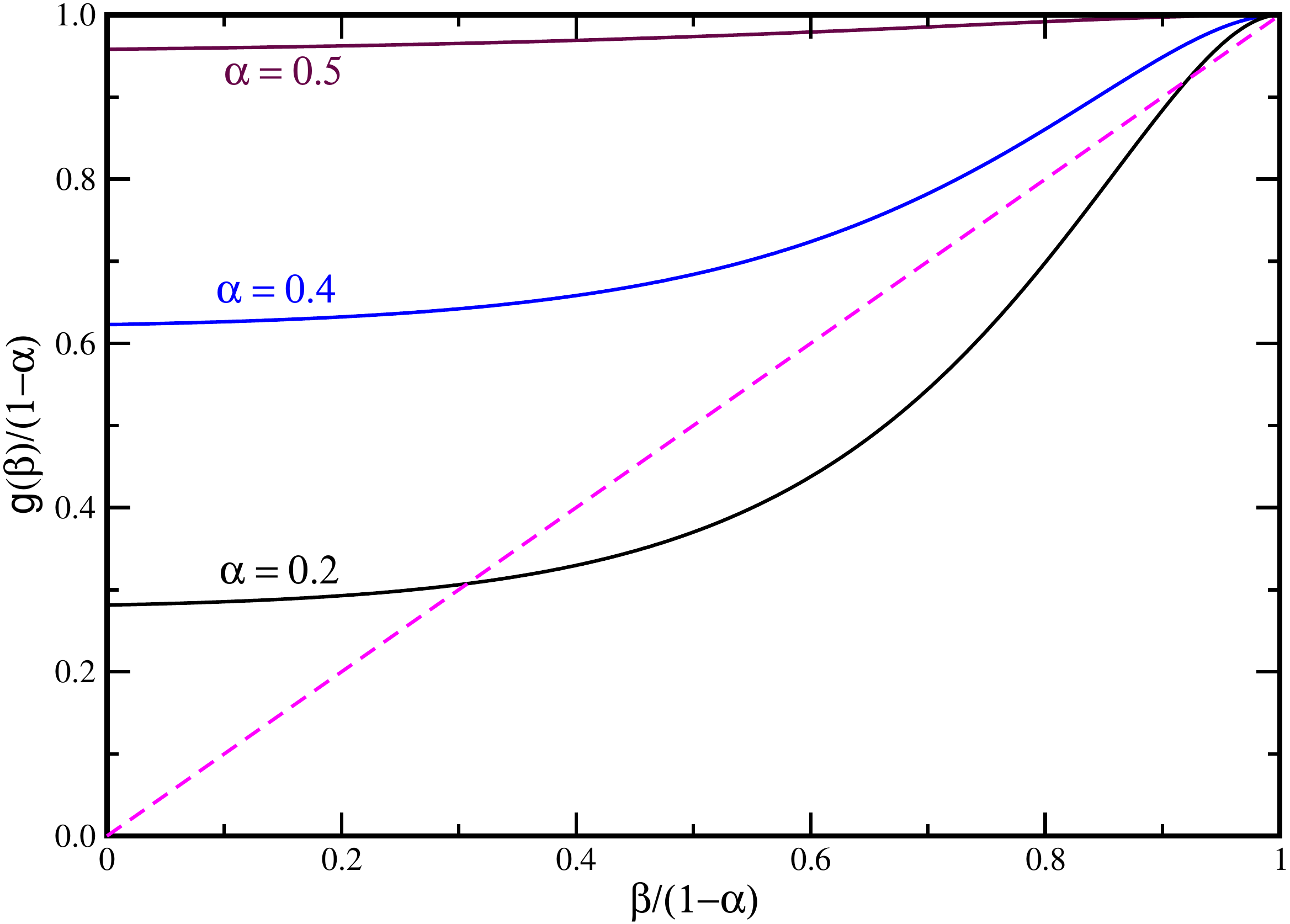}}
\caption{
\label{fig:gbetaK3}
{\bf Function $g(\beta)$ at $K= K^\prime=3$ for the infinite
ER random network with mean degree $c=7.5$.}
The initial fraction of protected nodes is $p=1$, and
the probability $\alpha$ is $\alpha=0.2$, $0.4$ or $0.5$.
Both $g(\beta)$ and $\beta$ are
normalized by  $(1-\alpha)$ so that all the curves meet at the point $\beta/(1-\alpha)=1$.
The solutions of the equation $\beta = g(\beta)$ can be read off from the
crossing points of the $g(\beta)$ curve with the dashed line representing
$\beta=\beta$.
}
\end{center}
\end{figure}\noindent
%
%\clearpage

The function $g(\beta)$ has the following expression
\begin{equation}
g(\beta) = 1 - \alpha - p \sum\limits_{k\geq K} Q(k) \sum\limits_{s = K-1}^{k-1}
C_{k-1}^{s} (1-\alpha-\beta)^s \beta^{k-1-s} \; .
\end{equation}
Some representative curves of $g(\beta)$ are shown in Fig.~\ref{fig:gbetaK3}
for the infinite ER random network of mean degree $c=7.5$.
The first derivative of $g(\beta)$ with respective to $\beta$ is
\begin{equation}
  \frac{ \partial g(\beta)}{\partial \beta}
 =
p \sum\limits_{k \geq K} (k-1) Q(k) C_{k-2}^{K-2}
\beta^{k-K}  (1-\alpha - \beta)^{K-2} \; .
 \label{eq:betaderive}
 \end{equation}
At $\beta = 1-\alpha$ this derivative is equal to $0$. There exists a
threshold value $\alpha_t < 1$ such that if $\alpha \geq \alpha_t$ then
$\frac{\partial g(\beta)}{\partial \beta} < 1$  for $0 \leq \beta \leq 1-\alpha$.
Therefore for $\alpha \in [\alpha_t, 1]$ the equation
$\beta = g(\beta)$ has only the unique solution $\beta = 1-\alpha$.

% figure S4
\begin{figure}[t]
\begin{center}
\resizebox{8.5cm}{!}{\includegraphics{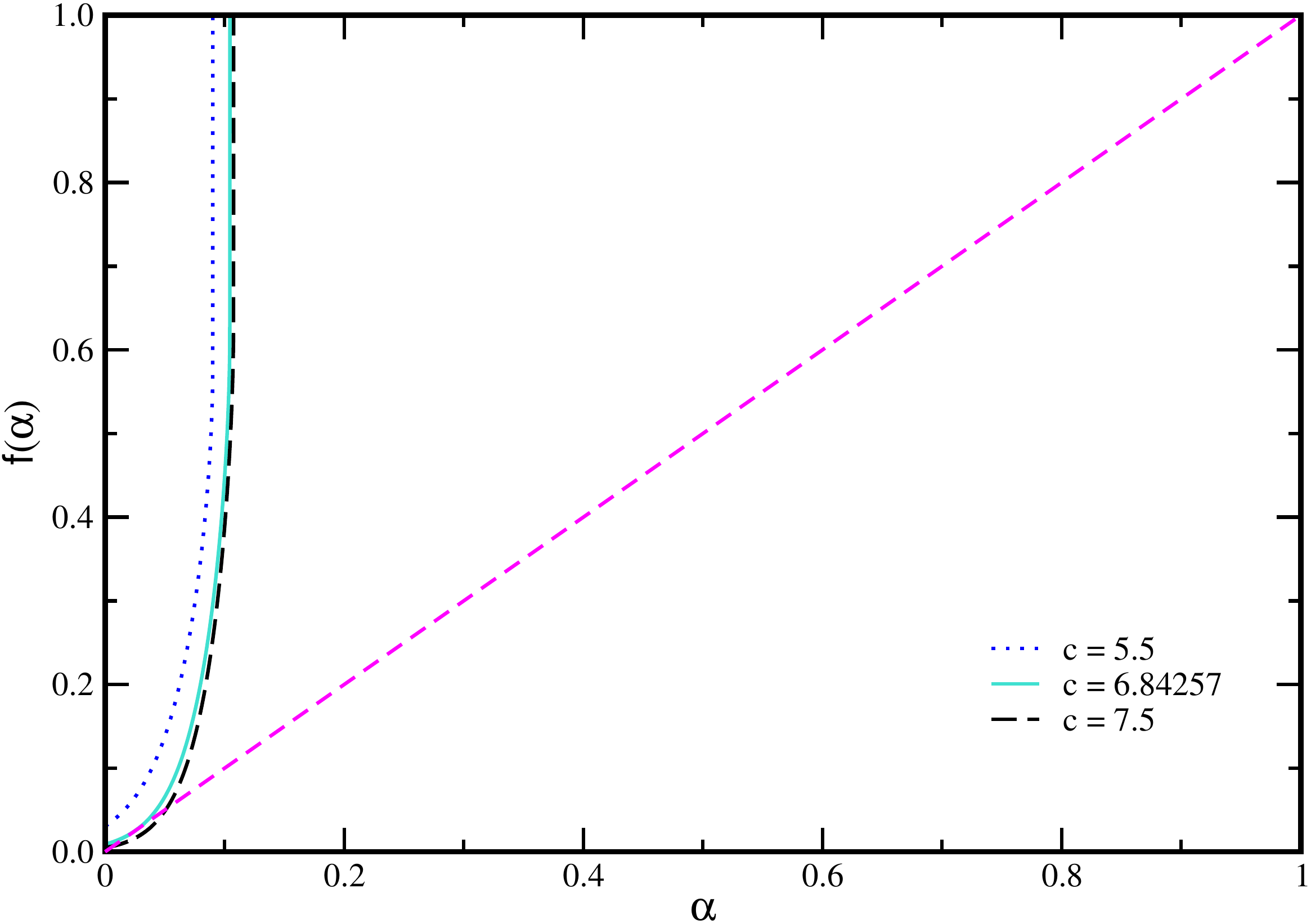}}
\caption{
\label{fig:falphaK3}
{\bf Function $f(\alpha)$ at $K= K^\prime = 3$
for the infinite ER random network.}
The initial fraction of protected nodes is set to $p=1$,
and the mean degree is $c=5.5$ (dotted line),
$c=6.68257$ (solid line), or $c=7.5$ (long-dashed line).
The function $f(\alpha)=1$ for $\alpha$ sufficiently close to $1$.
When $c<6.68257$ the equation $\alpha=f(\alpha)$ has only the solution $\alpha=1$.  When $c>6.68257$, there are three solutions of $\alpha=f(\alpha)$ as can be
read off from the crossing points of $f(\alpha)$ with the dashed line representing
$\alpha=\alpha$.
}
\end{center}
\end{figure}\noindent
%
%\clearpage

Due to the fact that $\left.\frac{\partial g(\beta)}{\partial \beta}\right|_{\beta =
1-\alpha} = 0$, if for some values of $\alpha< \alpha_t$
the equation $\beta = g(\beta)$ has another solution with $\beta  \neq 1-\alpha$, then
at this solution the sum $\alpha + \beta$ must be strictly less than $1$
(the behaviour of $g(\beta)$ at $\beta/(1-\alpha) \approx 1$ is demonstrated
in Fig.~\ref{fig:gbetaK3}).
Since $\beta = 1-\alpha$ for $\alpha> \alpha_t$, then the
function $f(\alpha)=1$ for $\alpha\in [\alpha_t, 1]$ (see
Fig.~\ref{fig:falphaK3} for some representative curves of $f(\alpha)$
for the infinite ER random network). If another stable solution of
$\alpha = f(\alpha)$ exists, the value of $\alpha$ at this
 solution must be strictly less than $\alpha_t$, and the value of
 $\alpha+\beta$ must be strictly less than $1$.

According to Eq.~(1), the normalized size of the protected core
$n_\mm{p\mhyphen core} = 0$ if $\alpha + \beta=1$.
On the other hand,
if $\alpha + \beta < 1$ then $n_\mm{p\mhyphen core}$ will
be strictly positive. We therefore conclude that, for $K\geq 3$,
 if the $(K,K^\prime)$-protected
core percolation transition occurs in a network, the normalized size of the
$(K,K^\prime)$-protected core will have a finite jump at the transition point.

\subsection*{(c) $K=1$}

\begin{figure}[t]
\begin{center}
\resizebox{8.5cm}{!}{\includegraphics{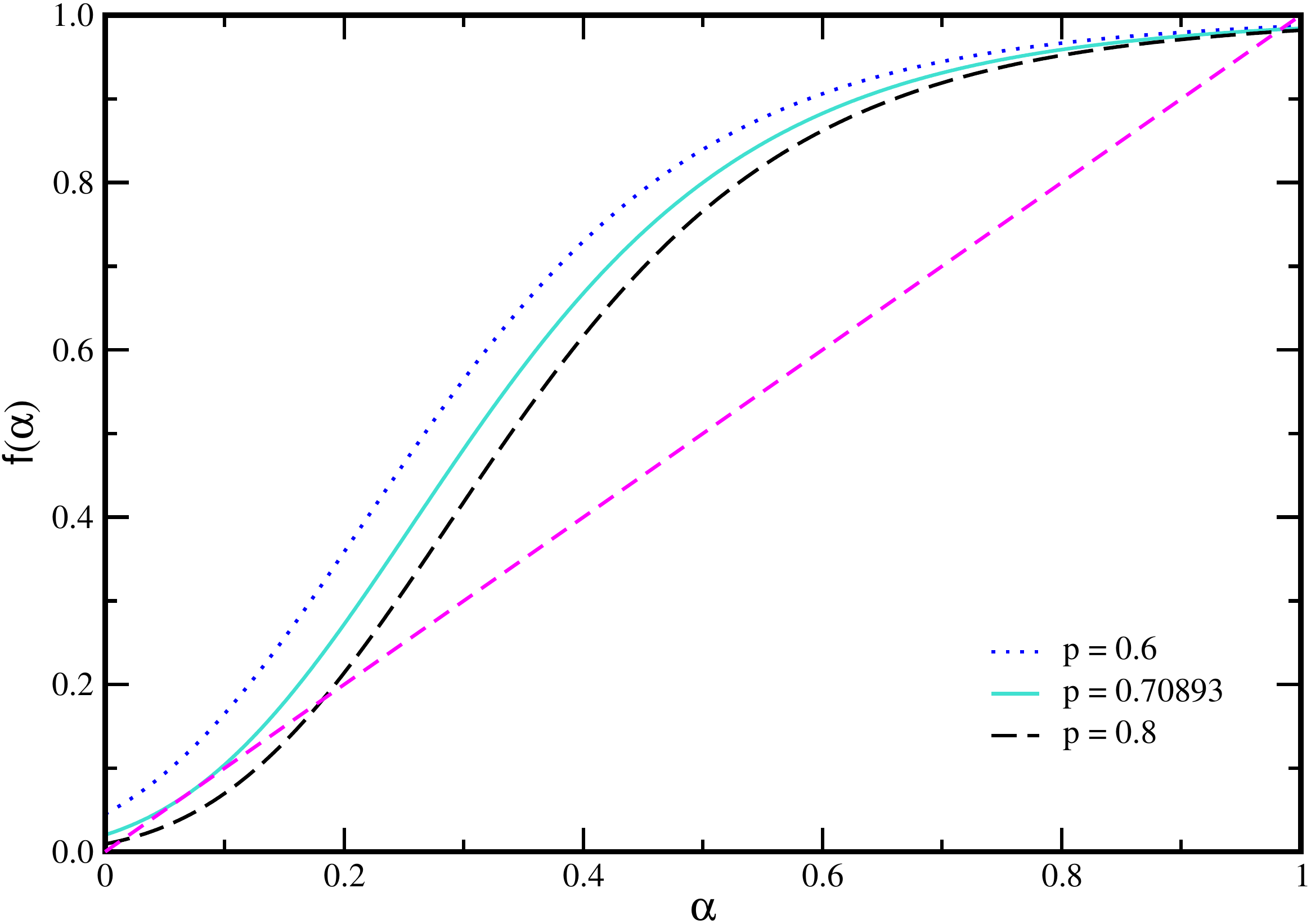}}
\caption{
\label{fig:falphaK1Kp2}
{\bf Function $f(\alpha)$ at $K=1$ and $K^\prime = 2$
for the infinite ER random network with mean degree $c=4$.}
The initial fraction of protected nodes is $p=0.6$ (dotted line),
$p=0.70893$ (solid line), or $p=0.8$ (long-dashed line).
When $p<0.70893$ the equation $\alpha=f(\alpha)$ has only one solution,
 located at $\alpha \approx 1$.  When $p>0.70893$, there are three solutions of $\alpha=f(\alpha)$ as can be
read off from the crossing points of $f(\alpha)$ with the dashed line representing
$\alpha=\alpha$.
}
\end{center}
\end{figure}

In this case, a protected node will spontaneously become unprotected
if it has no protected neighbour. From the mean field equations (2)-(5) we obtain
that
\begin{eqnarray}
\beta & = & 1-\alpha - p \sum\limits_{k\geq 1} Q(k) (1-\alpha)^{k-1} \; ,
\label{eq:betaK1}\\
\gamma & = & p \sum\limits_{k\geq 1} Q(k) \beta^{k-1} \; , \\
\eta & = & \alpha + \beta + \gamma \; .
\end{eqnarray}
The function $f(\alpha)$, namely the expression
(\ref{eq:falpha}) with $\beta, \gamma, \eta$
determined from $\alpha$ through the above three equations,
is a smooth function of $\alpha$ (some example curves of
$f(\alpha)$ are shown in Fig.~\ref{fig:falphaK1Kp2} for the infinite
 ER random network with mean degree $c=4$).
We can easily verify that $f(0)>0$.

% figure S5
\noindent
%
%\clearpage

Denote  $\alpha_{\rm max}$ as the largest root of the
equation
\begin{equation}
\alpha = 1- p \sum\limits_{k\geq 1} Q(k) (1-\alpha)^{k-1} \;
\end{equation}
in the interval of $0 \leq \alpha \leq 1$. If $p Q(1) = 0$,
then $\alpha_{\rm max} = 1$; if $0 < p Q(1) < 1$ then $0 < \alpha_{\rm max} < 1$;
if $p Q(1) = 1$ then $\alpha_{\rm max}=0$.
The last case is not interesting:
$p Q(1) = 1$ means that $Q(1)=1$ (all the nodes have only one neighbour)
and $p=1$ (all the nodes are initially in the protected state),
 then the states of
the nodes will not change with time. We assume that $p Q(1) < 1$ in the following
discussions (i.e., $\alpha_{\rm max}>0$).

When
$0\leq \alpha < \alpha_{\rm max}$ the value of $\beta$ as obtained by
Eq.~(\ref{eq:betaK1}) is positive. The value of $\beta$ reduces to zero
at $\alpha=\alpha_{\rm max}$. We can verify
that $f(\alpha_{\rm max}) < \alpha_{\rm max}$. To prove this statement,
let us first write $f(\alpha)$ as
\begin{eqnarray}
f(\alpha) & = & 1- p \sum\limits_{k\geq 1} Q(k) (1-\alpha)^{k-1}
\nonumber \\
& & - p \sum\limits_{k\geq K^\prime} Q(k)
\sum\limits_{s=K^\prime -1}^{k-1} C_{k-1}^s \bigl[ (\alpha + \beta)^{k-1-s}
-\beta^{k-1-s} \bigr]
\sum\limits_{r=K^\prime -1 }^{s} C_{s}^r (1-\alpha-\beta-\gamma)^r \gamma^{s-r} \nonumber \\
& &
- (1-p) \sum\limits_{k\geq K^\prime} Q(k) \sum\limits_{s=K^\prime -1}^{k-1}
C_{k-1}^s (1-\eta)^s \eta^{k-1-s}
\; .
\end{eqnarray}
Since the third and fourth term on the r.h.s. of the above expression are both negative,
we obtain that
 $f(\alpha) <1- p \sum_{k\geq 1}Q(k) (1-\alpha)^{k-1}$. It then follows that
 $f(\alpha_{\rm max}) < \alpha_{\rm max}$.

Because $f(\alpha) > \alpha$ at $\alpha=0$ and $f(\alpha)<\alpha$ at $\alpha=
\alpha_{\rm max}$, the curve $f(\alpha)$ will intersect with the rectilinear line
$\alpha=\alpha$ either once or three times
in the interval $\alpha \in (0, \alpha_{\rm max})$. This fact is demonstrated
clearly in Fig.~\ref{fig:falphaK1Kp2}. In other words, the
equation $\alpha = f(\alpha)$ either has only a single solution in the
interval $\alpha \in (0, \alpha_{\rm max})$ or has three solutions in this
interval. In the latter case,
since the function $f(\alpha)$ is a smooth function,
there must be a finite gap between the two stable solutions, with
 one stable solution located at $\alpha \approx 0$ (corresponding to
 a value of $n_\mm{p\mhyphen core}$ well above zero)
 and the other
stable solution at $\alpha \approx \alpha_{\rm max}$ (corresponding to
 a value of $n_\mm{p\mhyphen core}\approx 0$).

\subsection*{(d) $K=0$}

In this case,
an initially protected node will never spontaneously become
unprotected. It can
only be induced to the unprotected by a neighbouring unprotected node. This case is
actually very similar to the case of $K=1$.
For example the mean field equation for $\beta$ is the same as Eq.~(\ref{eq:betaK1}).
The only difference is that $\gamma=0$ in this limiting case.

We can follow the same theoretical arguments developed for the case of $K=1$ to prove
that, if  a $(0, K^\prime)$-protected core percolation transition occurs in a
network, the normalized size of the protected core at the transition point must
has a finite jump from nearly zero to a positive value well above zero.

\clearpage

{\bf Supplementary Note 5}

Here we list the explicit mean field equations for computing the
the sizes of $(0,1)$-protected core,
$(1,1)$-protected core, $(0,2)$-protected core, and $(1, 2)$-protected core,
respectively.

\subsection*{(a) $K=1, K^\prime =1$ (minimal spontaneous transition,
no inducing effect)}

In this case, an unprotected node is not able to induce its protected
neighbours. An initially protected node will spontaneously
become unprotected only if
all its neighbours are initially unprotected.

The fraction of protected nodes in the final state is
\begin{equation}
n_\mm{p\mhyphen core}(1,1) = p \Bigl[
1- \sum\limits_{k\geq 0} P(k) (1-p)^k \Bigr] \; .
\end{equation}
The normalized size $g_\mm{p\mhyphen core}(1,1)$ of the giant component of
protected nodes is expressed as
\begin{equation}
g_\mm{p\mhyphen core}(1,1) = p \Bigl[
1- \sum\limits_{k\geq 0} P(k) \mu^k \Bigr] \; ,
\label{eq:g11}
\end{equation}
where $\mu$ is the probability that, starting from a protected node $i$, the node $j$
reached by following a randomly chosen link $(i,j)$ is not belonging to the giant connected component of protected nodes if the link $(i,j)$  is absent.
The expression for $\mu$ is
\begin{equation}
\label{eq:xi11}
\mu = 1- p + p  \sum\limits_{k\geq 1} Q(k) \mu^{k-1}  \; .
\end{equation}

If $p \sum_{k\geq 2}  (k-1) Q(k) \leq 1$, Eq.~(\ref{eq:xi11}) has only a single
solution $\mu=1$ (no giant component of protected nodes). Equation (\ref{eq:xi11}) has
a stable solution with $\mu<1$ when $p \sum_{k\geq 2}  (k-1) Q(k) > 1$,
corresponding to the existence of a giant connected component of protected nodes.

\subsection*{(b) $K=0, K^\prime =1$ (no spontaneous transition and no inducing effect)}

Since an initially protected node will never spontaneously become
unprotected (because $K=0$) nor be induced to the unprotected state (because
$K^\prime = 1$), there is no any state evolution in the system.
The fraction of protected nodes in the final steady state is just $p$.
The normalized size $g_\mm{p\mhyphen core}(0,1)$ of the giant connected component of protected nodes has the same expression as Eq.~(\ref{eq:g11}), with the probability
$\mu$ determined by Eq.~(\ref{eq:xi11}).

\subsection*{(c) $K=1, K^\prime =2$ (minimal spontaneous transition and
minimal inducing effect)}

In this case, if an unprotected node has only a single protected neighbour,
it can induce this node to the unprotected state. An
initially protected node will spontaneously become
 unprotected only if all
of its neighbours are unprotected.
The fraction of protected nodes at the steady state is
\begin{equation}
n_\mm{p\mhyphen core}(1, 2)
= p \sum\limits_{k} P(k)
\bigl[ (1 -\alpha)^k - \beta^k \bigr] \; .
\end{equation}

The expressions for $\alpha$, $\beta$, $\gamma$ and $\eta$ are, respectively
\begin{eqnarray}
\alpha &= & (1-p) \sum\limits_{k\geq 1} Q(k) \eta^{k-1}
+ p \sum\limits_{k\geq 1} Q(k) \bigl[ (\alpha+\beta+\gamma)^{k-1}
- (\beta+\gamma)^{k-1} \bigr] \; , \\
\beta & = & 1  - (1-p) \sum\limits_{k\geq 1} Q(k) \eta^{k-1} \nonumber \\
& & \quad\quad\quad
+
p \sum\limits_{k\geq 1} Q(k) \bigl[
(\beta+\gamma)^{k-1} - (\alpha + \beta +\gamma)^{k-1}
- (1-\alpha)^{k-1} \bigr] \; , \\
\gamma & = &  p \sum\limits_{k\geq 1} Q(k) \beta^{k-1} \; , \\
\eta & = & 1  -
p \sum\limits_{k\geq 1} Q(k) \bigl[
 (1-\alpha)^{k-1} -\beta^{k-1} \bigr] \; .
\end{eqnarray}

The fraction of nodes in the giant component of the protected core is
\begin{equation}
g_\mm{p\mhyphen core}(1,2) = p \sum\limits_{k} P(k) \bigl[
(1-\alpha)^k - (\chi + \beta)^k \bigr] \; ,
\label{eq:g12}
\end{equation}
where $\chi$ is the probability that a neighbouring node $j$ of
a protected node $i$ is also in the protected state but does not belong to the
giant component of the protected core if the link $(i, j)$ is absent.
The expression for $\chi$ is
\begin{equation}
\label{eq:chi}
\chi = p \sum\limits_{k\geq 1} Q(k) (\chi + \beta )^{k-1} \; .
\end{equation}

\subsection*{(d)  $K=0, K^\prime =2$ (no spontaneous transition
but with minimal inducing effect)}

An initially protected node will never spontaneously become unprotected
(because $K=0$).
However, if an unprotected node has only a single protected neighbour,
it can induce this neighbour to the unprotected state (because $K^\prime = 2$).

The fraction of protected nodes at the steady state is
\begin{equation}
n_\mm{p\mhyphen core}(0,2)
= p \sum\limits_{k} P(k)
 (1 -\alpha)^k  \; .
\end{equation}
The expressions for $\alpha$, $\beta$, $\gamma$ and $\eta$ are
\begin{eqnarray}
\alpha &=&
  (1-p) \sum\limits_{k\geq 1} Q(k) \eta^{k-1}
  + p \sum\limits_{k\geq 1} Q(k) \bigl[ (\alpha+\beta)^{k-1}
- \beta^{k-1} \bigr] \; ,
\\
\beta & = & 1 - (1-p) \sum\limits_{k\geq 1} Q(k) \eta^{k-1}
+ p \sum\limits_{k\geq 1} Q(k) \bigl[
\beta^{k-1} - (\alpha + \beta)^{k-1}
- (1-\alpha)^{k-1} \bigr] \; , \\
\gamma & = & 0 \; , \\
\eta  &= & 1- p \sum\limits_{k\geq 1} Q(k) (1-\alpha)^{k-1} \; .
\end{eqnarray}

The fraction of nodes in the giant component of the protected core is
\begin{equation}
g_\mm{p\mhyphen core}(0,2)
= p \sum\limits_{k} P(k)
\bigl[ (1 -\alpha)^k - (\chi+\beta)^k \bigr] \; ,
\end{equation}
where $\chi$ is the probability that a neighbouring node $j$ of
a protected node $i$ is also in the protected state but does not belong to the
giant component of the protected core if the link $(i,j)$ is absent.
The expression for $\chi$ is also given by Eq.~(\ref{eq:chi}).

\clearpage

{\bf Supplementary Note 6}

We discuss here the scaling behaviour of the normalized size of the
$(K, K^\prime)$-protected core near the transition point.

We predict that, when the mean degree $c$ is only slightly
beyond the protected core percolation point $c^*$, the deviation between
the normalized size
$n_\mm{p\mhyphen core}$ of the protected core
and the value of $n_\mm{p\mhyphen core}^*$ at the transition point
follows the
scaling behaviour
\begin{equation}
\label{eq:nscale}
n_{\rm p-core} - n_{\rm p-core}^{*} \propto (c-c^*)^{1/2} \; .
\end{equation}
If the mean degree $c$ is fixed but the initial fraction $p$ of protected nodes changes,
the corresponding scaling behaviour is
\begin{equation}
\label{eq:nscale2}
n_{\rm p-core} - n_{\rm p-core}^{*} \propto (p - p^*)^{1/2} \; ,
\end{equation}
where $p^*$ is the critical initial fraction of protected nodes at the
transition.
Here we give a brief derivation of Eq.~(\ref{eq:nscale}). Equation
(\ref{eq:nscale2}) can be derived similarly.

Denote the value of $\alpha$ at the protected core percolation transition point as
 $\alpha=\alpha^*$, and the corresponding excess degree distribution $Q(k)$ as
 $Q^*(k)$. At the
transition point the following two properties hold for the function $f(\alpha)$
 as defined in Eq.~(\ref{eq:falpha}):
\begin{equation}
\alpha^* = f(\alpha^*) \; , \quad\quad\quad\quad
\left. \frac{\partial f(\alpha)}{\partial \alpha}\right|_{\alpha=\alpha^*}
 =  1 \; .
\end{equation}
The function $f(\alpha)$ also depends on  $Q(k)$.
As the network slightly changes, then $Q(k) \rightarrow Q^*(k)+\delta Q(k)$
and $\alpha \rightarrow \alpha^* + \delta \alpha$. The equation $\alpha = f(\alpha)$ can be
expanded as
\begin{eqnarray}
\alpha^* + \delta \alpha
& = & f(\alpha^*) + \frac{\partial f}{\partial \alpha}
 \delta \alpha
+\sum\limits_{k \geq 1} \frac{\partial f}{\partial Q(k)} \delta Q(k) \nonumber \\
& & + \frac{1}{2} \frac{\partial^2 f}{\partial \alpha^2} (\delta \alpha)^2
+ \sum\limits_{k \geq 1} \frac{\partial^2 f}{\partial \alpha \partial Q(k)}
\delta Q(k) \delta \alpha
 + \frac{1}{2} \sum\limits_{k,k^\prime  \geq 1}
\frac{\partial^2 f}{\partial Q(k)\partial Q(k^\prime)} \delta Q(k) \delta Q(k^\prime) \nonumber \\
& & +\ {\rm higher}\ {\rm order}\ {\rm terms} \; .
\end{eqnarray}
All the derivatives in the above equation are calculated at the transition point.

Because $\frac{\partial f}{\partial \alpha} =1$ at $\alpha= \alpha^*$, the above
equation is a quadratic equation of $\delta \alpha$ if we keep only the lowest-order terms:
\begin{equation}
\frac{1}{2} \frac{\partial^2 f}{\partial \alpha^2}
 (\delta \alpha)^2 + \sum_{k \geq 1}
 \frac{\partial f}{\partial Q(k)} \delta Q(k) = 0 \; .
\end{equation}
This expression leads to
\begin{equation}
\label{eq:deltaalpha}
\delta \alpha \propto
\Bigl|\sum_{k \geq 1}  \frac{\partial f}{\partial Q(k)} \delta Q(k) \Bigr|^{1/2} \; .
\end{equation}

The change in mean degree is
\begin{equation}
c - c^* \approx \sum_{k\geq 1} k \delta P(k) \approx c^* \sum_{k\geq 1} \delta Q(k) \; ,
\end{equation}
which scales linearly with the change of
the probability distribution $Q(k)$. In other words,
 $\delta Q(k) \propto (c-c^*)$. Because
$n_\mm{p\mhyphen core}-n_\mm{p\mhyphen core}^*$ is proportional to
 $\delta \alpha$ for small $\delta \alpha$,
 then Eq.~(\ref{eq:deltaalpha}) leads to the
 scaling behaviour shown in Eq.~(\ref{eq:nscale}).

\clearpage

{\bf Supplementary Note 7}

We offer some details on the numerical and analytical
calculations for  Erd\"{o}s-R\'{e}nyi (ER) random networks,
random regular (RR) networks, and scale-free (SF) random networks.

In our computer simulations, random networks characterized by a given
degree distribution are mainly generated by the configuration model
[45].
First each node $i$ of the network is assigned a
degree $k_i$ following the degree distribution $P(k)$, with a
 tiny restriction that the sum of node degrees must be even.
After each node $i$ has been assigned a degree $k_i$,
we attach $k_i$ `stubs' to each node $i$. We then randomly
pair two stubs to form a link
 (self-connections and multiple links between the
same pair of nodes are not allowed). We repeat this  process until all the
stubs have been used up.
An initial connection pattern of the random network with $N$ nodes and $M$ links is then formed. The links are then shuffled many times to further randomize
the connection pattern (typically $20 \times M$ link-shuffling trials are performed
for each random network). More details about the network generation process can be
found in [46, 47, 48, 49].

In our computer simulations, pseudo-random numbers are
generated by the random number generators of the TRNG library
[50].

We set $K^\prime = K$ and $p=1$ in all the following
theoretical calculations of this
supplementary note.

\subsection*{Erd\"{o}s-R\'{e}nyi (ER) random networks}

% figure S6
\begin{figure}[b]
\begin{center}
\resizebox{8.5cm}{!}{\includegraphics{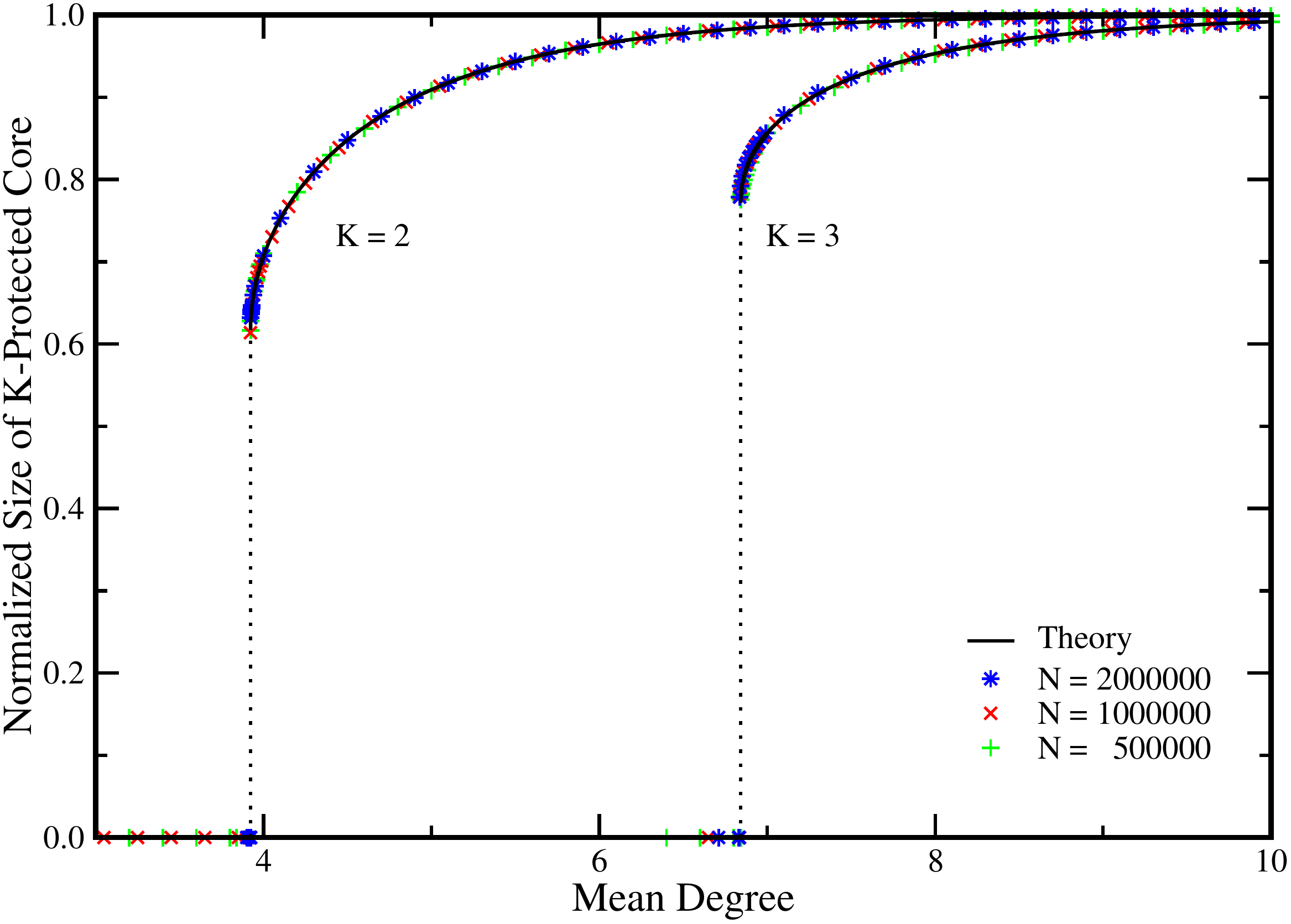}}
\caption{\label{fig:KJammCoreER}
{\bf Normalized size of the $K$-protected core for the ER random network.}
The value of $K$ is set to $K=2$ or $K=3$.
Symbols are simulation results on three network samples of size
$N=5 \times 10^5$, $10^6$, and $2\times 10^6$, respectively, while
lines are theoretical predictions at $N=\infty$.
}
\end{center}
\end{figure}\noindent

%\clearpage

For ER random networks in the limit of $N\rightarrow +\infty$,
the degree distribution is a Poisson distribution:
\begin{equation}
P(k) = \frac{e^{-c} c^k}{k!} \; ,
\end{equation}
where $c$ is the mean degree.
The excess degree distribution $Q(k)$ is also a Poisson distribution:
\begin{equation}
Q(k) = \frac{e^{-c} c^{k-1}}{(k-1)!} \; , \quad \quad (k\geq 1) \; .
\end{equation}

For the $2$-protected core percolation problem ($K=K^\prime=2$),
the expressions for
$\alpha$, $\beta$, $\gamma$, and $n_\mm{p \mhyphen core}$
are, respectively,
\begin{eqnarray}
\alpha  & = & e^{-c (1-\alpha - \beta - \gamma)}
+e^{-c (1-\beta)} - e^{-c (1-\beta-\gamma)} \; ,
\label{eq:alpha2ER} \\
\beta  & = & 1 + e^{-c (1-\beta - \gamma)} - e^{-c \alpha} -
e^{-c (1-\alpha - \beta - \gamma)} \; ,
\label{eq:beta2ER} \\
 \gamma  & = &  c (1-\alpha -\beta) e^{-c (1-\beta)} \; ,
\label{eq:gamma2ER} \\
 n_\mm{p \mhyphen core} & = & e^{-c \alpha} - e^{-c (1-\beta)} \bigl[1+
c(1-\alpha-\beta)\bigr] \; .
\label{eq:ncore2ER}
\end{eqnarray}
The $2$-protected core percolation transition occurs at
the critical mean degree $c^* \approx 3.9229$
(see Fig.~3 and Fig.~\ref{fig:KJammCoreER}).

For the $3$-protected core percolation problem
($K=K^\prime = 3$), the expressions for
$\alpha$, $\beta$, $\gamma$ and $n_\mm{p \mhyphen core}$
are, respectively,
\begin{eqnarray}
\alpha  & = & e^{-c (1-\alpha - \beta - \gamma)}
\bigl[1+ c (1-\alpha - \beta - \gamma) \bigr]
+e^{-c (1-\beta)} \bigl[1+ c(1-\alpha-\beta) \bigr] \nonumber \\
& & \quad \quad
 - e^{-c (1-\beta-\gamma)} \bigl[1 + c (1-\alpha -\beta - \gamma) \bigr]
 \; ,
\label{eq:alpha3ER} \\
\beta & = &  1 - e^{-c \alpha}
  - e^{-c (1-\alpha - \beta - \gamma)} \bigl[
1+ c (1-\alpha - \beta - \gamma)\bigr] \nonumber \\
& & \quad\quad
 + e^{-c (1-\beta - \gamma)} \bigl[1+
c(1-\alpha - \beta - \gamma) \bigr]
 \; ,
\label{eq:beta3ER} \\
\gamma  &= & \frac{c^2 (1-\alpha -\beta)^2}{2} e^{-c (1-\beta)} \; ,
\label{eq:gamma3ER} \\
n_\mm{p \mhyphen core}  &=&  e^{-c \alpha}
 - e^{-c (1-\beta)}  \bigl[1+
c(1-\alpha-\beta)
+ \frac{c^2 (1-\alpha -\beta)^2}{2} \bigr]
 \; .
\label{eq:ncore3ER}
\end{eqnarray}
The $3$-protected core percolation transition occurs at
the critical mean degree $c^* \approx 6.8426$
(see Fig.~\ref{fig:KJammCoreER}).

\subsection*{Regular random (RR) networks}

In a random regular network each node has exactly $k_0$ links.
If $k_0  \geq K$,
the $K$-protected
core will contain all the nodes in the network.
We consider the situation of randomly deleting a fraction $\rho$ of the
links. Then each node on average has $c = k_0 (1-\rho)$ links.
The degree distribution of the network is
\begin{equation}
\label{eq:PkRR}
P(k) = C_{k_0}^{k} (1-\rho)^k \rho^{(k_0-k)}
  \quad \quad (k \leq k_0) \; ,
\end{equation}
while the excess degree distribution $Q(k)$ is given by
\begin{equation}
\label{eq:QkRR}
Q(k) = C_{k_0-1}^{k-1}
 (1-\rho)^{k-1}
\rho^{(k_0-k)}
  \quad \quad ( 1 \leq k \leq k_0) \; .
\end{equation}

For the $2$-protected core percolation problem, we have
\begin{eqnarray}
\alpha & = &
\bigl[\rho+(1-\rho) (\alpha + \beta + \gamma) \bigr]^{k_0 - 1}
 + \bigl[ \rho + (1-\rho) \beta \bigr]^{k_0 - 1}
  - \bigl[ \rho + (1-\rho) (\beta + \gamma ) \bigr]^{k_0-1}
 \; , \\
\beta & = & 1- \bigl[ \rho+(1-\rho) (1-\alpha) \bigr]^{k_0-1}
 -\bigl[ \rho+(1-\rho) (\alpha+\beta+\gamma) \bigr]^{k_0-1} \nonumber \\
 & & \quad\quad
 + \bigl[ \rho+(1-\rho)(\beta+\gamma) \bigr]^{k_0-1}
 \; , \\
\gamma & = &  (1-\alpha-\beta) (1-\rho)
(k_0 - 1) \bigl[ \rho + (1-\rho) \beta\bigr]^{k_0 - 2} \; .
\label{eq:gammaRR}
\end{eqnarray}
If each node has $k_0=4$ neighbours, then a phase transition
occurs at $\rho \approx 0.2303$ (corresponding to critical mean degree
$c^* \approx 3.0789$), at which the fraction of protected nodes jumps from
$0$ to $n_\mm{p \mhyphen core}^* \approx 0.7725$.
If each node has $k_0=6$
neighbours, the phase transition occurs at
$\rho \approx  0.4380$ (corresponding
to critical mean degree $c^* \approx 3.3721$), at which the normalized size of
the $2$-protected core
jumps from $0$ to $n_\mm{p \mhyphen core}^* \approx 0.7079$. The comparison between
theory and simulations is shown in Fig.~5.

\subsection*{Random scale-free (SF)  networks}

Two types of random SF networks are considered in this work, namely purely
SF networks and asymptotically  SF networks.

\subsubsection*{Purely scale-free network with a minimal degree
$k_{\rm min}$}

% figure S7
\begin{figure}[b]
\begin{center}
\resizebox{8.5cm}{!}{\includegraphics{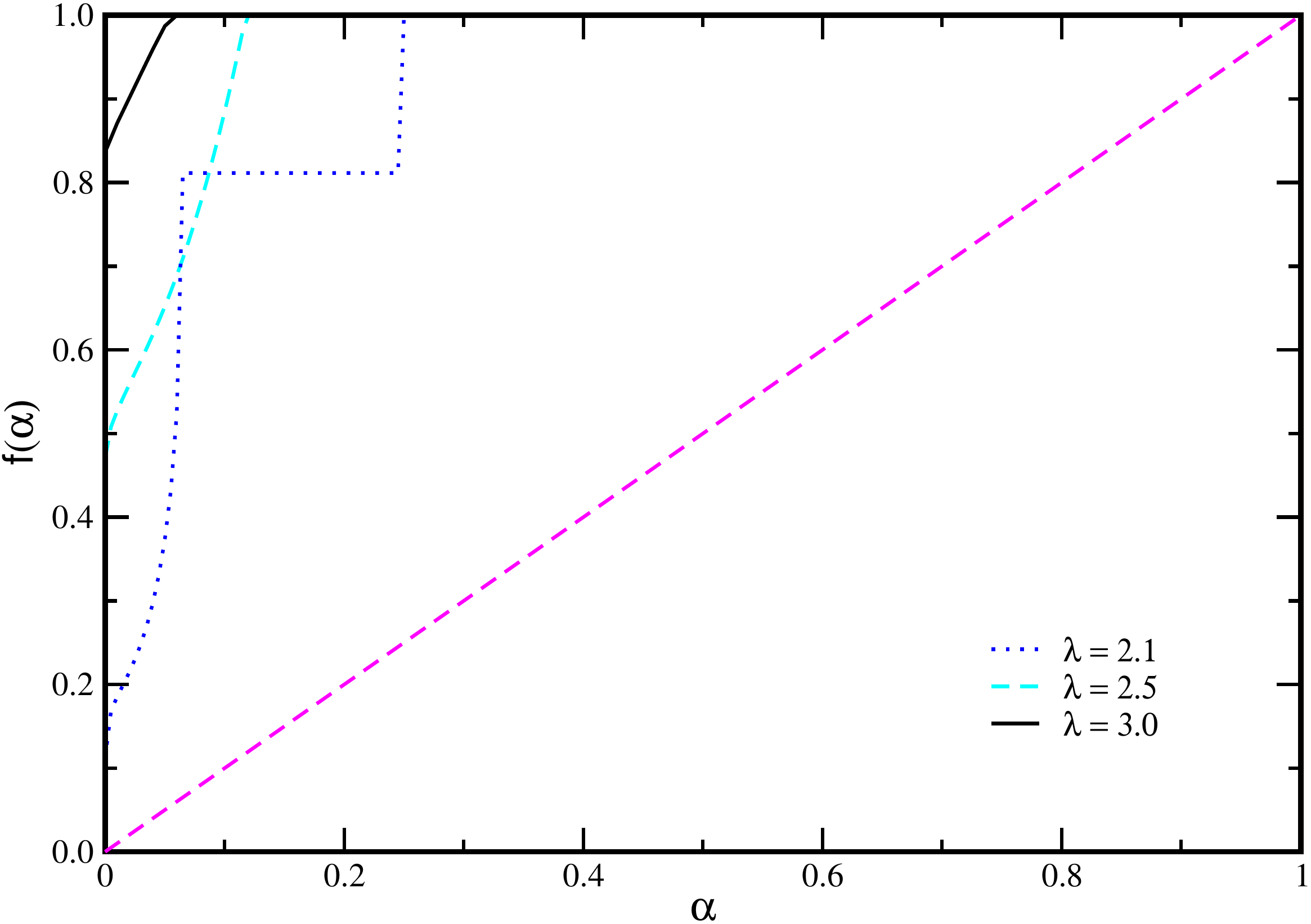}}
\caption{
\label{fig:sfkmin1}
{\bf Function $f(\alpha)$ at $K=K^\prime = 2$ for the purely SF random
network.}
The minimal degree is $k_{\rm min}=1$ and the network size $N=\infty$. The
degree exponent is set to $\lambda = 2.1$ (dotted line), $2.5$ (dashed line) or
$3.0$ (solid line).
The equation $\alpha = f(\alpha)$ has only a single solution
$\alpha = 1$ for any degree exponent $\lambda > 2$.
}
\end{center}
\end{figure}\noindent
%
%\clearpage

This type of SF networks is characterized by the degree distribution
\begin{equation}
P_0(k) = \frac{k^{-\lambda}}{\sum_{m\geq k_{\rm min}} m^{-\lambda}}
\quad\quad (k\geq k_{\rm min}) \; .
\end{equation}
There are two control parameters, the degree exponent $\lambda$ and the
minimal degree $k_{\rm min}$. If $\lambda \leq 2$, the mean degree of the network
diverges in the thermodynamic limit of $N\rightarrow \infty$. We
therefore assume that $\lambda > 2$ hereafter.
The minimal degree $k_{\rm min} \geq 1$.
 The excess degree distribution $Q_0(k) \equiv k P_0(k)/\sum_{k\geq k_{\rm min}} k P_0(k)$ is expressed  as
\begin{equation}
Q_0(k) = \frac{k^{1-\lambda}}{\sum_{m\geq k_{\rm min}} m^{1-\lambda}} \; .
\end{equation}

For the special case of $k_{\rm min}=1$, Fig.~\ref{fig:sfkmin1} shows the
curve of $f(\alpha)$ at several different fixed $\lambda$ values. Because $\alpha = f(\alpha)$
has only a single solution $\alpha = 1$, there is no $K$-protected core percolation
transition in the system for all $K \geq 2$.

When $k_{\rm min} \geq K$, the whole network forms a $K$-protected core.
If a randomly chosen fraction $\rho$ of the links are removed from the network, a
$K$-protected core percolation transition occurs when the mean degree $c$
of the remaining network
is decreased to certain threshold value $c^*$.
After a fraction $\rho$ of the links are removed,
the degree distribution $P(k)$ of the remaining network becomes
\begin{equation}
P(k) = \frac{1}{\sum_{m\geq k_{\rm min}} m^{-\lambda}}
 \sum\limits_{m\geq \max(k, k_{\rm min})} m^{-\lambda}
 C_{m}^{k} (1-\rho)^{k} \rho^{m-k} \; ,
 \end{equation}
while the corresponding excess degree distribution $Q(k)$ is given by
\begin{equation}
Q(k) = \frac{1}{\sum_{m\geq k_{\rm min}} m^{1-\lambda}}
 \sum\limits_{m\geq \max(k, k_{\rm min})} m^{1-\lambda}
 C_{m-1}^{k-1} (1-\rho)^{k-1} \rho^{m-k} \; .
\quad  \quad (k \geq 1)
 \end{equation}

For the case of $K=2$, the expressions for the
probabilities $\alpha$, $\beta$, and $\gamma$ are
\begin{eqnarray}
\alpha & = &
\frac{1}{\sum_{m\geq k_{\rm min}} m^{1-\lambda}}
\sum\limits_{k\geq k_{\rm  min}} k^{1-\lambda}
\Bigl[ \bigl(\rho + (1-\rho) (\alpha + \beta + \gamma)\bigr)^{k-1}
+\bigl(\rho+(1-\rho) \beta \bigr)^{k-1} \nonumber \\
& & \quad\quad \quad \quad\quad\quad\quad \quad\quad\quad\quad\quad
\quad \quad
-\bigl(\rho+(1-\rho) (\beta+\gamma) \bigr)^{k-1}
\Bigr] \; , \\
\beta & = &
1- \frac{1}{\sum_{m\geq k_{\rm min}} m^{1-\lambda}}
\sum\limits_{k\geq k_{\rm  min}} k^{1-\lambda}
\Bigl[ \bigl(\rho + (1-\rho) (1-\alpha )\bigr)^{k-1}
+\bigl(\rho+(1-\rho) (\alpha+ \beta +\gamma) \bigr)^{k-1} \nonumber \\
& & \quad\quad \quad \quad \quad\quad \quad\quad \quad\quad\quad\quad \quad\quad\quad \quad
-\bigl(\rho+(1-\rho) (\beta+\gamma) \bigr)^{k-1}
\Bigr] \; ,  \\
\gamma & = & \frac{(1-\alpha - \beta) (1-\rho)}{
\sum_{m\geq k_{\rm min}} m^{1-\lambda}} \sum\limits_{k\geq k_{\rm  min}} k^{1-\lambda}
(k-1) \bigl(\rho+ (1-\rho) \beta\bigr)^{k-2}
\; .
\end{eqnarray}
And the normalized size of the $2$-protected core is
\begin{eqnarray}
n_\mm{p \mhyphen core} &=& \frac{1}{\sum_{m\geq k_{\rm min}} m^{-\lambda}}
\sum\limits_{k \geq k_{\rm min}} m^{-\lambda}
\Bigl[ \bigl(\rho + (1-\rho) (1-\alpha) \bigr)^{k}
- \bigl( \rho+ (1-\rho) \beta \bigr)^{k}  \nonumber \\
& & \quad\quad\quad\quad\quad
- (1-\rho) (1-\alpha-\beta) k \bigl( \rho + (1-\rho) \beta \bigr)^{k-1}
\Bigr] \; .
\end{eqnarray}
As long as $\lambda > 2$,
the summations in the above several equations converge  even when
the network size $N=\infty$.

For networks of finite size $N$, we generate a scale-free degree distribution as follows [46]:
(0) Initialize an integer $n=0$ and initialize the degree to be
$k=k_{\rm min}$, and
initialize the candidate node set $U$ as containing all the $N$ nodes.
(1) Set $n_k$ to be the integer that is closest to the real value
$N P_0(k)$ (if $n_k=0$, then set $n_k=1$, and if $n+n_k > N$, then set $n_k=N-n$);
perform the updating $n\leftarrow n+n_k$, and choose $n_k$ different nodes from the candidate set $U$ and  assign the degree $k$ to each of them (and then remove these nodes from set $U$).
(2) Set $k\leftarrow k+1$, and go back to step $(1)$ if $n < N$.

Through this construction, the degree distribution of the network is scale-free with a maximal
degree $k_{\rm max}$, whose value scales with $N$ as
$k_{\rm max} \approx k_{\rm min} N^{1/(\lambda-1)}$ [46].

After each node has been assigned a degree, we can construct a random SF network by
the configuration model. Notice that when $\lambda < 3$ there are intrinsic
degree correlations in a random SF network [42, 43, 46, 47, 48, 49].

% figure S8
\begin{figure}[t]
\begin{center}
\resizebox{8.5cm}{!}{\includegraphics{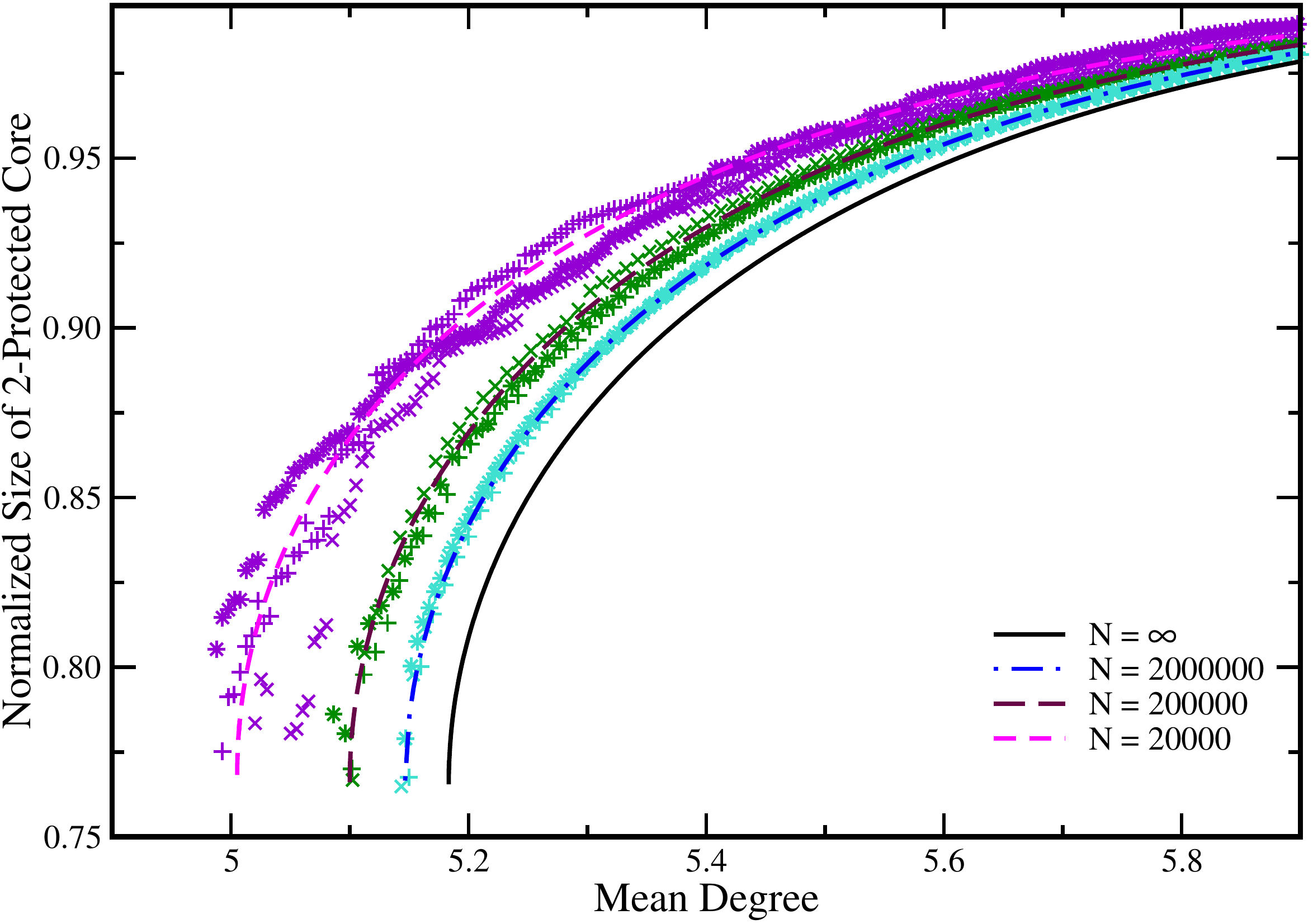}}
\caption{\label{fig:KJammCoreSFK2}
{\bf Normalized size of $2$-protected core for SF random networks
with $\lambda =3$ and
$k_{\rm min}=4$.}
A fraction $\rho$ of the links are randomly chosen and
removed from the network.
At each value of $N$, the line is theoretical prediction based on the degree distribution $P(k)$ while symbols are simulation results on  three independent
network instances with the same $P(k)$.
The networks are obtained through the
configuration model.
}
\end{center}
\end{figure}\noindent
%\clearpage

Some results on random SF networks with $\lambda=3$ and $k_{\rm min}=4$ are shown in
Fig.~\ref{fig:KJammCoreSFK2}.
For $N\rightarrow \infty$, a
 $2$-protected core percolation transition occurs when the
 fraction of removed links $\rho\approx 0.2692$
 (corresponding to critical mean degree $c^* \approx 5.1832$), with
 a jump of $n_\mm{p \mhyphen core}$ from $0$ to $n_\mm{p \mhyphen core}^*
 \approx 0.7655$. For finite networks, however, the theory predicts that the
 $2$-protected core actually
 is formed at much lower values of mean degree. For example,
Fig.~\ref{fig:KJammCoreSFK2} demonstrates that,
the critical degree is $c^* \approx 5.0052$ (for $N=2\times 10^4$),
 $c^*\approx 5.0999$ (for $N=2\times 10^5$),
$c^* \approx 5.1465$ (for $N=2 \times 10^6$), respectively.
These predictions on finite-$N$ systems are confirmed by simulation results on single network instances.

% figure S9
\begin{figure}[t]
\begin{center}
\resizebox{8.5cm}{!}{\includegraphics{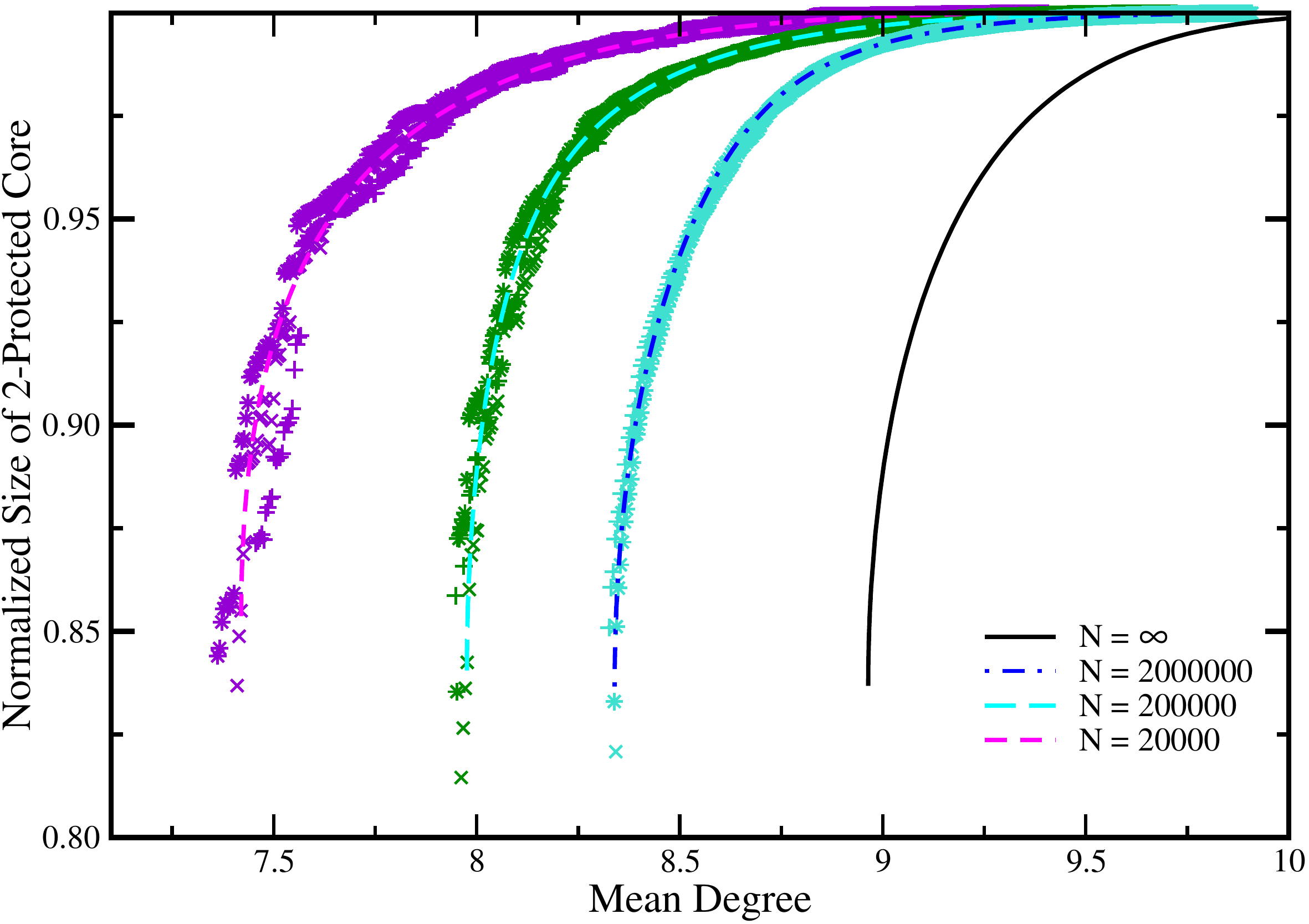}}
\caption{
\label{fig:sfsimulation}
{\bf Normalized size of $2$-protected core for SF random networks
with $\lambda =2.5$ and
$k_{\rm min}=4$.}
A fraction $\rho$ of the links are randomly chosen and
removed from the network.
At each value of $N$, the line is theoretical prediction based on the degree distribution $P(k)$ while symbols are simulation results on  three independent
network instances with the same $P(k)$. The networks are obtained through the
configuration model.
}
\end{center}
\end{figure}\noindent
%\clearpage

Figure~\ref{fig:sfsimulation} shows the comparison
between theory and simulations on $2$-protected core percolation
for random SF networks with minimal degree $k_{\rm min}=4$
and degree exponent $\lambda = 2.5$. Similar to the results shown in
Fig.~\ref{fig:KJammCoreSFK2}, the $2$-protected
core percolation transition is discontinuous, and there are
also strong finite-size effects.

\subsubsection*{Static model}

Random SF networks can also be constructed from the static model [41].
In the static model, each node $i \in \{1, 2, \ldots, N\}$ has a weight
$w_i = \frac{i^{-\xi}}{\sum_{j=1}^{N} j^{-\xi}}$,
where $ 0 \leq \xi < 1$ is a control parameter. To create a link, two nodes
$i$ and $j$
are chosen independently from the set of $N$ nodes, and the probability that
node $i$ and node $j$ being chosen is equal to $w_i \times w_j$; if
nodes $i$ and $j$ are different and the link $(i, j)$ has not been created before,
then a link between $i$ and $j$ is set up. By repeating this connection process,
a total number of $M = (c/2) N$ links are connected between pairs of
nodes, with $c$ being the mean degree of the network. The resulting network
has a scale-free degree distribution $P(k) \propto k^{-\lambda}$ for $k \gg 1$,
with degree exponent $\lambda = 1 + 1/ \xi $ [41].
In the limit of $N\rightarrow \infty$, an analytic expression for $P(k)$ is
given by [42]
\begin{equation}
P(k) = \frac{[c (1-\xi)]^k}{\xi \ k! }
\int_{1}^{\infty} {\rm d} t \  e^{-c (1-\xi) t} t^{k-1-1/\xi} \; .
\end{equation}
At $N\rightarrow \infty$ the excess degree distribution $Q(k)$ is given by
\begin{equation}
\label{eq:staticQk}
Q(k) = \Bigl(\frac{1}{\xi}-1\Bigr)
\frac{[c (1-\xi)]^{k-1}}{ (k-1)! }
\int_{1}^{\infty} {\rm d} t \ e^{-c (1-\xi) t} t^{k-1-1/\xi} \; .
\end{equation}
If we set $\xi = 0$ in the static model we then obtain
ER random networks.
For $\xi < 0.5$, the degree-degree correlations of neighbouring
nodes in the network are negligible. But as $\xi$ increases from $0.5$
(therefore $\lambda$ is below $3.0$), the degree-degree correlations become
more and more pronounced [42, 43].

% figure S10
\begin{figure}[t]
\begin{center}
\resizebox{8.5cm}{!}{\includegraphics{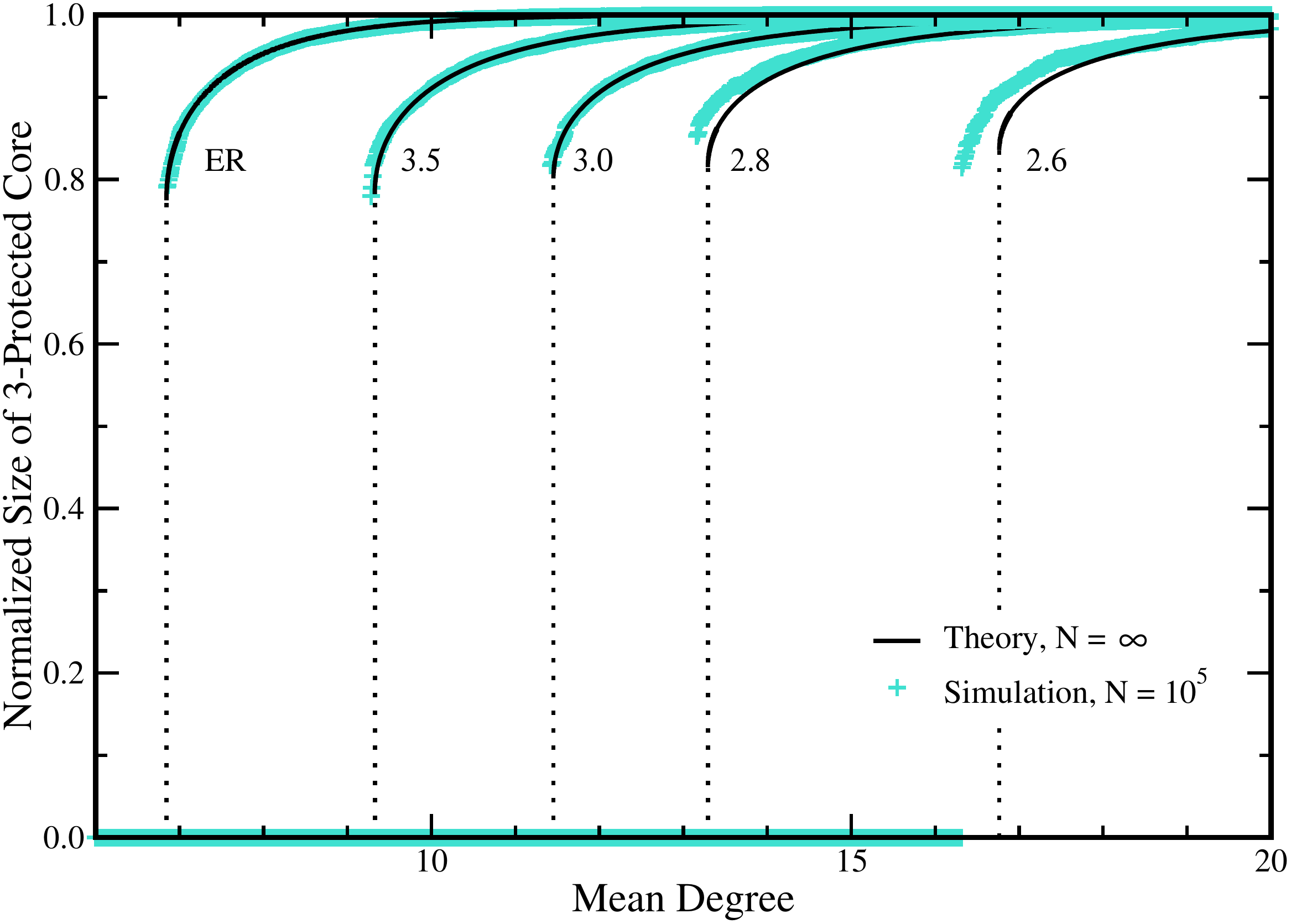}}
\caption{
\label{fig:sfstaticK3}
{\bf Normalized $3$-protected core size for ER
  networks and SF networks.} The degree exponents of the SF networks
  are $\lambda = 3.5, 3.0, 2.8, 2.6$ (from left to right).
  Lines are analytic predictions for infinite system ($N=\infty$),
  while symbols are simulation results obtained on a single network instance with
  $N=10^5$ nodes.
  The SF and ER networks are all generated through the static model.
}
\end{center}
\end{figure}\noindent
%
%\clearpage

For the case of $K=2$, using Eq.~(\ref{eq:staticQk}) we obtain the
follow expressions for $\alpha$, $\beta$, $\gamma$, and the normalized
$2$-protected core size:
\begin{eqnarray}
\alpha & = & \Bigl(\frac{1}{\xi}-1\Bigr)
\Bigl\{ E_{\frac{1}{\xi}}\bigl [ c (1-\xi) (1-\alpha -\beta -\gamma)\bigr]
+E_{\frac{1}{\xi}}\bigl[ c (1-\xi) (1-\beta) \bigr] \nonumber \\
& & \quad \quad\quad
-E_{\frac{1}{\xi}}\bigl[ c (1-\xi) (1-\beta - \gamma) \bigr]
\Bigr\} \; , \\
\beta & = & 1 - \Bigl(\frac{1}{\xi}-1\Bigr)
\Bigl\{ E_{\frac{1}{\xi}}\bigl[ c (1-\xi) \alpha \bigr]
+E_{\frac{1}{\xi}}\bigl[ c (1-\xi) (1-\alpha - \beta - \gamma) \bigr]
\nonumber \\
& & \quad\quad\quad\quad
-E_{\frac{1}{\xi}}\bigl[ c (1-\xi) (1-\beta - \gamma) \bigr]
\Bigr\} \; , \\
\gamma & = & \frac{(1-\alpha - \beta) c (1-\xi)^2}{\xi}
E_{\frac{1-\xi}{\xi}}\bigl[ c (1-\xi) (1-\beta)\bigr]
\;
, \\
n_\mm{p \mhyphen core} & = & \frac{1}{\xi}E_{\frac{1+\xi}{\xi}}\bigl[
c(1-\xi) \alpha \bigr]
-\frac{1}{\xi} E_{\frac{1+\xi}{\xi}} \bigl[
c(1-\xi) (1-\beta) \bigr] \nonumber \\
& & \quad\quad\quad
- \frac{ c (1-\xi) (1-\alpha - \beta)}{\xi}
E_{\frac{1}{\xi}}\bigl[ c (1-\xi) (1-\beta) \bigr] \; .
\end{eqnarray}
In the above equations, $E_a(x)$ is the generalized
 exponential integral defined as
$E_a(x) \equiv \int_{1}^{\infty} {\rm d} t e^{- x t} t^{-a}$.
In the numerical calculations, the value of $E_a(x)$ is calculated by converting it
into an incomplete gamma function and then using the GNU Scientific Library (gsl,
{\tt http://www.gnu.org/software/gsl/}).

For the case of $K=3$, the explicit expressions for
$\alpha$, $\beta$, $\gamma$, and $n_\mm{p \mhyphen core}$ are, respectively,
\begin{eqnarray}
\alpha & = &
\Bigl(\frac{1}{\xi}-1\Bigr)
\Bigl\{ E_{\frac{1}{\xi}}\bigl[c (1-\xi) (1-\alpha -\beta -\gamma)\bigr]
+E_{\frac{1}{\xi}}\bigl[ c (1-\xi) (1-\beta) \bigr] \nonumber \\
& & \quad\quad
 -E_{\frac{1}{\xi}}\bigl[ c (1-\xi) (1-\beta - \gamma) \bigr]
\Bigr\}  + \frac{(1-\alpha - \beta) c (1-\xi)^2}{\xi}
E_{\frac{1-\xi}{\xi}}\bigl[ c (1-\xi) (1-\beta) \bigr] \nonumber \\
& &
+ \frac{(1-\alpha - \beta -\gamma) c (1-\xi)^2}{\xi}
\Bigl\{
E_{\frac{1-\xi}{\xi}}\bigl[ c (1-\xi) (1-\alpha -\beta -\gamma) \bigr] \nonumber \\
& & \quad\quad\quad\quad\quad\quad\quad\quad\quad\quad\quad\quad\quad\quad
-E_{\frac{1-\xi}{\xi}}\bigl[ c (1-\xi) (1-\beta -\gamma) \bigr]
\Bigr\}  \; , \\
\beta &  = & 1 - \Bigl(\frac{1}{\xi}-1\Bigr)
\Bigl\{ E_{\frac{1}{\xi}}\bigl[c (1-\xi) \alpha  \bigr]
+E_{\frac{1}{\xi}}\bigl[ c (1-\xi) (1-\alpha - \beta -\gamma) \bigr]
-E_{\frac{1}{\xi}}\bigl[ c (1-\xi) (1-\beta - \gamma) \bigr]
\Bigr\} \nonumber \\
& &  - \frac{(1-\alpha - \beta -\gamma) c (1-\xi)^2}{\xi}
\Bigl\{
E_{\frac{1-\xi}{\xi}}\bigl[ c (1-\xi) (1-\alpha -\beta -\gamma) \bigr]
\nonumber \\
& & \quad\quad\quad\quad\quad\quad\quad\quad\quad\quad\quad\quad\quad\quad
-E_{\frac{1-\xi}{\xi}}\bigl[ c (1-\xi) (1-\beta -\gamma) \bigr]
\Bigr\}  \; , \\
\gamma & = & \frac{(1-\alpha - \beta)^2 c^2 (1-\xi)^3}{2 \xi}
E_{\frac{1-2 \xi}{\xi}}\bigl[ c (1-\xi) (1-\beta) \bigr] \; , \\
n_\mm{p \mhyphen core} & = &
\frac{1}{\xi} \Bigl\{
E_{\frac{1+\xi}{\xi}}\bigl[ c (1-\xi) \alpha \bigr]
-E_{\frac{1+\xi}{\xi}}\bigl[ c (1-\xi) (1-\beta) \bigr] \Bigr\}
-\frac{(1-\alpha -\beta) c (1-\xi)}{\xi}
E_{\frac{1}{\xi}} \bigl[ c (1-\xi) (1-\beta) \bigr] \nonumber \\
& & - \frac{ (1-\alpha - \beta)^2 c^2 (1-\xi)^2}{2 \xi}
E_{\frac{1-\xi}{\xi}}\bigl[c (1-\xi) (1-\beta) \bigr] \; .
\end{eqnarray}
Figure~\ref{fig:sfstaticK3} shows additional numerical results on the
normalized $3$-projected core size for SF networks generated through the
static model.

\clearpage

{\bf Supplementary note 8}

We offer more details on the numerical simulations in lattice systems.

% figure S11
\begin{figure}[b]
  \begin{center}
  \subfigure[]{
    \resizebox{85mm}{!}{\includegraphics{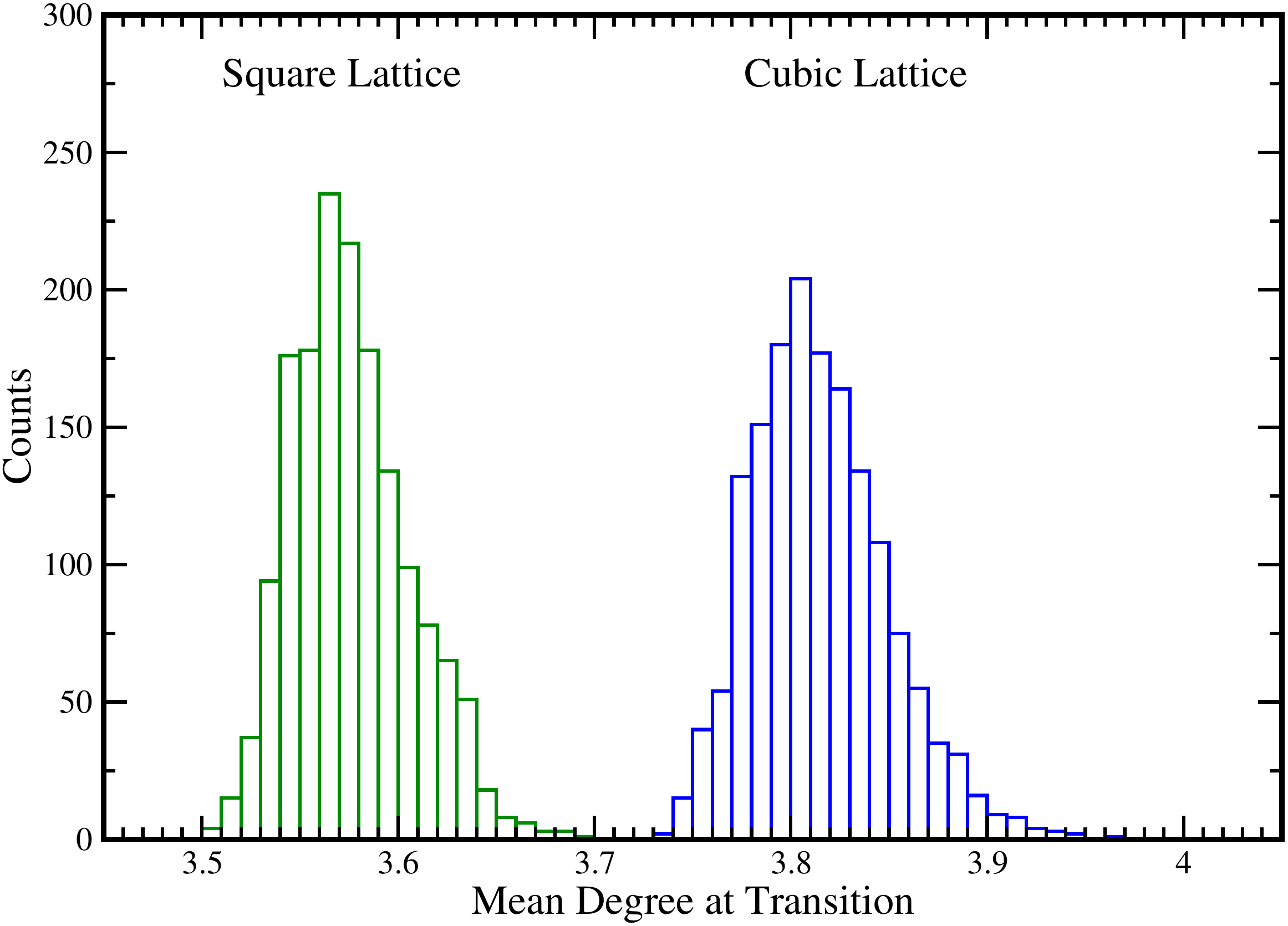}}
  }
%  \hspace{1.0cm}
  \subfigure[]{
    \resizebox{85mm}{!}{\includegraphics{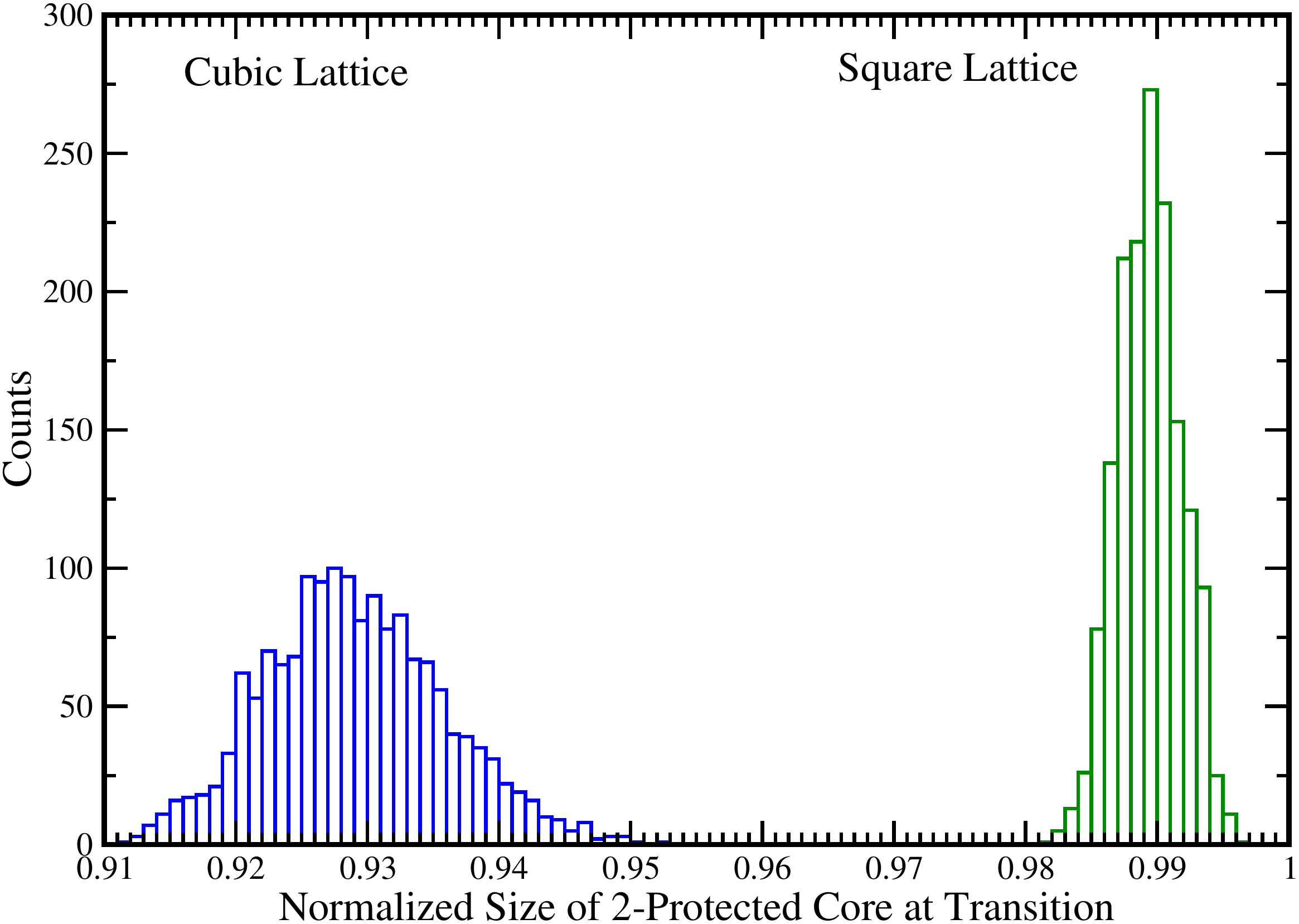}}
  }
  \caption{
   \label{fig:Histogram}
{\bf Fluctuations of the $2$-protected core percolation transition point in
2D square lattice and 3D cubic lattice.}
Histograms of mean network degree $c^*$ (a) and normalized sizes of the
$2$-protected core $n_{p-core}^*$ (b) at the transition point
are obtained by simulations on
$1600$ independent network instances of the 2D periodic
square lattice and the 3D periodic cubic lattice of  size $N =
10^6$.
}
\end{center}
\end{figure}\noindent
%
%\clearpage

We consider $D$-dimensional hypercubic lattices of  side length $L$ and
periodic boundary conditions. The total number of
nodes in the lattice is $N=L^D$, and each
node has $k_0= 2 D$ links.  After a randomly chosen fraction $\rho$ of the
links are deleted from the network, the degree distribution
 $P(k)$ and the excess degree distribution $Q(k)$ of the remaining
network are described by Eq.~(\ref{eq:PkRR}) and Eq.~(\ref{eq:QkRR}), respectively.
Therefore the theoretical predictions of $K$-protected
core percolation in the lattice systems are
identical to those of the random regular network systems.

Some simulation results for the $D=2$ (side length $L=1000$) and $D=3$ (side length $L=100$)
lattices are described in Fig.~5. Our mean field
theory correctly predicts the normalized
$2$-protected core size once the percolation transition occurs, but it fails to predict the
transition point. With a given value of $N$,
the $2$-protected core percolation transition points $c^*$ actually fluctuate considerably among different
network instances (obtained by removing a
randomly chosen subset of the whole links). Figure~\ref{fig:Histogram}
shows the fluctuations of the value of $c^*$ among
$1600$ independent network instances with $N = 10^6$ nodes,
and the associated fluctuations of the normalized $2$-protected core sizes
$n_{\rm p-core}^*$.

% figure S12
\begin{figure}[t]
\begin{center}
\resizebox{8.5cm}{!}{\includegraphics{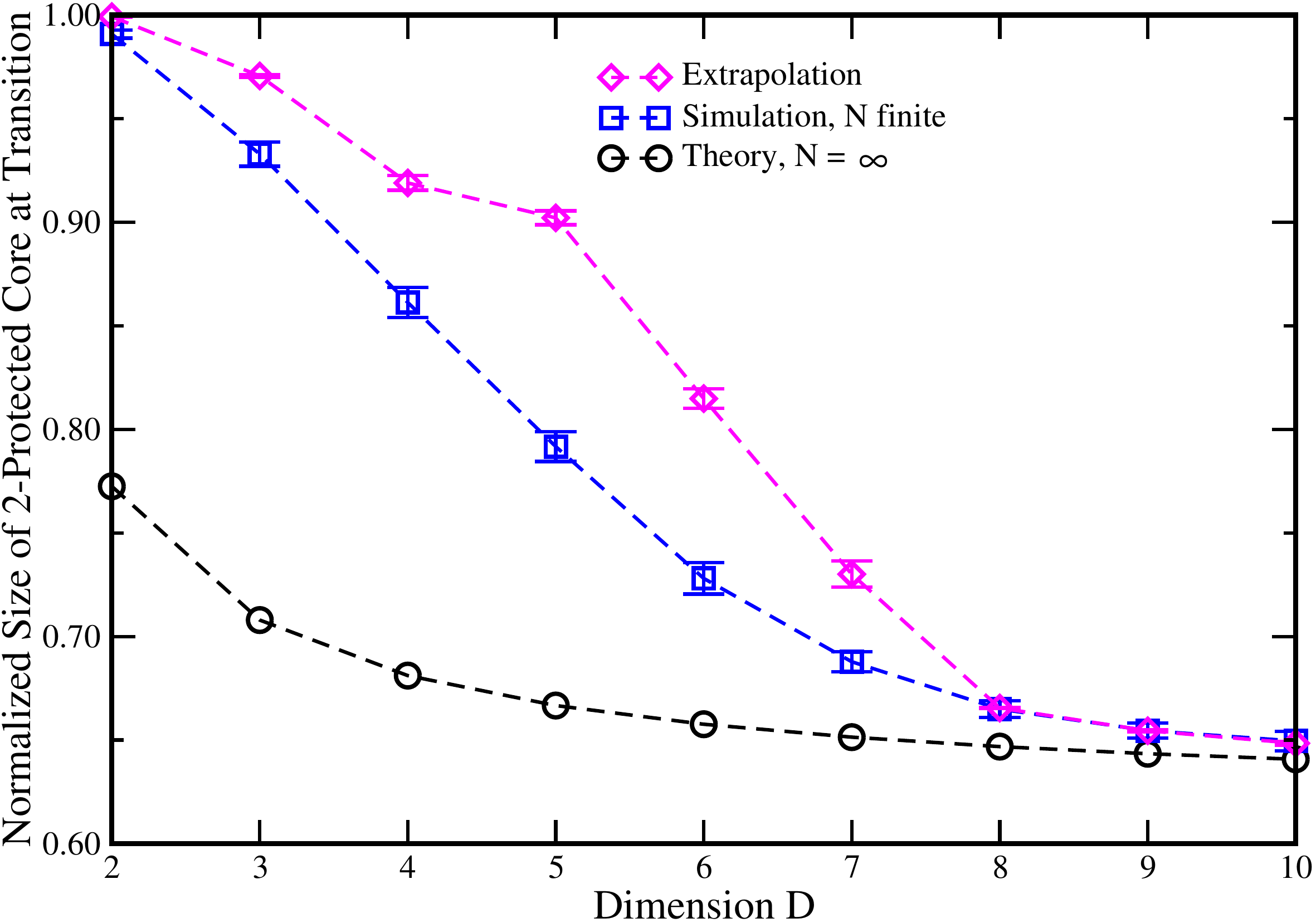}}
\caption{ \label{fig:Dimension2}
{\bf The normalized size of the $2$-protected core at the
transition point in  $D$-dimensional hypercubic lattices.}
Each node of the network originally has $k_0 = 2 D$ neighbours.
A randomly chosen fraction $\rho$ of
the links are deleted from the network.
The lattice data (squares) were obtained by simulating on
$1600$ independent network instances (the number of nodes in the networks is $N=L^D$,
with $L=1414, 126, 38, 18, 11, 8, 6, 5, 4$ for $D=2$ to $D=10$, respectively); the diamond symbols
are the extrapolated simulation value in the $N\rightarrow \infty$ limit.
The theoretical predictions (circle symbols)
on the $2$-protected core percolation transition point for
 infinite large system ($N = \infty$) are obtained by
 using Eqs.~(\ref{eq:PkRR})--(\ref{eq:gammaRR}).
}
\end{center}
\end{figure}\noindent
%
%\clearpage

\begin{figure}[t]
  \begin{center}
  \subfigure[]{
  \resizebox{85mm}{!}{\includegraphics{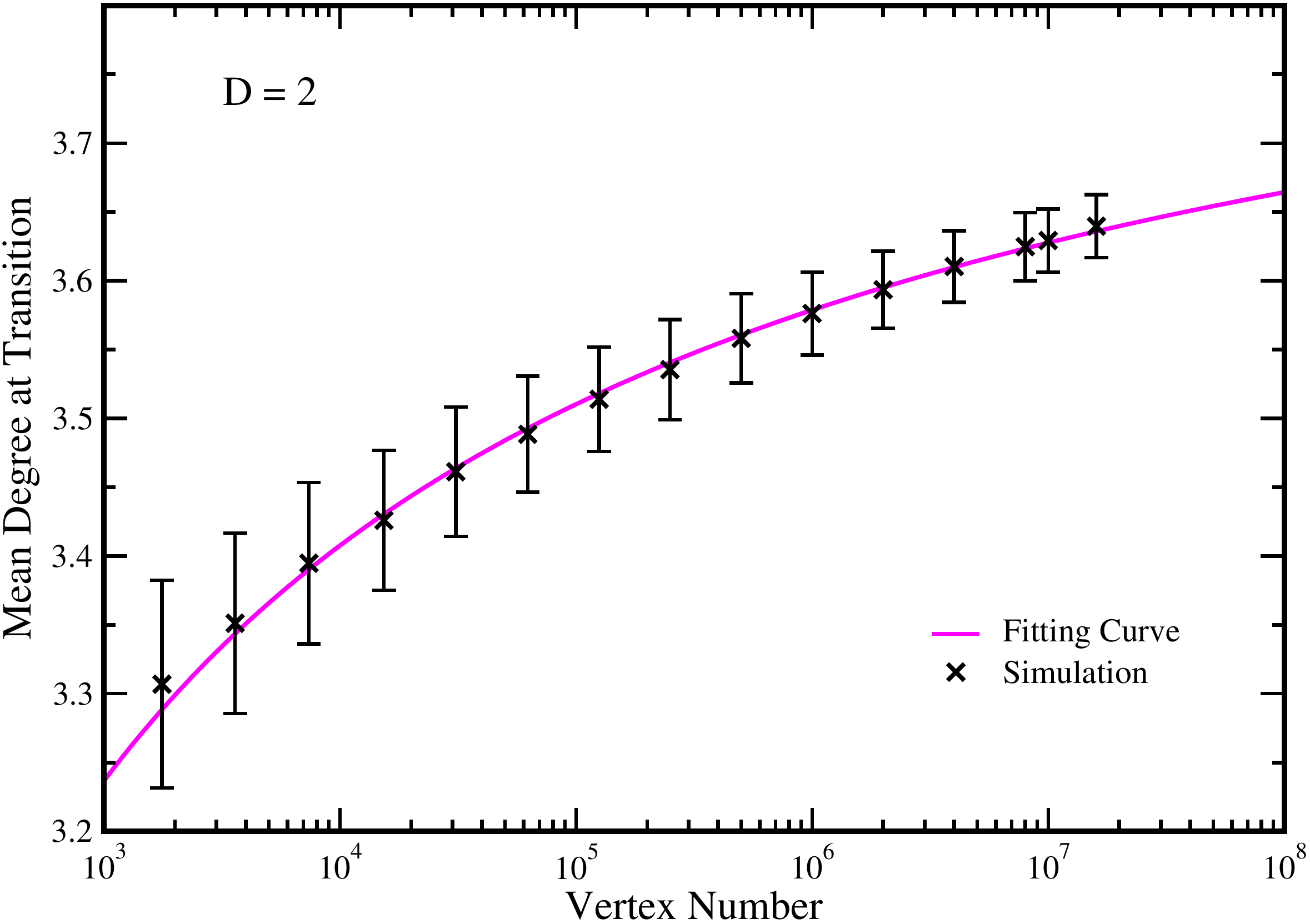}}
  }
%  \hspace{1cm}
  \subfigure[]{
  \resizebox{85mm}{!}{\includegraphics{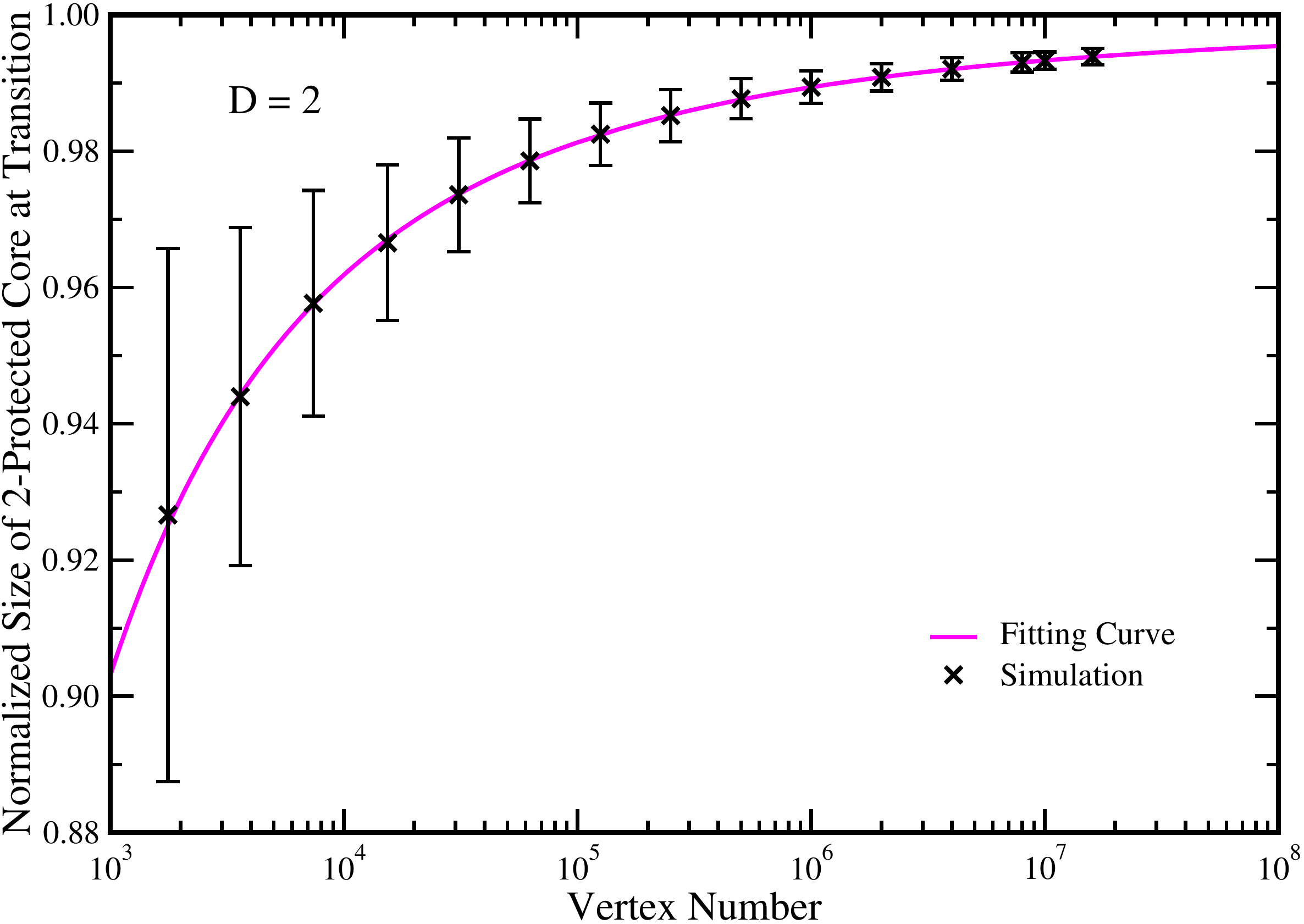}}
  }
\caption{ \label{fig:2D}
{\bf Extrapolation of $2$-protected core percolation transition point to the $N\rightarrow \infty$
limit for square lattices.}
(a) Average value  $\overline{c^*}$ of the mean degree $c^*$ at
 the $2$-protected core transition.
 (b) Average value $\overline{n_{\rm p-core}^*}$ of
 the normalized $2$-protected core size $n_{\rm p-core}^*$ at the percolation transition.
 Each data point is obtained by averaging over
 $1600$ independent network instances. The fitting cure is
 $y = a_1 - b_1/\ln (x)$, with $a_1=3.921 \pm 0.005$ and $b_1=4.73\pm 0.06$
 (a)
 and $y=a_2-b_2/\ln^\eta (x)$ with $a_2=0.9993\pm 0.0001$, $b_2=54.6\pm 3.6$,
 $\eta = 3.28\pm 0.03$ (b).
 The fitting parameters $a_1$ and $a_2$ are regarded as the
 value of $c^*$ and the value of $n_{\rm p-core}^*$ in
 the $N \rightarrow  \infty$ system (they are shown in Fig.~6 and
 Fig.~\ref{fig:Dimension2}).
}
\end{center}
\end{figure}

\begin{figure}[t]
  \begin{center}
  \subfigure[]{
  \resizebox{85mm}{!}{\includegraphics{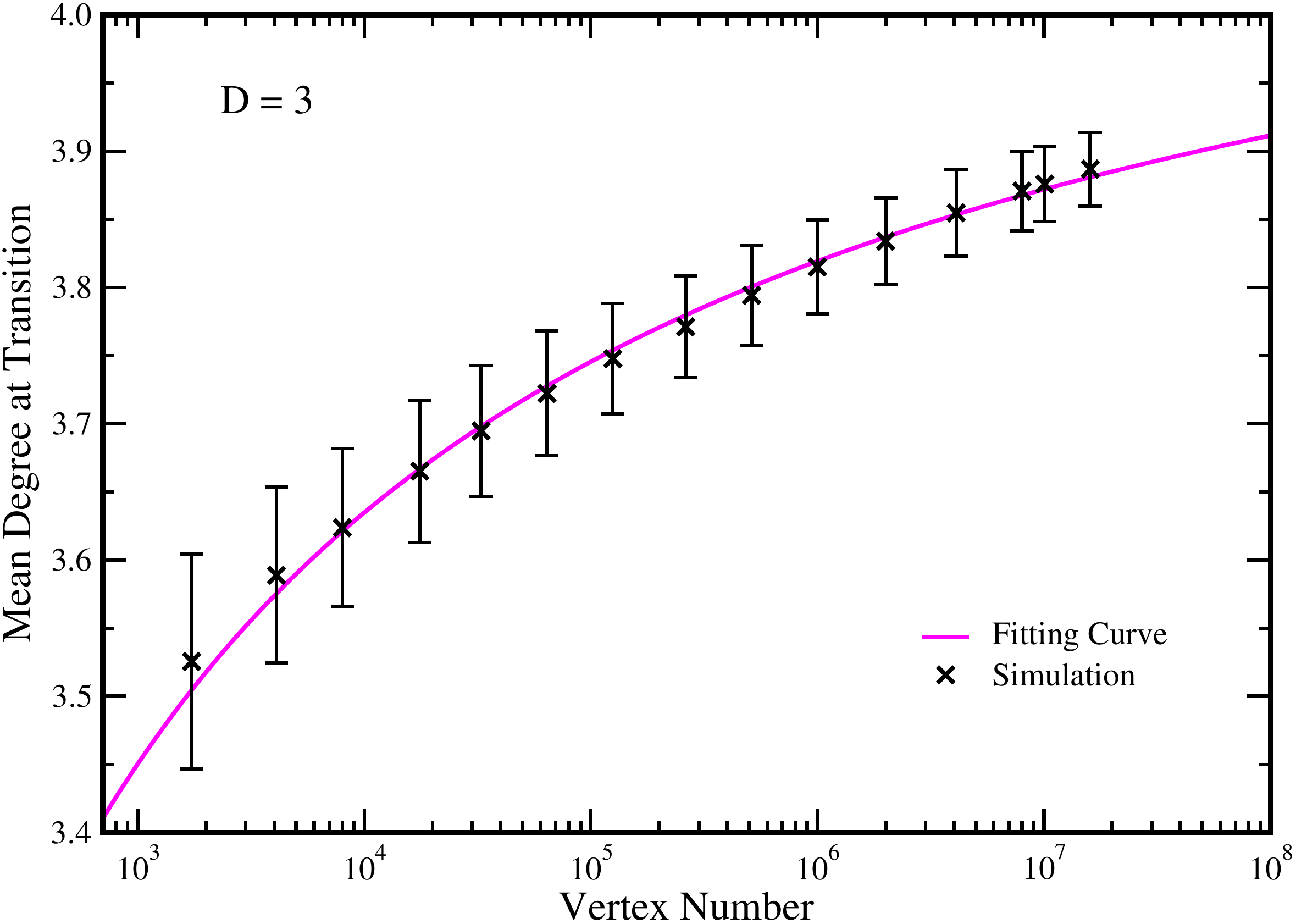}}
  }
%  \hspace{1cm}
  \subfigure[]{
  \resizebox{85mm}{!}{\includegraphics{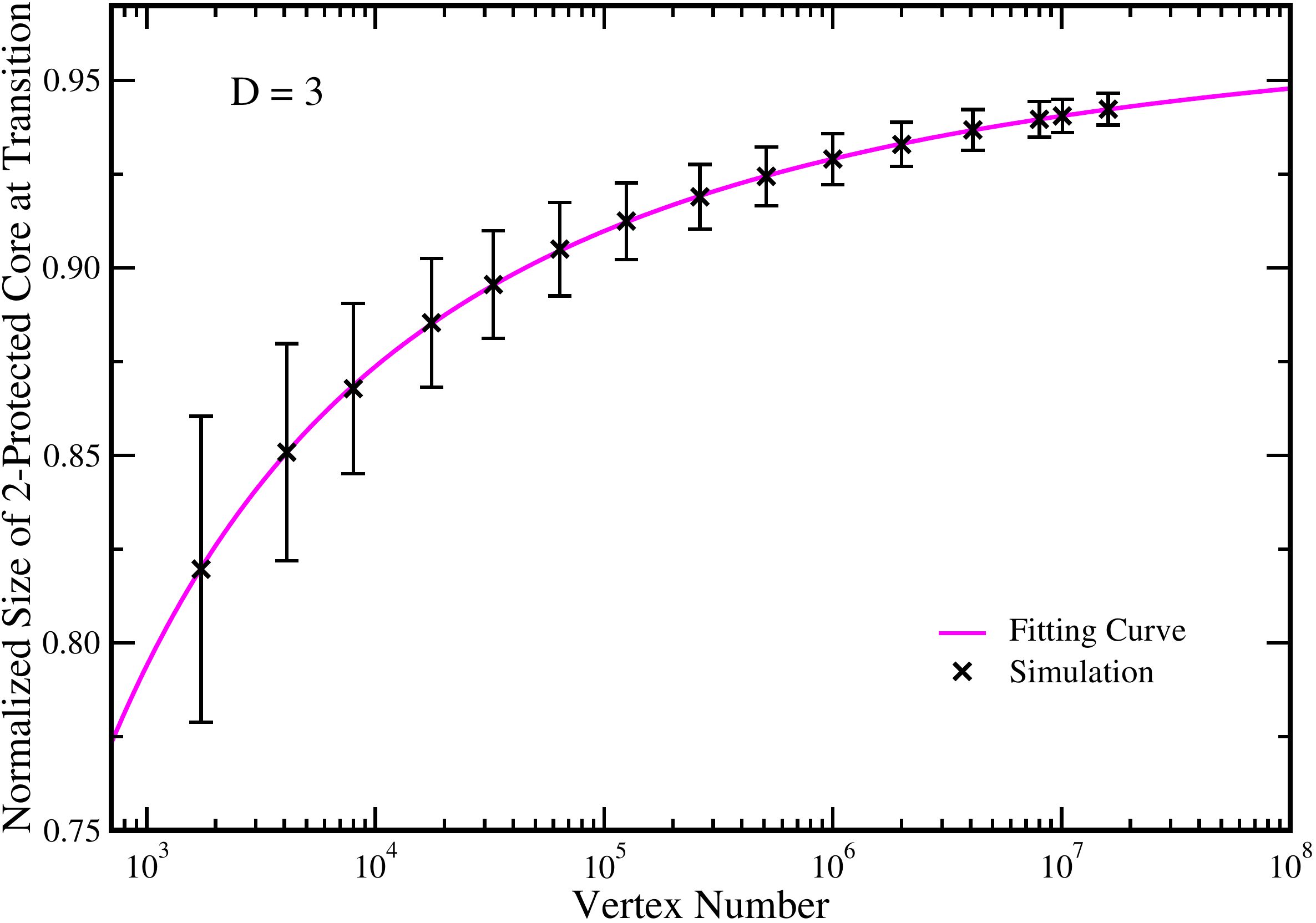}}
  }
\caption{ \label{fig:3D}
{\bf Extrapolation of $2$-protected core percolation transition point to the $N\rightarrow \infty$
limit for cubic lattices.}
(a) Average value  $\overline{c^*}$ of the mean degree $c^*$ at
 the $2$-protected core transition.
 (b) Average value $\overline{n_{\rm p-core}^*}$ of
 the normalized $2$-protected core size $n_{\rm p-core}^*$ at the percolation transition.
Each data point is obtained by averaging over
 $1600$ independent network instances.
The fitting cure is  $y = a_1 - b_1/\ln (x)$, with $a_1=4.188 \pm 0.007$ and $b_1=5.1\pm 0.1$ (a)
 and $y=a_2-b_2/\ln^\eta (x)$ with $a_2=0.9707\pm 0.0006$, $b_2=9.99\pm 0.47$,
 $\eta = 2.09\pm 0.02$ (b).
}
\end{center}
\end{figure}

\begin{figure}[t]
  \begin{center}
  \subfigure[]{
  \resizebox{85mm}{!}{\includegraphics{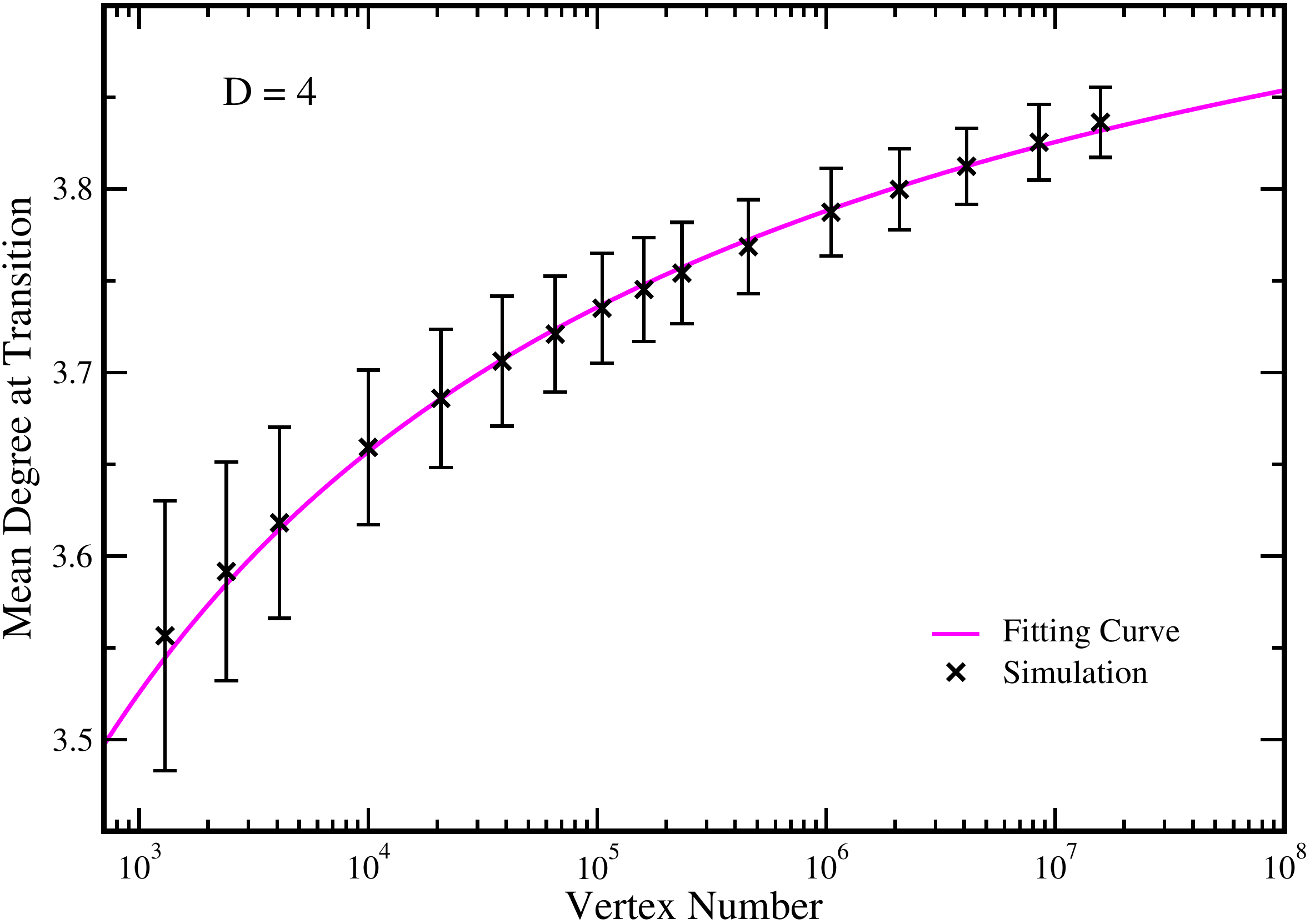}}
  }
%  \hspace{1cm}
  \subfigure[]{
  \resizebox{85mm}{!}{\includegraphics{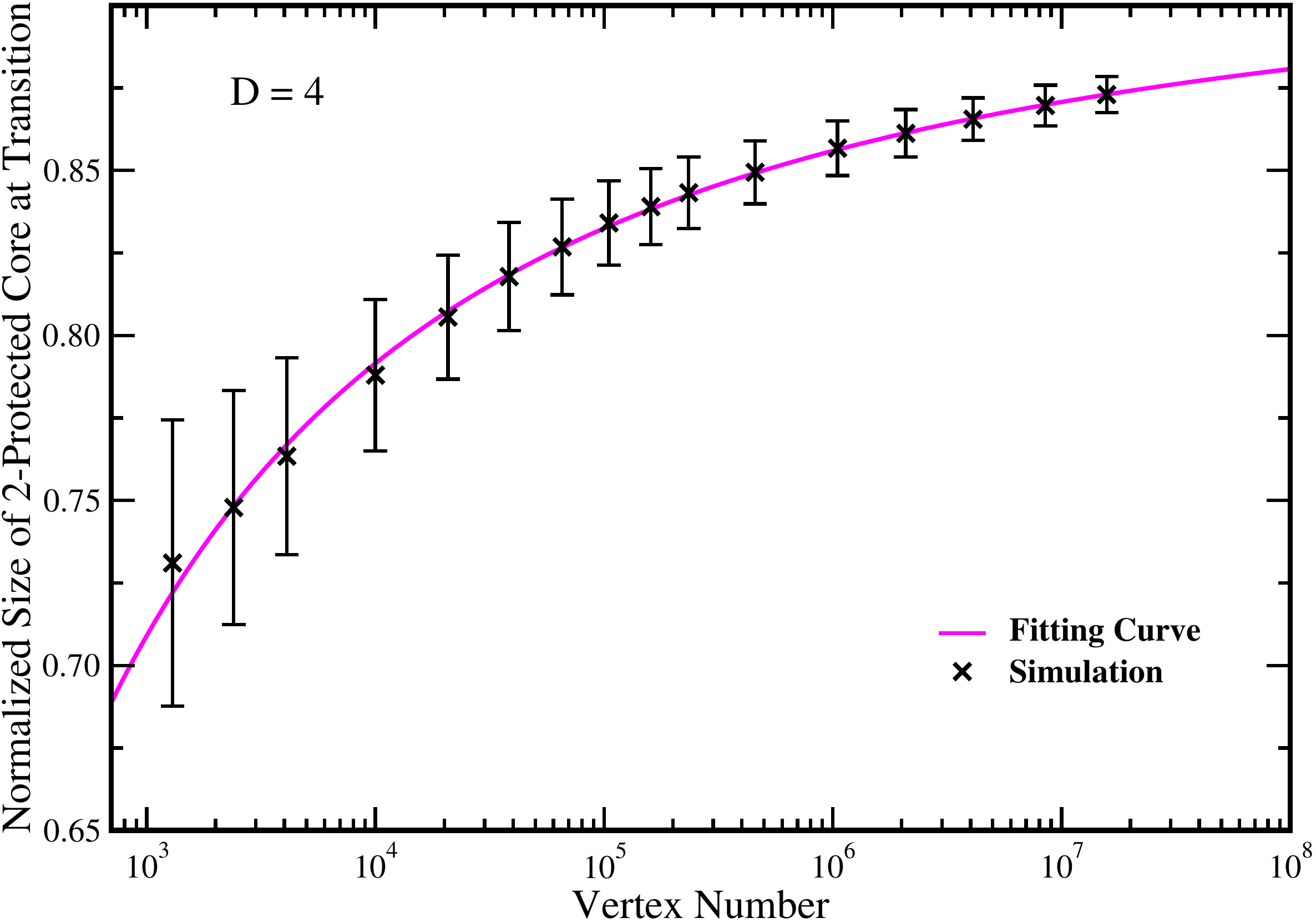}}
  }
\caption{ \label{fig:4D}
{\bf Extrapolation of $2$-protected core percolation transition point to the $N\rightarrow \infty$
limit for $4$-dimensional hypercubic  lattices.}
(a) Average value  $\overline{c^*}$ of the mean degree $c^*$ at
 the $2$-protected core transition.
 (b) Average value $\overline{n_{\rm p-core}^*}$ of
 the normalized $2$-protected core size $n_{\rm p-core}^*$ at the percolation transition.
Each data point is obtained by averaging over
 $1600$ independent network instances.
The fitting cure is  $y = a_1 - b_1/\ln (x)$, with $a_1=4.051 \pm 0.004$ and $b_1=3.63 \pm 0.05$ (a)
 and $y=a_2-b_2/\ln^\eta (x)$ with $a_2=0.919\pm 0.004$, $b_2=6.0\pm 0.9$,
 $\eta = 1.74\pm 0.08$ (b).
}
\end{center}
\end{figure}

\begin{figure}[t]
  \begin{center}
  \subfigure[]{
  \resizebox{85mm}{!}{\includegraphics{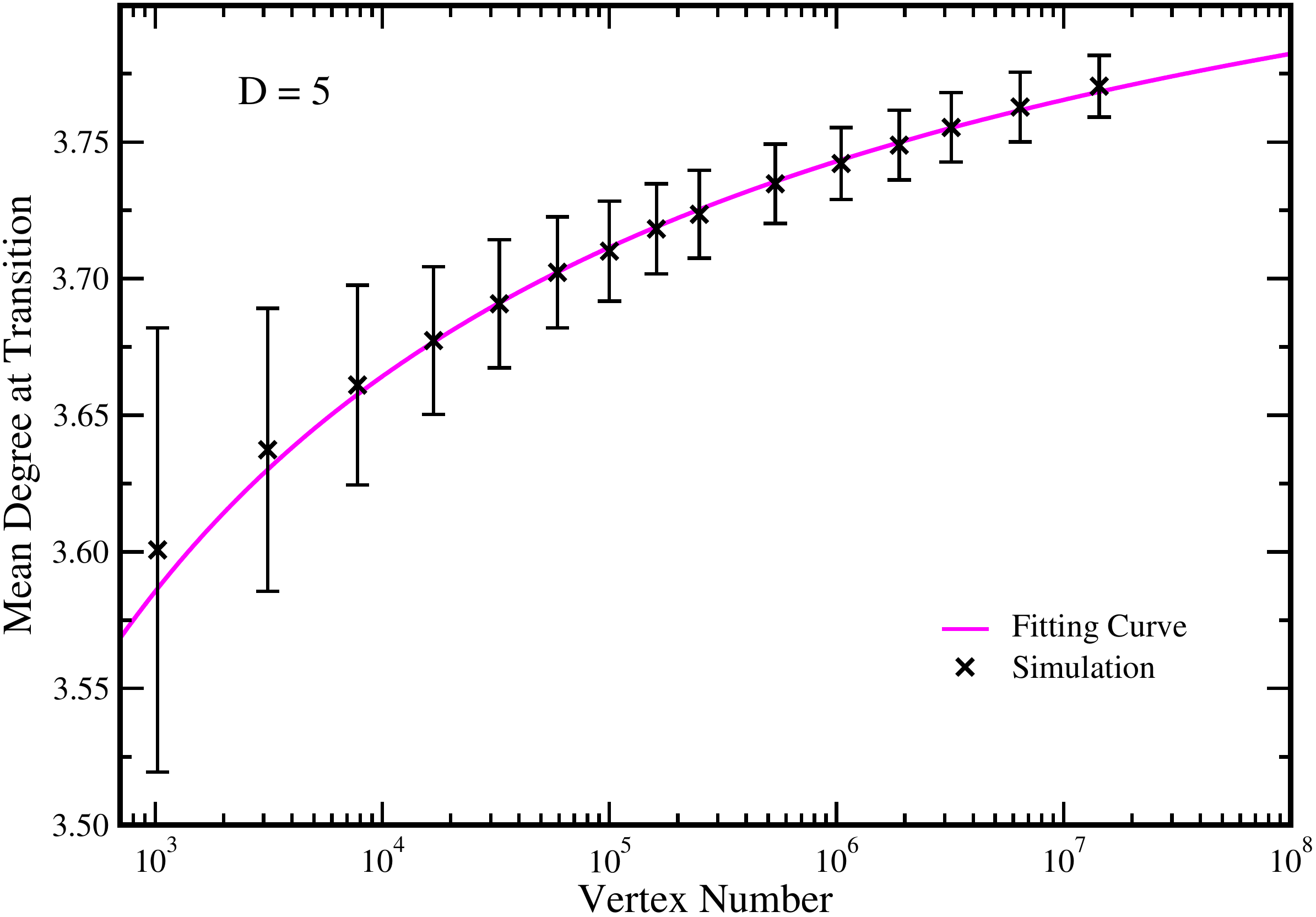}}
  }
%  \hspace{1cm}
  \subfigure[]{
  \resizebox{85mm}{!}{\includegraphics{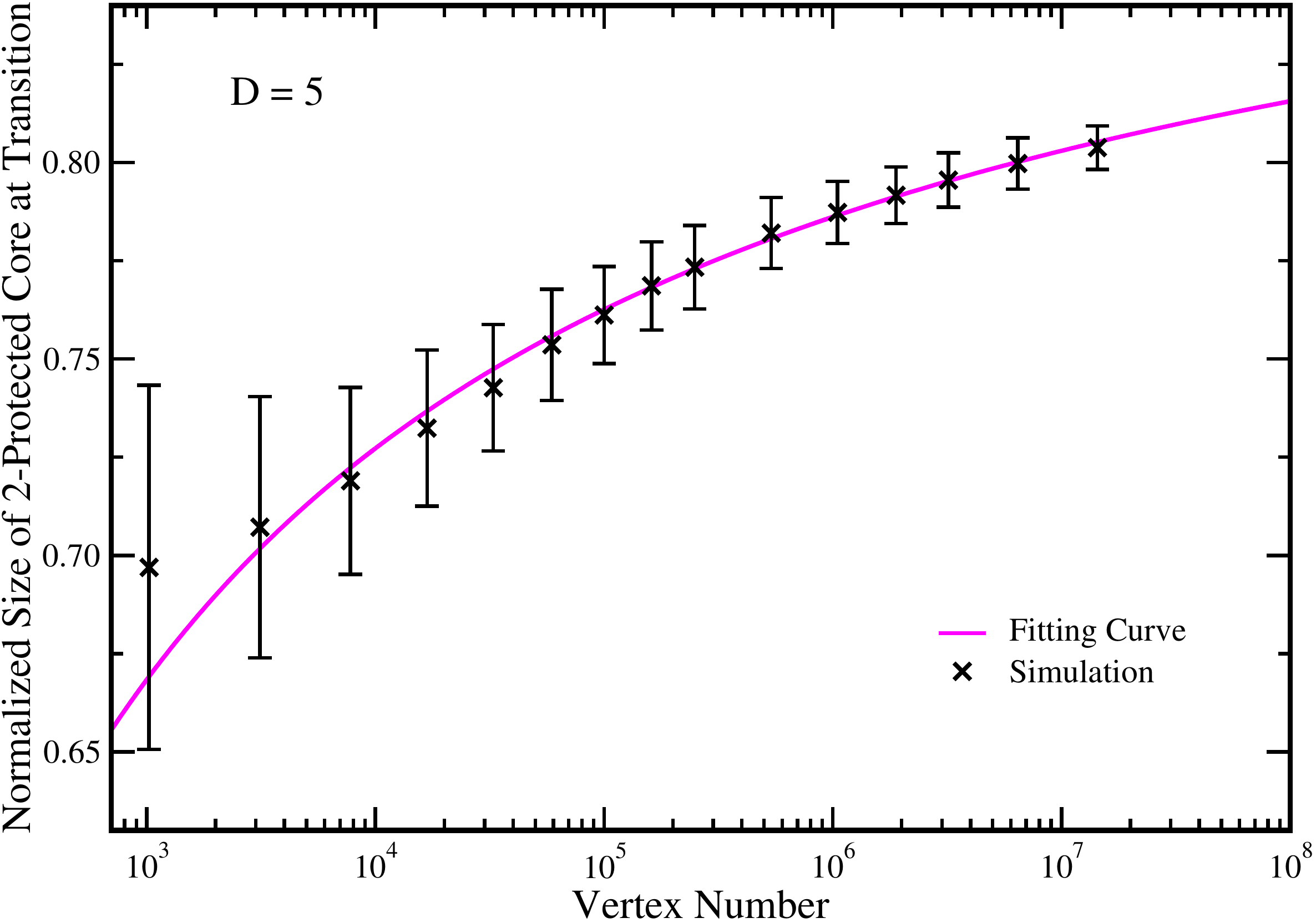}}
  }
\caption{ \label{fig:5D}
{\bf Extrapolation of $2$-protected core percolation transition point to the $N\rightarrow \infty$
limit for $5$-dimensional hypercubic  lattices.}
(a) Average value  $\overline{c^*}$ of the mean degree $c^*$ at
 the $2$-protected core transition.
 (b) Average value $\overline{n_{\rm p-core}^*}$ of
 the normalized $2$-protected core size $n_{\rm p-core}^*$ at the percolation transition.
Each data point is obtained by averaging over
 $1600$ independent network instances.
The fitting cure is  $y = a_1 - b_1/\ln (x)$, with $a_1=3.900 \pm 0.003$ and $b_1=2.17 \pm 0.03$ (a)
 and $y=a_2-b_2/\ln(x)$ with $a_2=0.902\pm 0.003$, $b_2=1.61\pm 0.05$ (b).
}
\end{center}
\end{figure}

\begin{figure}[t]
  \begin{center}
  \subfigure[]{
  \resizebox{85mm}{!}{\includegraphics{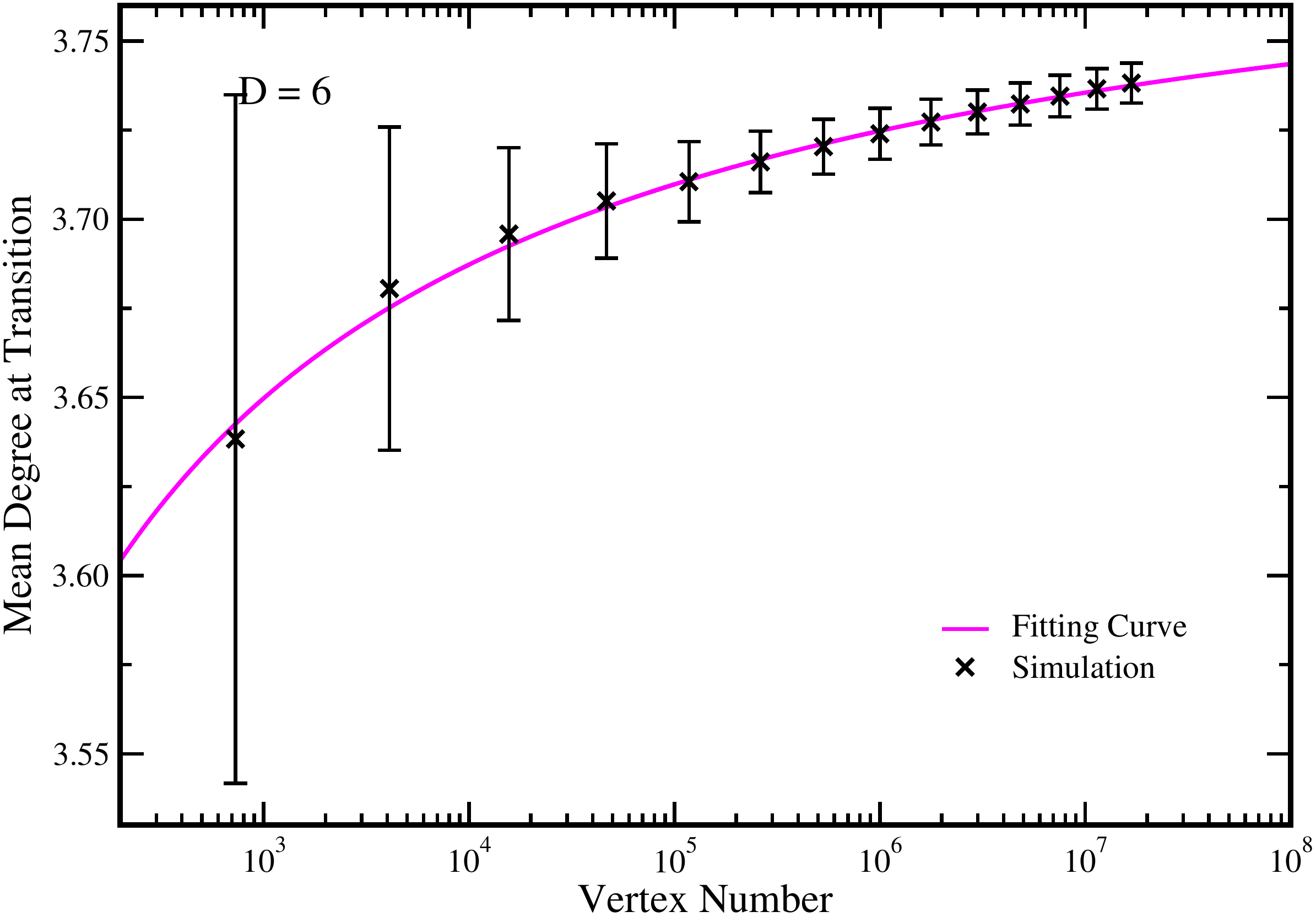}}
  }
%  \hspace{1cm}
  \subfigure[]{
  \resizebox{85mm}{!}{\includegraphics{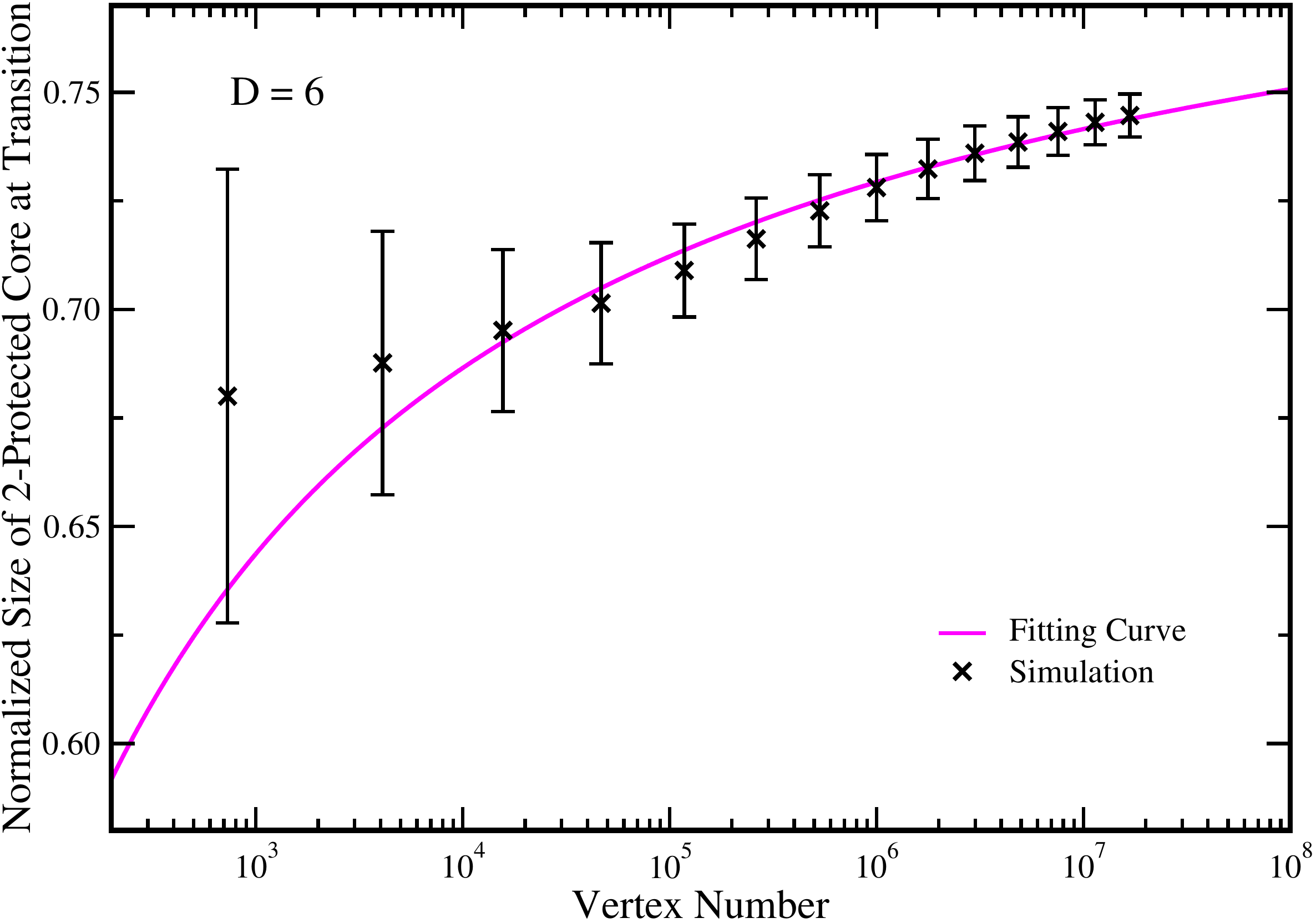}}
  }
\caption{ \label{fig:6D}
{\bf Extrapolation of $2$-protected core percolation transition point to the $N\rightarrow \infty$
limit for $6$-dimensional hypercubic  lattices.}
(a) Average value  $\overline{c^*}$ of the mean degree $c^*$ at
 the $2$-protected core transition.
 (b) Average value $\overline{n_{\rm p-core}^*}$ of
 the normalized $2$-protected core size $n_{\rm p-core}^*$ at the percolation transition.
Each data point is obtained by averaging over
 $1600$ independent network instances.
The fitting cure is  $y = a_1 - b_1/\ln (x)$, with $a_1=3.800 \pm 0.002$ and $b_1=1.04\pm 0.02$ (a)
 and $y=a_2-b_2/\ln(x)$ with $a_2=0.815\pm 0.005$, $b_2=1.18\pm 0.07$
 (b).
}
\end{center}
\end{figure}

\begin{figure}[t]
  \begin{center}
  \subfigure[]{
  \resizebox{85mm}{!}{\includegraphics{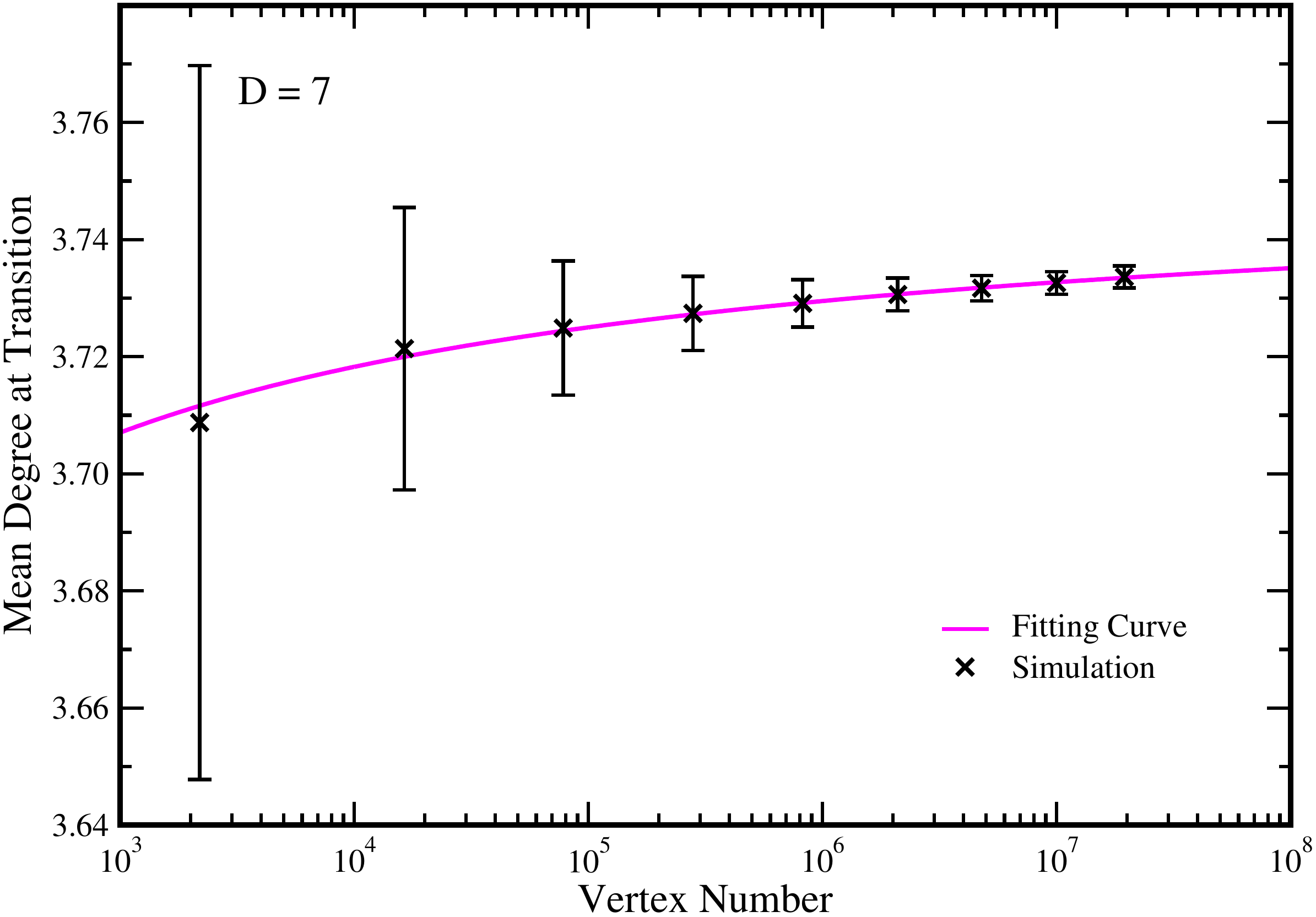}}
  }
 % \hspace{1cm}
  \subfigure[]{
  \resizebox{85mm}{!}{\includegraphics{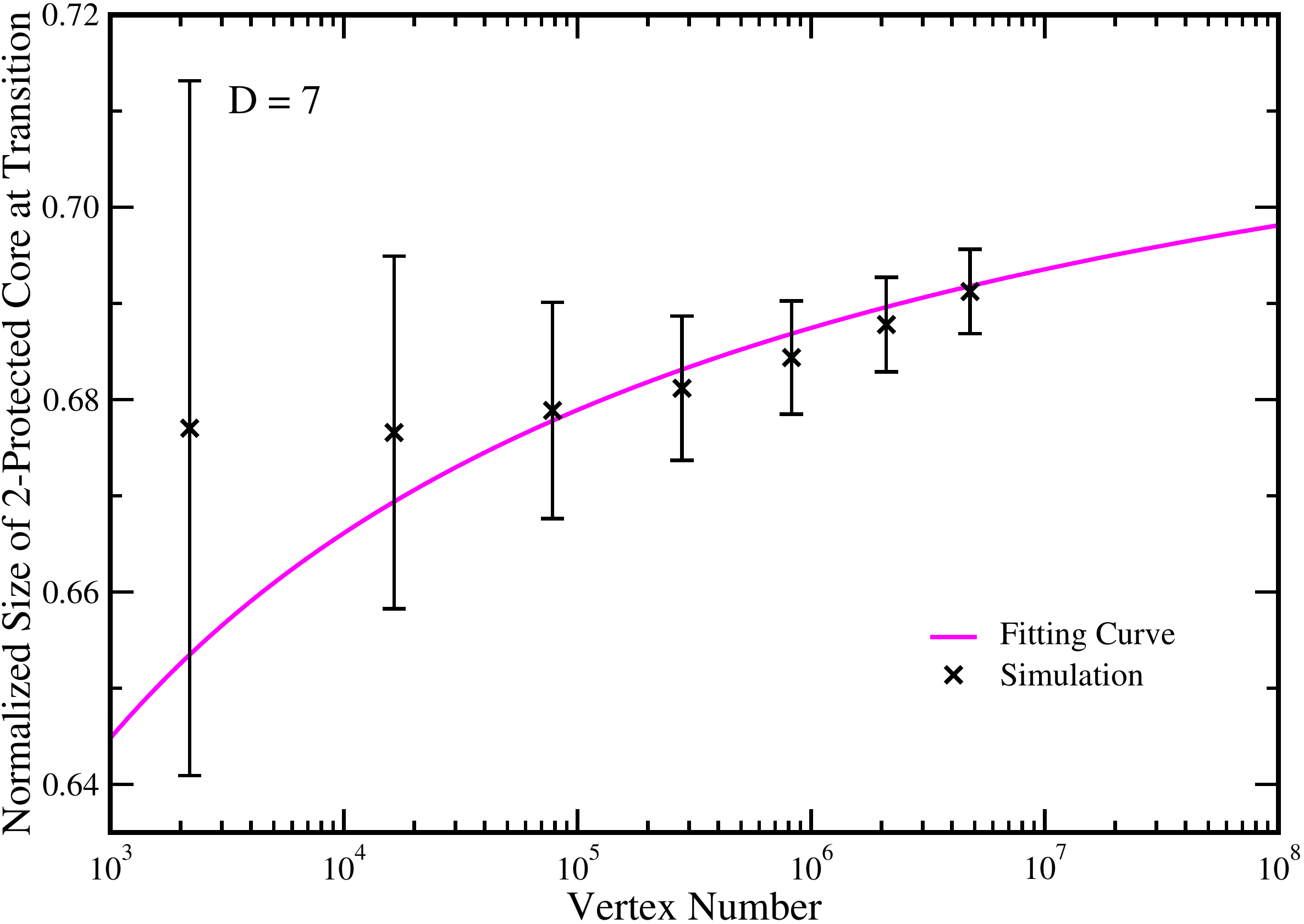}}
  }
\caption{ \label{fig:7D}
{\bf Extrapolation of $2$-protected core percolation transition point to the $N\rightarrow \infty$
limit for $7$-dimensional hypercubic  lattices.}
(a) Average value  $\overline{c^*}$ of the mean degree $c^*$ at
 the $2$-protected core transition.
 (b) Average value $\overline{n_{\rm p-core}^*}$ of
 the normalized $2$-protected core size $n_{\rm p-core}^*$ at the percolation transition.
Each data point is obtained by averaging over
 $1600$ independent network instances.
The fitting cure is
$y = a_1 - b_1/\ln (x)$, with $a_1=3.75197 \pm 0.0006$ and
$b_1=0.3104 \pm 0.0098$ (a)
 and $y=a_2-b_2/\ln(x)$ with $a_2=0.730 \pm 0.006$, $b_2=0.59\pm 0.09$
(b).
}
\end{center}
\end{figure}

\begin{figure}[t]
  \begin{center}
  \subfigure[]{
  \resizebox{85mm}{!}{\includegraphics{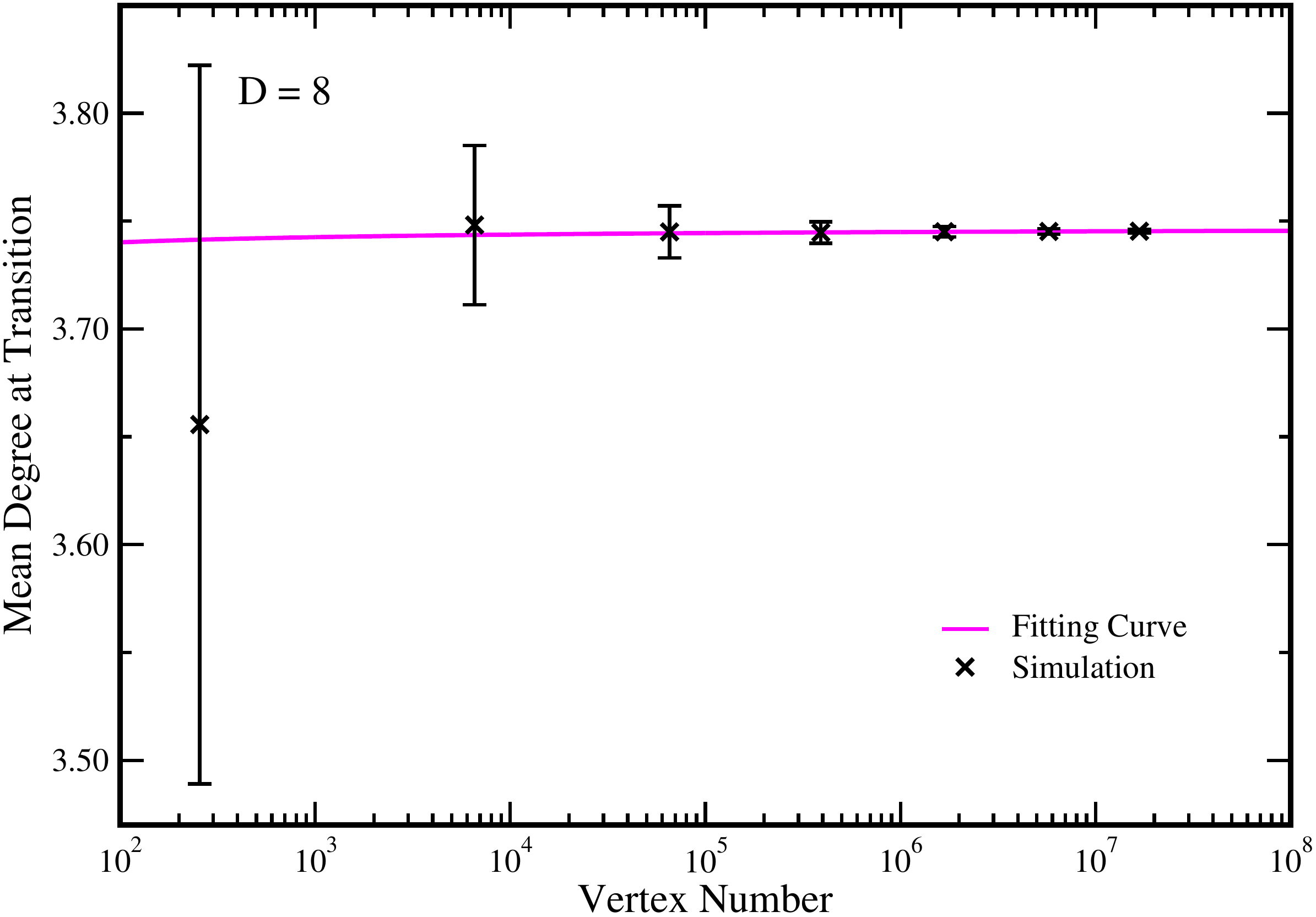}}
  }
%  \hspace{1cm}
  \subfigure[]{
  \resizebox{85mm}{!}{\includegraphics{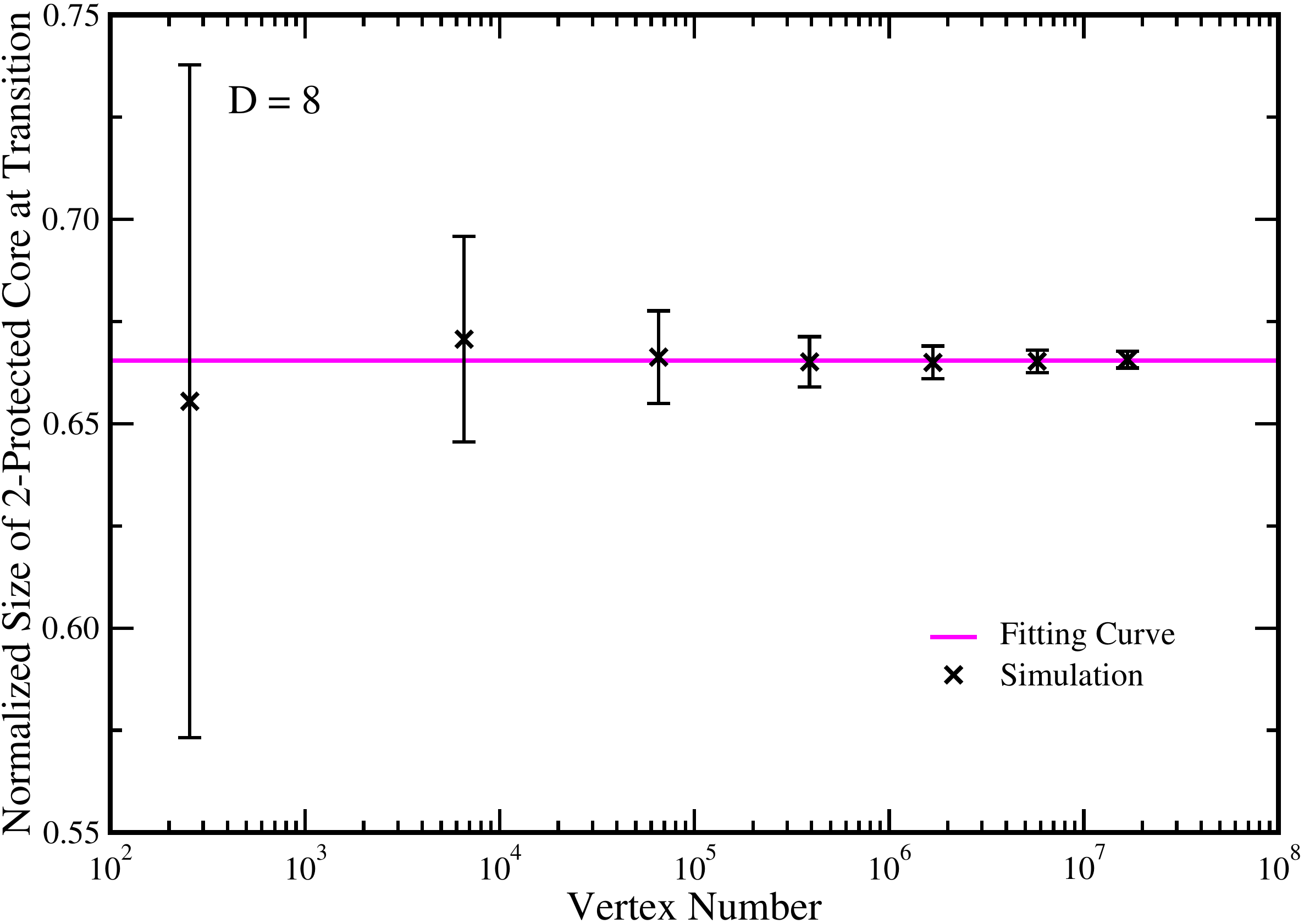}}
  }
\caption{ \label{fig:8D}
{\bf Extrapolation of $2$-protected core percolation transition point to the $N\rightarrow +\infty$
limit for $8$-dimensional hypercubic  lattices.}
(a) Average value  $\overline{c^*}$ of the mean degree $c^*$ at
 the $2$-protected core transition.
 (b) Average value $\overline{n_{\rm p-core}^*}$ of
 the normalized $2$-protected core size $n_{\rm p-core}^*$ at the percolation transition.
Each data point is obtained by averaging over
 $1600$ independent network instances.
The fitting cure is  $y = a_1 - b_1/\ln (x)$, with $a_1=3.747 \pm 0.002$ and $b_1=0.03 \pm 0.04$ (a)
 and $y=a_2$ with $a_2=0.6655 \pm 0.0002$ (b).
}
\end{center}
\end{figure}

\begin{figure}[t]
  \begin{center}
  \subfigure[]{
  \resizebox{85mm}{!}{\includegraphics{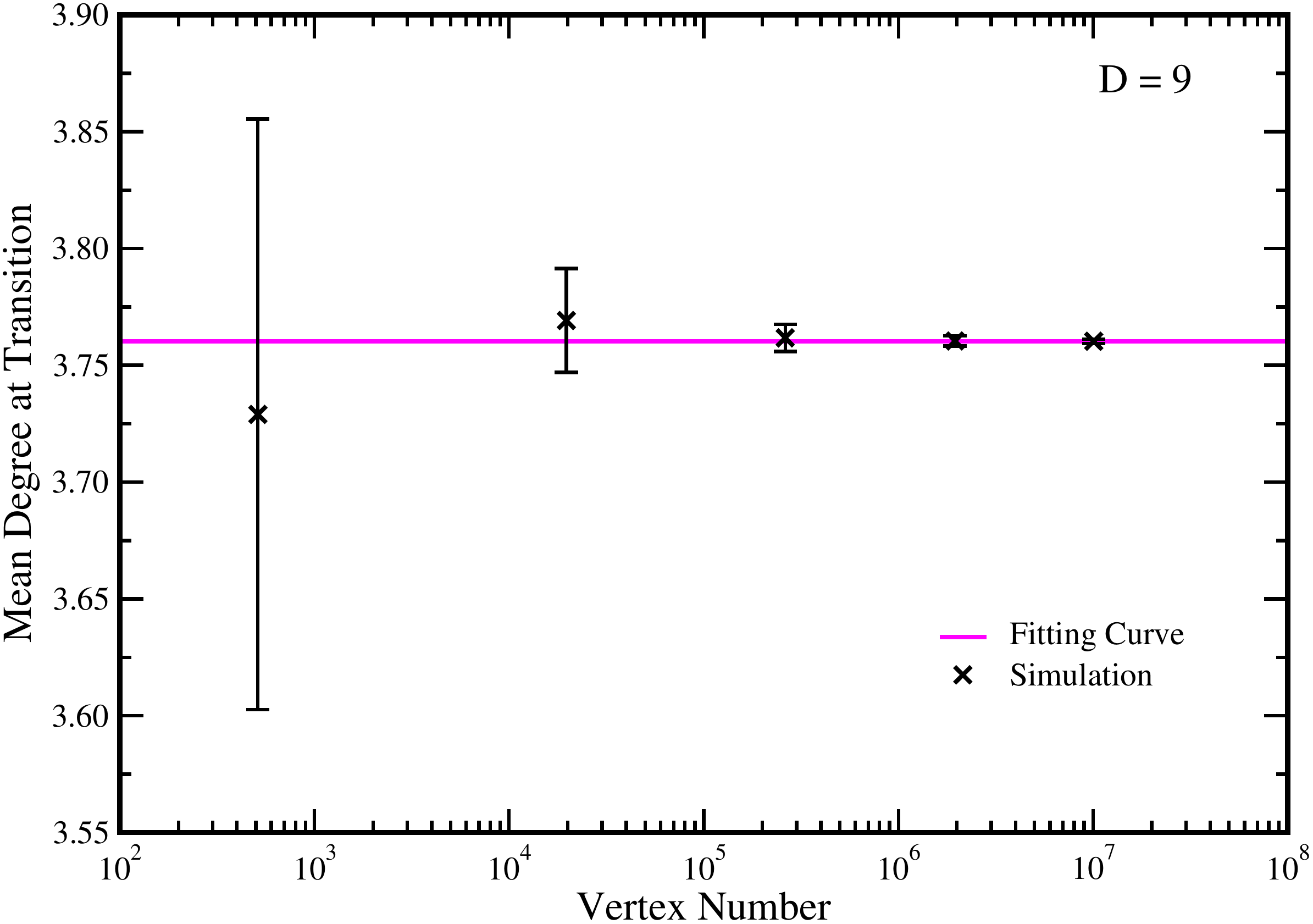}}
  }
 % \hspace{1cm}
  \subfigure[]{
  \resizebox{85mm}{!}{\includegraphics{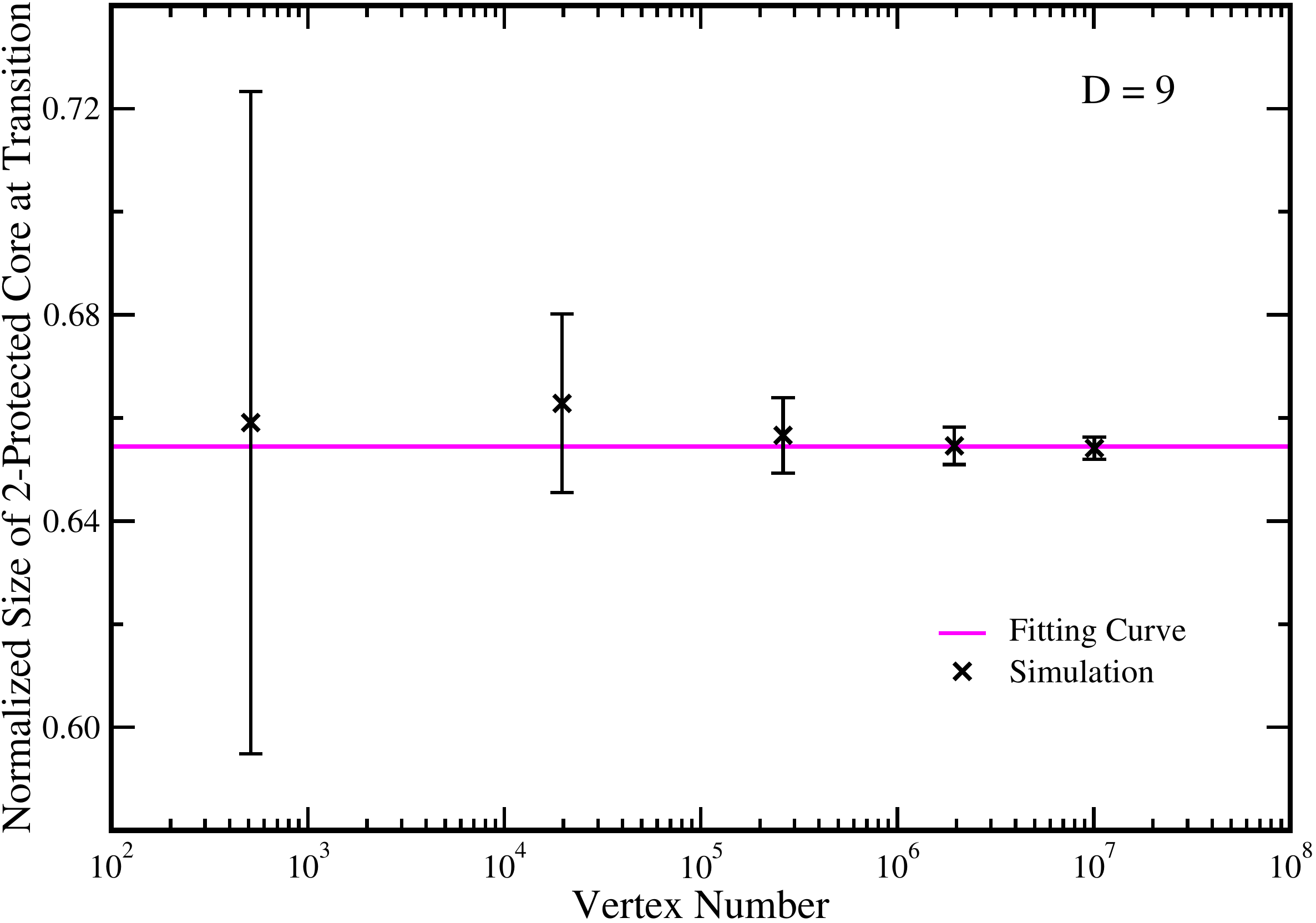}}
  }
\caption{ \label{fig:9D}
{\bf Extrapolation of $2$-protected core percolation transition point to the $N\rightarrow \infty$
limit for $9$-dimensional hypercubic  lattices.}
(a) Average value  $\overline{c^*}$ of the mean degree $c^*$ at
 the $2$-protected core transition.
 (b) Average value $\overline{n_{\rm p-core}^*}$ of
 the normalized $2$-protected core size $n_{\rm p-core}^*$ at the percolation transition.
Each data point is obtained by averaging over
 $1600$ independent network instances.
The fitting cure is  $y = a_1$, with $a_1=3.7604 \pm 0.0002$ (a)
 and $y=a_2$ with $a_2=0.6545\pm 0.0005$ (b).
}
\end{center}
\end{figure}

\begin{figure}[t]
  \begin{center}
  \subfigure[]{
  \resizebox{85mm}{!}{\includegraphics{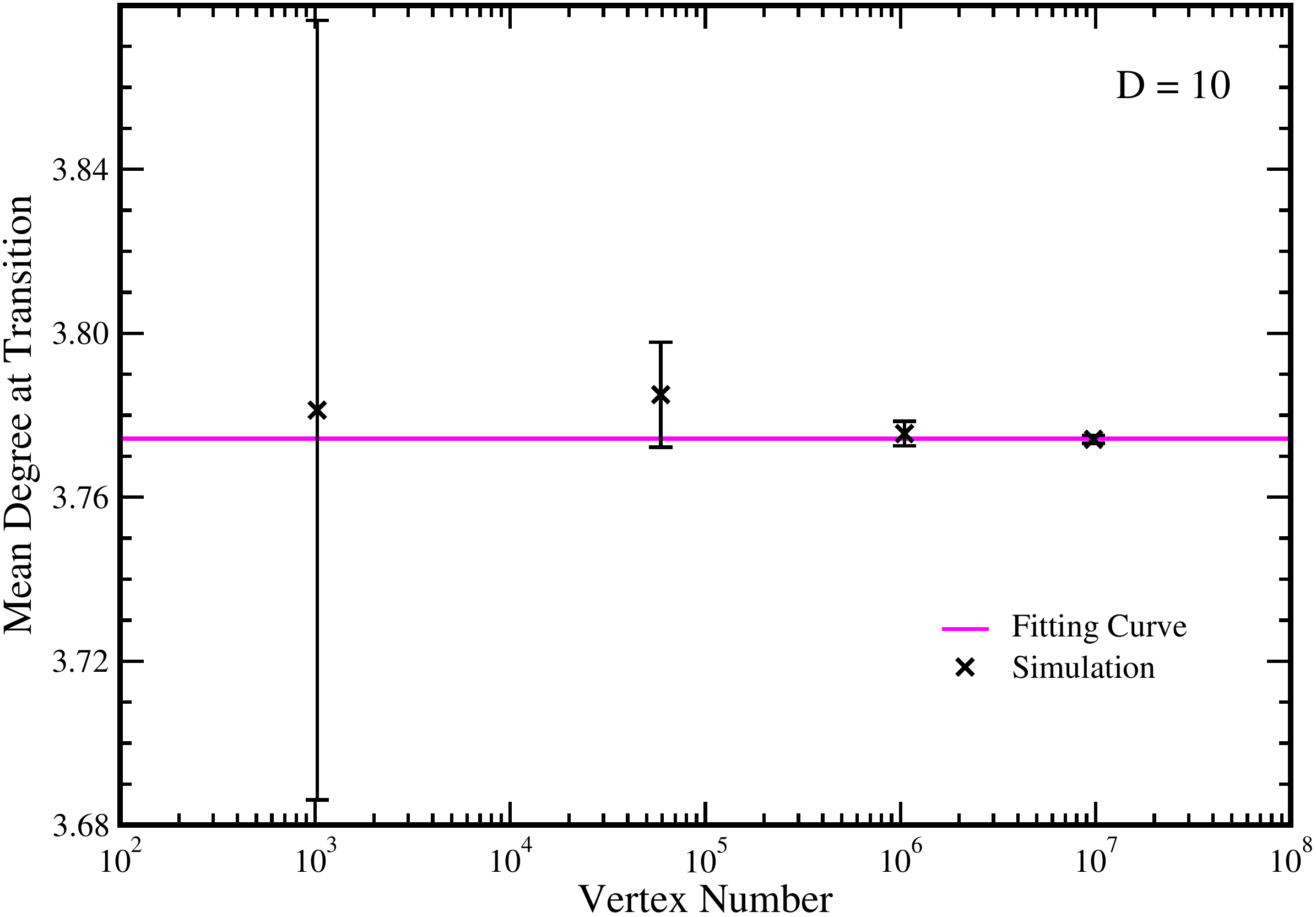}}
  }
%  \hspace{1cm}
  \subfigure[]{
  \resizebox{85mm}{!}{\includegraphics{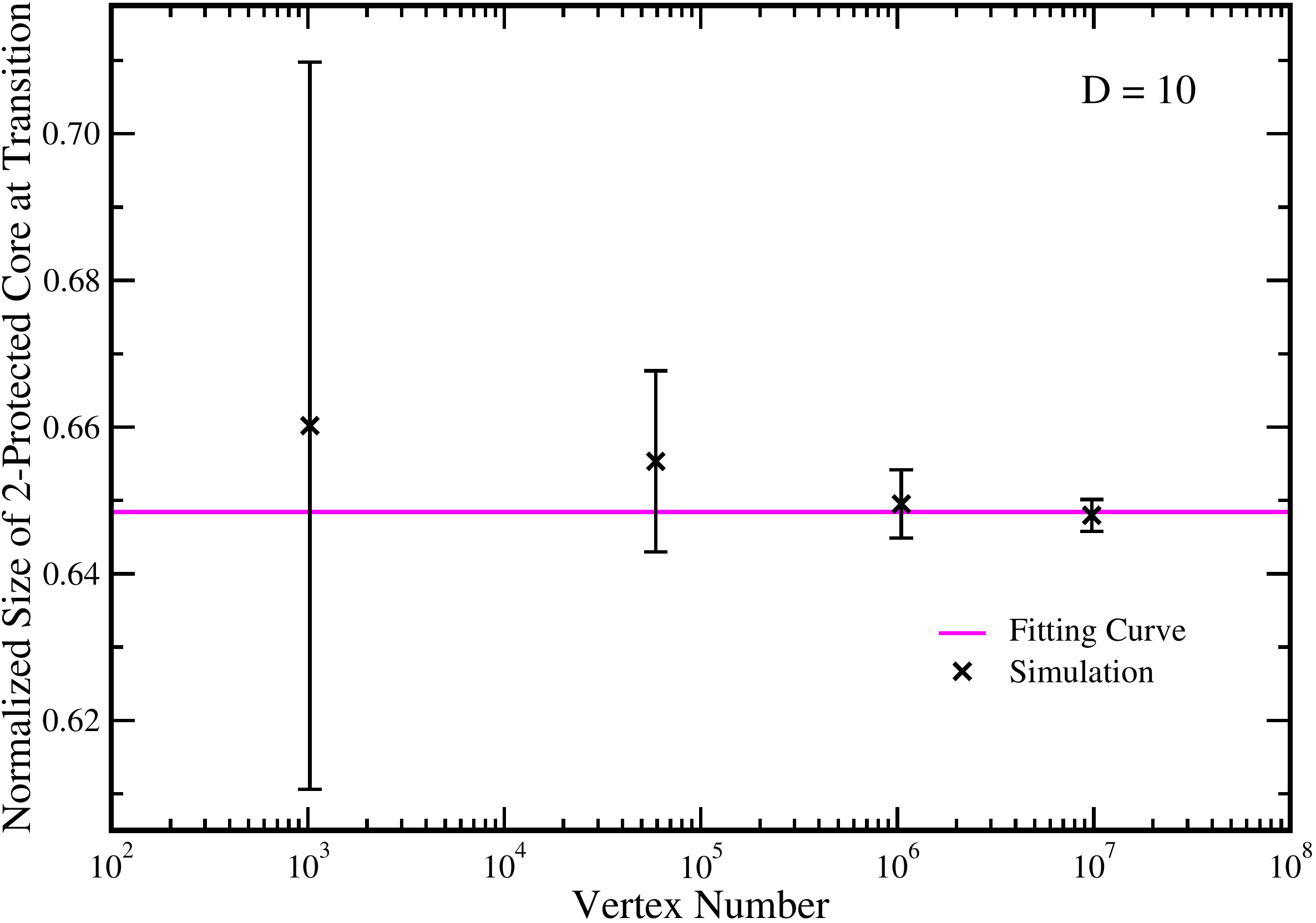}}
  }
\caption{ \label{fig:10D}
{\bf Extrapolation of $2$-protected core percolation transition point to the $N\rightarrow +\infty$
limit for $10$-dimensional hypercubic  lattices.}
(a) Average value  $\overline{c^*}$ of the mean degree $c^*$ at
 the $2$-protected core transition.
 (b) Average value $\overline{n_{\rm p-core}^*}$ of
 the normalized $2$-protected core size $n_{\rm p-core}^*$ at the percolation transition.
Each data point is obtained by averaging over
 $1600$ independent network instances.
The fitting cure is  $y = a_1$, with $a_1=3.7743 \pm 0.0005$ (a)
 and $y=a_2$ with $a_2=0.6484\pm 0.0008$ (b).
}
\end{center}
\end{figure}

The mean value $\overline{c^*}$ of the $2$-protected core percolation transition point $c^*$
is obtained by averaging over $1600$ independent network instances.
The value of $\overline{c^*}$ changes
with network size $N$  for $D \leq 7$ (see Figs.~\ref{fig:2D}--\ref{fig:8D} for the dimensions from
$D=2$ to $D=8$). For $D\leq 8$,
it appears that $\overline{c^*}$ approaches a limiting value $c^*_{\infty}$
as follows
\begin{equation}
\overline{c^*} = c^*_{\infty} - \frac{a}{\ln N} \; ,
\end{equation}
where $a$ is a dimension-dependent constant. From the above fitting formula we obtain the value
of $c^*_{\infty}$ in the thermodynamic limit $N\rightarrow \infty$. As shown in
Fig.~6, the value of $c^*_{\infty}$
is markedly different from the mean value $\overline{c^*}$ of finite systems with $N\approx
2\times 10^6$, especially for dimension $D \leq 6$.
For $D\geq 9$ the average value of the $2$-protected core transition point $c^*$ does not
change much with system size $N$ (see Fig.~\ref{fig:9D} and Fig.~\ref{fig:10D}).
When $D\rightarrow \infty$, we expect the
behaviors of the lattice systems to be the same as
 the random regular networks.

We also extrapolate the average values of $n_{\rm p-core}^*$ at the $2$-protected
core percolation transition point to $N\rightarrow \infty$, see Figs.~\ref{fig:2D}--\ref{fig:10D}.
As shown in Fig.~\ref{fig:Dimension2},
finite-size effects are most significant for $D=2, 3$. As $D$ increases the
results obtained from the finite-dimensional lattice
systems become more and more closer
to the theoretical
predictions obtained from random regular networks.

% figure S13
\noindent
%
%\clearpage

% figure S14
\noindent
%

%\clearpage

% figure S15
\noindent
%

%\clearpage

% figure S16
\noindent
%

%\clearpage

% figure S17
\noindent
%

%\clearpage

% figure S18
\noindent
%

%\clearpage

% figure S19
\noindent
%

%\clearpage

% figure S20
\noindent
%

%\clearpage

% figure S21
\noindent
%

%\clearpage

\clearpage

{\bf Supplementary Note 9}

We test the performance of the mean-field theory on a set of
$37$ real-world networks listed in Tab.~\ref{tab:realnetwork}.
As shown in Tab.~\ref{tab:realnetwork2}, the
normalized $2$-protected core sizes $n_\mm{p\mhyphen core}$
calculated from the theory
are in good agreement with the
empirical results for $32$ of the networks. Such a
good performance is rather surprising to us,
since our mean field theory only uses the degree
distribution $P(k)$ as input and it completely ignores all the
possible higher order correlations (e.g., degree-degree
correlation, clustering,
modularity, etc.) in real-world networks.

We also realize  that the mean-field theory
does not perform very well for metabolic
networks and world-wide web (WWW).
In one data set (the network of hyperlinks between weblogs on politics [67],
our theory predicts large $2$-protected core size,
 while the real network does not have a $2$-protected core at all.
In other four cases (two WWW domain networks
[57, 66] and two metabolic networks [63]), the
theory predicts zero $2$-protected core size. Yet, these
 real-world networks do have $2$-protected cores
 containing $20\%-75\%$ of the nodes.

The above findings raise a fundamental question: Beside the degree distribution of
the network, which other network characteristics also have significant
influences to the size of the protected core? Addressing this and other
related questions deserves a systematic study and we leave it as future work.

\clearpage

{\bf Supplementary References}

\begin{enumerate}[label={[\arabic*]}]
\setcounter{enumi}{44}

\item
 {Sinclair, A.}
\newblock \emph{{Algorithms for Random Generation and Counting:
  a Markov Chain Approach}} ({Birkh\"{a}user},
  {Boston, MA},  {1993}).

\item {Zhou, H.~J.} \& {Lipowsky, R.}
\newblock {Dynamic pattern evolution on scale-free networks}.
\newblock \emph{{Proc. Natl. Acad. Sci. USA}}
  \textbf{{102}}, {10052--10057}
  ({2005}).

\item  {Zhou, H.~J.} \&  {Lipowsky, R.}
\newblock {Activity patterns on random scale-free networks:
  global dynamics arising from local majority rules}.
\newblock \emph{{J. Stat. Mech.: Theor. Exp.}}
  {P01009} ({2007}).

\item
 {King, O.~D.}
\newblock {Comment on "subgraphs in random networks"}.
\newblock \emph{{Phys. Rev. E}} \textbf{{70}},
  {058101} ({2004}).

\item
 {{Klein-Hennig}, H.} \&  {Hartmann, A.~K.}
\newblock {Bias in generation of random graphs}.
\newblock \emph{{Phys. Rev. E}} \textbf{{85}},
  {026101} ({2012}).

\item
 {Bauke, H.} \&  {Mertens, S.}
\newblock {Random numbers for large-scale distributed monte
  carlo simulations}.
\newblock \emph{{Phys. Rev. E}} \textbf{{75}},
  {066701} ({2007}).

\item
 {Balaji, S.},  {Babu, M.~M.},
   {Iyer, L.~M.},  {Luscombe, N.~M.} \&
   {Aravind, L.}
\newblock {Comprehensive analysis of combinatorial regulation
  using the transcriptional regulatory network of yeast}.
\newblock \emph{{J. Mol. Biol.}}
  \textbf{{360}}, {213--227}
  ({2006}).

\item
 {Milo, R.} \emph{et~al.}
\newblock {Network motifs: Simple building blocks of complex
  networks}.
\newblock \emph{{Science}} \textbf{{298}},
  {824--827} ({2002}).

\item
 {{Gama-Castro}, S.} \emph{et~al.}
\newblock {Regulondb (version 6.0): gene regulation model of
  {\em escherichia coli} k-12 beyond transcription, active (experimental)
  annotated promoters and textpresso navigation}.
\newblock \emph{{Nucleic Acids Res.}}
  \textbf{{36}}, {D120--124}
  ({2008}).

\item
 {Norlen, K.},  {Lucas, G.},
   {Gebbie, M.} \&  {Chuang, J.}
\newblock {Eva: Extraction, visualization and analysis of the
  telecommunications and media ownership network}.
\newblock In \emph{{Proceedings of International
  Telecommunications Society 14th Biennial Conference}}
  ({Seoul, Korea},  {2002}).

\item
 {{van Duijn}, M. A.~J.},  {Zeggelink, E. P.~H.},
   {Huisman, M.},  {Stokman, F.~N.} \&
   {Wasseur, F.~W.}
\newblock {Evolution of sociology freshmen into a friendship
  network}.
\newblock \emph{{J. Math. Sociol.}}
  \textbf{{27}}, {153--191}
  ({2003}).

\item
 {Milo, R.} \emph{et~al.}
\newblock {Superfamilies of evolved and designed networks}.
\newblock \emph{{Science}} \textbf{{503}},
  {1538--1542} ({2004}).

\item
 {Leskovec, J.},  {Lang, K.~J.},
   {Dasgupta, A.} \&  {Mahoney, M.~W.}
\newblock {Community structure in large networks: Natural
  cluster sizes and the absence of large well-defined clusters}.
\newblock {arXiv:0810.1355} ({2008}).

\item
 {Richardson, M.},  {Agrawal, R.} \&
   {Domingos, P.}
\newblock {Trust management for the semantic web}.
\newblock \emph{{Lect. Notes Comput. Sci.}}
  \textbf{{2870}}, {351--368}
  ({2003}).

\item
 {Dunne, J.~A.},  {Williams, R.~J.} \&
   {Martinez, N.~D.}
\newblock {Food-web structure and network theory: The role of
  connectance and size}.
\newblock \emph{{Proc. Natl. Acad. Sci. USA}}
  \textbf{{99}}, {12917--12922}
  ({2002}).

\item
 {Martinez, N.~D.}
\newblock {Artifacts or attributes? effects of resolution on the
  little rock lake food web}.
\newblock \emph{{Ecol. Monographs}}
  \textbf{{61}}, {367--392}
  ({1991}).

\item
 {Christian, R.~R.} \&  {Luczkovich, J.~J.}
\newblock {Organizing and understanding a winter's seagrass
  foodweb network through effective trophic levels}.
\newblock \emph{{Ecol. Modelling}}
  \textbf{{117}}, {99--124}
  ({1999}).

\item
 {Bianconi, G.},  {Gulbahce, N.} \&
   {Motter, A.~E.}
\newblock {Local structure of directed networks}.
\newblock \emph{{Phys. Rev. Lett.}}
  \textbf{{100}}, {118701}
  ({2008}).

\item
 {Jeong, H.},  {Tombor, B.},
   {Albert, R.},  {Oltvai, Z.~N.} \&
   {Barab\'{a}si, A.-L.}
\newblock {The large-scale organization of metabolic networks}.
\newblock \emph{{Nature}} \textbf{{407}},
  {651--654} ({2000}).

\item
 {Watts, D.~J.} \&  {Strogatz, S.~H.}
\newblock {Collective dynamics of 'small-world' netowrks}.
\newblock \emph{{Nature}} \textbf{{393}},
  {440--442} ({1998}).

\item
 {Leskovec, J.},  {Kleinberg, J.} \&
   {Faloutsos, C.}
\newblock {Graphs over time: densification laws, shrinking
  diameters and possible explanations}.
\newblock In \emph{{Proceedings of the eleventh ACM SIGKDD
  international conference on Knowledge discovery in data mining}},
  {177--187} ({ACM, New York},
   {2005}).

\item
 {Albert, R.},  {Jeong, H.} \&
   {Barab\'{a}si, A.-L.}
\newblock {Internet: Diameter of the world-wide web}.
\newblock \emph{{Nature}} \textbf{{401}},
  {130--131} ({1999}).

\item
 {Adamic, L.~A.} \&  {Glance, N.}
\newblock {The political blogosphere and the 2004 u.s. election:
  divided they blog}.
\newblock In \emph{{Proceedings of the 3rd international
  workshop on Link discovery}}, {36--43}
  ({ACM, New York},  {2005}).

\item
 {Leskovec, J.},  {Kleinberg, J.} \&
   {Faloutsos, C.}
\newblock {Graph evolution: Densification and shrinking
  diameters}.
\newblock \emph{{ACM Transactions on Knowledge Discovery from
  Data}} \textbf{{1}}, {2}
  ({2007}).

\item
 {Opsahl, T.} \&  {Panzarasa, P.}
\newblock {Clustering in weighted networks}.
\newblock \emph{{Social Networks}}
  \textbf{{31}}, {155--163}
  ({2009}).

\item
 {Eckmann, J.-P.},  {Moses, E.} \&
   {Sergi, D.}
\newblock {Entropy of dialogues creates coherent structures in
  e-mail traffic}.
\newblock \emph{{Proc. Natl. Acad. Sci. USA}}
  \textbf{{101}}, {14333--14337}
  ({2004}).

\item
 {Song, C.},  {Qu, Z.},  {Blumm,
  N.} \&  {Barab\'{a}si, A.-L.}
\newblock {Limits of predictability in human mobility}.
\newblock \emph{{Science}} \textbf{{327}},
  {1018--1021} ({2010}).

\item
 {Freeman, S.} \&  {Freeman, L.}
\newblock \emph{Social Science Research Reports 46 (University
  of California, Irvine, CA)} ({1979}).

\item
 {Cross, R.} \&  {Parker, A.}
\newblock \emph{The Hidden Power of Social Networks}
  (Harvard Business School Press,
  Boston, MA, 2004).

\end{enumerate}

\end{document}